\documentclass[12pt, draftclsnofoot, onecolumn]{IEEEtran}
\usepackage{amsmath}
\usepackage{amssymb}
\usepackage{amsfonts}

\newcommand{\Rmnum}[1]{\expandafter\@slowromancap\romannumeral #1@}
\usepackage{booktabs}
\usepackage{graphicx}
\usepackage{subfigure}
\usepackage{epsfig}
\usepackage{cite}
\usepackage{xcolor}
\usepackage[shortlabels]{enumitem}
\usepackage{extarrows}
\usepackage[ruled,linesnumbered,lined]{algorithm2e}
\usepackage{algorithmic}
\usepackage{tensor}
\usepackage{mathtools}
\usepackage{bbm}
\usepackage[colorlinks,linkcolor=blue,anchorcolor=blue,citecolor=blue,urlcolor=black]{hyperref}
\begin{document}
\title{Federated Learning over Wireless {\color{black}Device-to-Device} Networks: Algorithms and Convergence Analysis}
\author{Hong Xing, Osvaldo Simeone, and Suzhi Bi
\thanks{Part of this paper has been presented at the IEEE International Workshop on Signal Processing Advances in Wireless Communications (SPAWC), May 2020 \cite{xing20FL}.}
\thanks{H. Xing and S. Bi are with the College of Electronic and Information Engineering, Shenzhen University, Shenzhen, 518060, China (e-mails: \{hong.xing,~bsz\}@szu.edu.cn).}
\thanks{O. Simeone is with the King’s Communications, Learning, and Information Processing (KCLIP) lab, Department of Engineering, King's College London, London, WC2R 2LS, U.K. (e-mail: osvaldo.simeone@kcl.ac.uk).}
}

\maketitle
\vspace{-.70in}
\begin{abstract}
The proliferation of Internet-of-Things (IoT) devices and cloud-computing applications over siloed data centers is motivating renewed interest in the collaborative training of a shared model by multiple individual clients via federated learning (FL). To improve the communication efficiency of FL implementations in wireless systems, recent works have proposed compression and dimension reduction mechanisms, along with digital and analog transmission schemes that account for channel noise, fading, and interference. The prior art has {\color{black}mainly} focused on star topologies consisting of distributed clients and a central server. In contrast, this paper studies FL over wireless device-to-device (D2D) networks by providing theoretical insights into the performance of digital and analog implementations of decentralized stochastic gradient descent (DSGD). First, we introduce generic digital and analog wireless implementations of communication-efficient DSGD algorithms, leveraging random linear coding (RLC) for compression and over-the-air computation (AirComp) for simultaneous analog transmissions. Next, under the assumptions of convexity and connectivity, we provide convergence bounds for both implementations. The results demonstrate the dependence of the optimality gap on the connectivity and on the signal-to-noise ratio (SNR) levels in the network. The analysis is corroborated by experiments on an image-classification task.
\end{abstract}


\IEEEpeerreviewmaketitle
\newtheorem{definition}{\underline{Definition}}[section]
\newtheorem{fact}{Fact}
\newtheorem{assumption}{Assumption}[section]
\newtheorem{theorem}{Theorem}[section]
\newtheorem{lemma}{Lemma}[section]
\newtheorem{proposition}{\underline{Proposition}}[section]
\newtheorem{corollary}[theorem]{\underline{Corollary}}
\newtheorem{example}{\underline{Example}}[section]
\newtheorem{remark}{Remark}[section]
\newtheorem{proof}{\underline{Proof}}
\newcommand{\mv}[1]{\mbox{\boldmath{$ #1 $}}}
\newcommand{\mb}[1]{\mathbb{#1}}
\newcommand{\mc}[1]{\mathcal{#1}}
\newcommand{\tr}{{\sf Tr}}
\newcommand{\Myfrac}[2]{\ensuremath{#1\mathord{\left/\right.\kern-\nulldelimiterspace}#2}}
\DeclarePairedDelimiter{\diagfences}{(}{)}
\newcommand{\diag}{\operatorname{diag}\diagfences}
\newcommand\Perms[2]{\tensor[^{#2}]P{_{#1}}}
\newcommand{\bigO}{\mathcal{O}}

\section{Introduction}\label{sec:Introduction}
With the proliferation of Internet-of-Things (IoT) {\color{black}devices} and cloud-computing applications over siloed data centres, {\em distributed learning} has become a critical enabler for artificial intelligence (AI) solutions  \cite{bekkerman11scaling,chang20distributed}. In distributed learning, multiple agents collaboratively train a machine learning model {\color{black}via the exchange of training data, model parameters and/or gradient vectors} over geographically distributed computing resources and data. {\em Federated learning (FL)} refers to distributed learning protocols that do not directly exchange the training data in an attempt to reduce the communication load and to limit privacy concerns \cite{kairouz19advances,li20challenges,zhou19EI,zhu20EI}. In conventional FL, multiple clients train a shared model by exchanging model-related parameters with a central node. This class of protocols hence relies on a  parameter-server (PS) architecture, which is typically realized in wireless settings via a base station-centric network topology \cite{alistarh17qsgd,bernstein18signSGD,abdi20analog,wu18error,basu19qsparse}.

{\color{black}There are important scenarios when there is no central coordinator acting as the parameter server (PS) for FL, and therefore distributed learning must rely on a peer-to-peer communication topology that encompasses device-to-device (D2D) links among individual learning agents over an arbitrary connectivity graph. For example, reference \cite{roy19BrainTorrent} demonstrated the effectiveness of a D2D FL framework across siloed medical centers. There are also other scenarios when PS-based architecture is undesirable due to coverage, privacy, implementation efficiency, or fault-tolerance considerations, and thus a D2D architecture becomes a preferred option thanks to its resilience and increased parallelism. An example is the ring AllReduce architecture for deep learning (DL) described in \cite{baidu17allreduce} that enables efficient GPU-and-GPU communications.}

\vspace{-.20in}
\subsection{Related Work}\label{subsec:Related Work}
The problem of alleviating the communication load in FL systems has been widely investigated, mostly under the assumption of noiseless, rate-limited links, and {\em star topologies}. Key elements of these solutions are compression and dimension-reduction operations that map the original model parameters or gradient vectors into representations defined by a limited number of bits and/or sparsity. Important classes of solutions include unbiased compressors (e.g., \cite{alistarh17qsgd,bernstein18signSGD,abdi20analog}) and biased compressors with error-feedback mechanisms (e.g., \cite{wu18error,basu19qsparse,amiri20fading}).

In a {\em D2D architecture}, devices can only exchange information with their respective neighbors, making consensus mechanisms essential to ensure agreement towards the common learning goal \cite{nedic18topology}. A well-known protocol integrating stochastic gradient (SGD) and consensus is {\em Decentralized Stochastic Gradient Descent (DSGD)} \cite{xin19decentralized,sun21DeFedAvg}. Similar to FL in star topologies, there have been lots of previous efforts aiming for accelerating consensus and removing communication overhead of the DSGD from algorithmic perspectives by, e.g., variance-reduction for large data heterogeneity among agents \cite{tang18Dsquare} and by compression \cite{koloskova19decentralized,koloskova20arbitrary,singh19sparq-sgd,vogels20low-rank}. The CHOCO-SGD algorithm proposed in \cite{koloskova19decentralized,koloskova20arbitrary,singh19sparq-sgd}, which combines the standard DSGD algorithm with biased compression, was studied for strongly convex and smooth objectives in \cite{koloskova19decentralized}, and for non-convex smooth objectives in \cite{koloskova20arbitrary}, combined with event-triggered protocols in \cite{singh19sparq-sgd}.
{\color{black}The authors in \cite{vogels20low-rank} introduced a compression scheme for model exchange between neighboring nodes that improves convergence while requiring no additional hyperparameters. All these prior works on decentralized FL assume either ideal or rate-limited {\em noiseless} D2D communications. Wireless implementations of these algorithms are non-trivial in general, because the goal of achieving consensus is compromised to different extents by channel impairments, such as fading, or packet losses.}

A large number of recent works have proposed communication strategies and multi-access protocols for {\em FL in wireless star topologies} \cite{zhu20broadband,cao21feel,chen19joint,elgabi21scalable}. At the physical layer, {\em over-the-air computation (AirComp)} was investigated in \cite{amiri20fading,ahn19distillation,zhu20broadband,cao21feel,guo21AGA} as a promising solution to support simultaneous transmissions by leveraging the waveform superposition property of the wireless medium. Unlike conventional digital communication over orthogonal transmission blocks, AirComp is based on analog, e.g., uncoded, transmission, which enables the estimate of aggregated statistics directly from the received baseband samples. This reduces the communication burden, relieving the network from the need to decode individual information separately for all participating devices. For example, the authors in \cite{guo21AGA} proposed an adaptive learning-rate scheduler and investigated the convergence of the resulting protocol.

The literature on {\em decentralized FL in wireless D2D architecture} is, in contrast, still quite limited. A DSGD based algorithm termed MATCHA was proposed in \cite{wang19MATCHA} by accounting for interference among nearby links. By sampling a matching decomposition of an interference graph, MATCHA schedules non-interfering communication links in parallel, among which the connectivity-critical links are activated with a higher probability. However, no attempt was made to take physical layer transmissions into account. {\color{black} A real-time implementation of decentralized FL systems was proposed in \cite{savazzi20industry} over industrial wireless networks, and the joint effects of model pruning, sparisification, and quantization were considered. The wireless transmission constraints under study were only suitable for digital transmissions. References \cite{ozfatura20decentralized} and \cite{shi21decentralized} both considered AirComp-based DSGD for decentralized FL over D2D networks, where precoding/decoding strategies for analog D2D transmissions, possibly combined with interference-free D2D scheduling policy, were developed. Compared with \cite{ozfatura20decentralized}, the work \cite{shi21decentralized} proved that the original performance of DSGD with gradient tracking is compromised by an error floor term due to the channel noise introduced by AirComp, while relying on the strong assumption that the intermediate information broadcast by the center of a ``star'' sub-graph can be ideally received by its neighbors. Furthermore, reference \cite{saha21decentralized} provided a theoretical analysis for the analog implementation of another fully decentralized optimization algorithm, {\it decentralized lazy mirror decent (DLMD)}, which accommodates convex non-smooth loss functions under channel noise and rate constraints. However, since the convergence rate is conditioned on an increasing power sequence over training iterations, the performance guarantee may be seriously compromised in power-constrained wireless D2D networks. In addition, how the proposed analog implementation of the DLMD scheme addresses the mismatch between the available channel uses and the typically larger dimension of the model has yet been investigated.}

\vspace{-.20in}
\subsection{Main Contributions}
{\color{black}This work investigates the impact of wireless communication constraints induced by blockages, pathloss, channel noise and fading on the convergence of DSGD-based FL algorithms for digital and analog transmission implementations. The analysis applies to general scheduling and power allocation policies, with the CHOCO-SGD chosen as the baseline DSGD algorithm due to its generality and flexibility in the choice of the compression operators. The main contributions are as follows:} 

1) We study general digital and analog wireless implementations of  DSGD algorithms that rely on dimension reduction-based compression via {\em random linear coding (RLC)} and enable broadcasting for digital transmission, as well as both broadcasting and AirComp for analog transmission. 

2) Under the assumptions of convexity and connectivity, we derive {\em convergence bounds} for the mentioned general class of digital wireless implementations that demonstrate the dependence of the optimality gap on the connectivity of the graph and on the {\color{black}model} estimation error due to compression. 

3) We also provide {\em convergence bounds} for the mentioned general class of analog wireless implementations of DSGD that quantify the impact of topology and channel noise. The analysis also reveals the role played by an {\em adaptive} consensus step size in combating the effect of accumulative channel noise in the analog implementation. To the best of our knowledge, this is the first time that an adaptive consensus step size is shown to be beneficial for convergence. 

4) We provide numerical experiments for image classification, confirming the benefits of the proposed adaptive consensus rate, and demonstrating the agreement between  analytical and empirical results.  

The conference version \cite{xing20FL} of this paper provided some preliminary experimental results on digital and analog DSGD, without offering any theoretical results.

The remainder of this paper is organized as follows. The system model is presented in Section \ref{sec:System Model}. Digital and analog transmission protocols are introduced in Section \ref{sec:Digital and Analog Transmisison Protocols}. The convergence analysis for both implementations is presented in Section \ref{sec:Convergence Analysis for Digital Transmission} and Section \ref{sec:Convergence Analysis For Analog Transmission}, respectively. Numerical performance results are described in Section \ref{sec:Numerical Experiments}, followed by conclusions in Section \ref{sec:Conclusions}.
 
\subsection{Notations}\label{subsec:Notations}
\indent We use the upper case boldface letters for matrices and lower case boldface letters for vectors. We also use $\|\cdot\|$ to denote the Euclidean norm of a vector or the spectral norm of a matrix, and $\|\cdot\|_F$ to denote the Frobenius norm of a matrix. {\color{black} We denote by $\vert\mc{V}\vert$ the cardinality of a set $\mc{V}$. The average of vectors \(\mv x_i\) over  \(i\in\mc{V}\) is defined as \(\bar{\mv x}=\tfrac{1}{\vert\mc{V}\vert}\sum_{i\in\mc{V}}\mv x_i\).} Notations $\tr(\cdot)$ and $(\cdot)^{T}$ denote the trace and the transpose of a matrix, respectively. $\mathbb{E}[\cdot]$ stands for the statistical expectation of a random variable. $\mv{I}$ represents an identity matrix with appropriate size, and \(\triangleq\) indicates a mathematical definition. \(\lambda_i(\cdot)\) denotes the $i$th largest eigenvalue of a matrix.

\section{System Model}\label{sec:System Model} 
\begin{figure}[htp]
	\centering
	\includegraphics[width=3.5in]{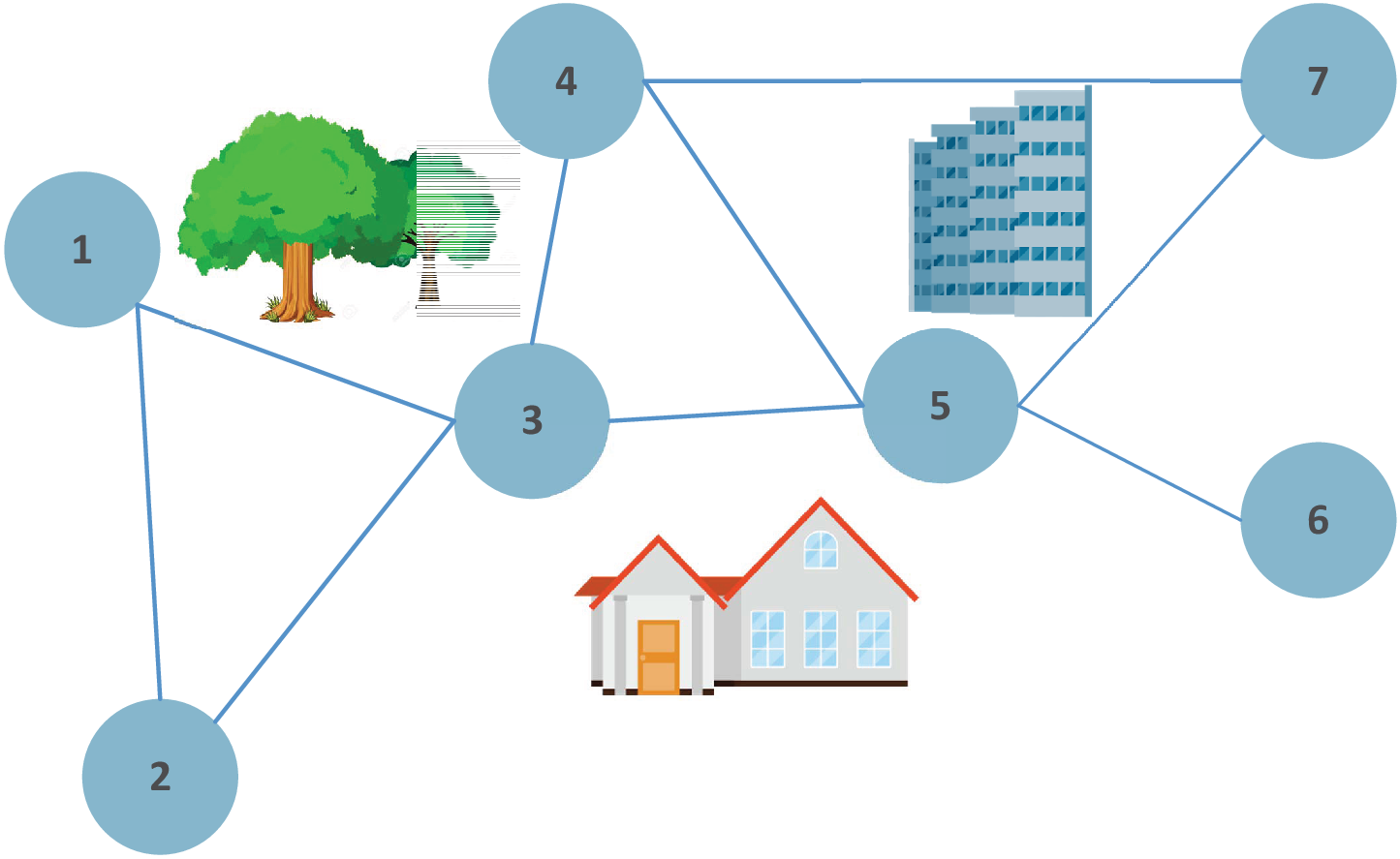}
	\caption{The connectivity graph \(\mc{G}(\mc{V},\mc{E})\) for a wireless D2D network.}\label{fig:system model}
		\vspace{-.30in}
\end{figure}

In this paper, we consider a FL problem in a decentralized setting as shown in Fig. \ref{fig:system model}, in which a set \(\mc{V}=\{1,\ldots,K\}\) of $K$ devices can only communicate with their respective neighbors over a wireless D2D network whose connectivity is characterized by an undirected graph \(\mc{G}(\mc{V},\mc{E})\), with \(\mc{V}\) denoting the set of nodes and \(\mc{E}\subseteq\{(i,j)\in\mathcal{V}\times\mathcal{V}\left.\vert\right.i\neq j\}\) the set of edges. The set of neighbors of node $i$ is denoted as \(\mc{N}_i=\{j\in\mc{V}\left.|\right.(i,j)\in\mc{E}\}\). Following the FL framework, each device has available a local data set, and all devices collaboratively train a machine learning model by exchanging model-related information without directly disclosing data samples to one another. 

\subsection{Learning Model}
Each device \(i\in\mc{V}\) has access to its local data set \(\mc{D}_i\), which may have non-empty intersection with the data set \(\mc{D}_j\) of any other device \(j\in\mc{V}\), \(i\neq j\). All devices share a common machine learning model class, which is parametrized by a vector \(\mv\theta\in\mb{R}^{d\times 1}\). As a typical example, the model class may consist of a neural network with a given architecture. The goal of the network is to solve the empirical risk minimization problem \cite{xin19decentralized,koloskova20arbitrary}
\begin{align*}
\mathrm{(P0)}:&~\mathop{\mathtt{Minimize}}_{\mv \theta}~~~F(\mv\theta)\triangleq\frac{1}{K}\sum\limits_{i\in\mathcal{V}}f_i(\mv \theta),
\end{align*} 
where \(f_i(\mv\theta)=\tfrac{1}{|\mc{D}_i|}\sum_{\mv\xi\in\mc{D}_i}l(\mv\theta,\mv\xi)\) is the local empirical risk function for device $i$ with \(l(\mv\theta; \mv\xi)\) denoting the loss accruing from parameter \(\mv\theta\) on data sample \(\mv\xi\in\mc{D}_i\), which may include the effect of regularization. 

{\color{black}Among the communication-efficient variants of the DSGD algorithm reviewed in Section \ref{subsec:Related Work}, we develop our analysis in the sequel based on CHOCO-SGD for the following reasons. First, CHOCO-SGD is flexible enough to support the general class of compression schemes satisfying the compression operator condition \cite{koloskova19decentralized}, allowing us to model both analog and digital implementations. Second, it provides state-of-the-art convergence guarantees for in noiseless and rate-limited communications, thus offering a suitable baseline to adapt to various wireless transmission protocols.} At the start of each iteration $t+1$, device $i\in\mc{V}$ has in its memory its current model iterate \(\mv\theta_i^{(t)}\), the corresponding estimated version \(\hat{\mv\theta}_i^{(t)}\) and the estimated iterates \(\hat{\mv\theta}_j^{(t)}\) for all its neighbors \(j\in\mc{N}_i\). We note that an equivalent version of the algorithm that requires less memory can be found in \cite[Algorithm 6]{koloskova19decentralized}, but we do not consider it here since it does not change the communication requirements. Furthermore, at each iteration $t$, device \(i\in\mc{V}\) first executes a local update step by SGD based on its data set \(\mc{D}_i\) as
\begin{align}
\mv\theta_i^{(t+\Myfrac{1}{2})}=\mv\theta_i^{(t)}-\eta^{(t)}\hat{\nabla}f_i(\mv\theta_i^{(t)}), \label{eq:local updates}
\end{align} where \(\eta^{(t)}\) denotes the learning rate, and \(\hat{\nabla}f_i(\mv\theta_i^{(t)})\) is an estimate of the exact gradient \({\nabla}f_i(\mv\theta_i^{(t)})\) obtained from a mini-batch \(\mc{D}_i^{(t)}\subseteq\mc{D}_i\) of the data set, i.e., \(\hat{\nabla}f_i(\mv\theta_i^{(t)})=\tfrac{1}{|\mc{D}_i^{(t)}|}\sum_{\mv\xi\in\mc{D}_i^{(t)}}\nabla l(\mv\theta_i^{(t)};\mv\xi)\). \footnote{\color{black}The gradient is estimated by averaging the sample-wise gradient \(\nabla l(\mv\theta_i^{(t)};\mv\xi)\) over a mini-batch $\mc{D}_i^{(t)}$ of the data set with replacement, and therefore the estimate \(\hat{\nabla}f_i(\mv\theta_i^{(t)})\) coincides with the true gradient \(\nabla f_i(\mv\theta_i^{(t)})\) when \(\mc{D}_i^{(t)}=\mc{D}_i\).}

Then, each device \(i\in\mc{V}\) compresses the difference \(\mv\theta_i^{(t+\Myfrac{1}{2})}-\hat{\mv\theta}_i^{(t)}\) between the locally updated model \eqref{eq:local updates} and the previously estimated iterate \(\hat{\mv\theta}_i^{(t)}\). The compressed difference \(\mc{C}^{(t)}(\mv\theta_i^{(t+\Myfrac{1}{2})}-\hat{\mv\theta}_i^{(t)})\) is then exchanged with the neighbors of node $i$. Assuming that communication is reliable --- an assumption that we will revisit in the rest of the paper---each device \(i\in\mc{V}\) updates the estimated model parameters \(\hat{\mv\theta}_j^{(t)}\) for itself and for its neighbors as
\begin{align}
\hat{\mv\theta}_j^{(t+1)}=\hat{\mv\theta}_j^{(t)}+\mc{D}^{(t)}\left(\mc{C}^{(t)}(\mv\theta_j^{(t+\Myfrac{1}{2})}-\hat{\mv\theta}_j^{(t)})\right),\; j\in\{i\}\cup\mc{N}_i,  \label{eq:local estimates}
\end{align} where \(\mc{D}^{(t)}(\cdot)\) is a decoding function.
Next, device $i\in\mc{V}$ executes a consensus update step by correcting the updated model \eqref{eq:local updates} using the estimated parameters \eqref{eq:local estimates} as
\begin{align}
\mv\theta_i^{(t+1)}=\mv\theta_i^{(t+\Myfrac{1}{2})}+\zeta^{(t)}\sum\limits_{j\in\mc{N}_i\cup\{i\}}w_{ij}\left(\hat{\mv\theta}_j^{(t+1)}-\hat{\mv\theta}_i^{(t+1)}\right), \label{eq:consensus updates}
\end{align} where \(\zeta^{(t)}\) is the consensus rate,  and the mixing matrix \(\mv W=\mv W^T\in\mb{R}^{K\times K}\) is selected to be \emph{doubly stochastic}, i.e., \([\mv W]_{ij}=w_{ij}\ge 0\), \(\mv W\mv 1 =\mv 1\), \(\mv 1^T\mv W=\mv 1^T\) and \(\|\mv W-\Myfrac{\mv 1\mv 1^T}{K}\|_2<1\). 
We postpone discussion regarding the compression operator \(\mc{C}^{(t)}(\cdot)\) and the decoding operator \(\mc{D}^{(t)}(\cdot)\) to Section \ref{subsec:compression}. The considered decentralized learning protocol is summarized in Algorithm \ref{alg:standard implementation}.
\begin{algorithm}[htp]
	\vspace{1pt}\caption{{\color{black}{Decentralized Learning with Noiseless  Communication}}\vspace{1pt}}\label{alg:standard implementation}
	\SetKwInOut{Input}{Input}
	\SetKwInOut{Output}{Output}
	\Input{Consensus step size \(\zeta^{(t)}\), SGD learning step size \(\eta^{(t)}\), connectivity graph \(\mc{G}(\mc{V},\mc{E})\) and mixing matrix \(\mv W\)}
	Initialize at each node \(i\in\mathcal{V}\): \(\mv\theta_i^{(0)}\), \(\hat{\mv\theta}_j^{(0)}=\mv 0\), \(\forall j\in\mathcal{N}_i\bigcup\{i\}\)\;
	\For{\(t=0, 1, \ldots,T-1\)}{
		\For(in parallel){\emph{each device \(i\in\mc{V}\)}}{
			update \(\mv\theta_i^{(t+\Myfrac{1}{2})}=\mv\theta_i^{(t)}-\eta^{(t)}\hat\nabla f_i(\mv\theta_i^{(t)})\)\;
			compress the difference \(\mv\theta_i^{(t+\Myfrac{1}{2})}-\hat{\mv\theta}_i^{(t)}\) to obtain \(\mc{C}^{(t)}(\mv\theta_i^{(t+\Myfrac{1}{2})}-\hat{\mv\theta}_i^{(t)})\)\;
			\For(in parallel){\emph{each neighboring  device \(j\in\mc{N}_i\)}}{
				send \(\mc{C}^{(t)}(\mv\theta_i^{(t+\Myfrac{1}{2})}-\hat{\mv\theta}_i^{(t)})\)\;
				receive \(\mc{C}^{(t)}(\mv\theta_j^{(t+\Myfrac{1}{2})}-\hat{\mv\theta}_j^{(t)})\)\;
			}
			update \(\hat{\mv\theta}_j^{(t+1)}=\hat{\mv\theta}_j^{(t)}+\mc{D}^{(t)}\left(\mc{C}^{(t)}(\mv\theta_j^{(t+\Myfrac{1}{2})}-\hat{\mv\theta}_j^{(t)})\right),\; \mbox{for}\  j\in\{i\}\cup\mc{N}_i\)\;
			udpate \(\mv\theta_i^{(t+1)}=\mv\theta_i^{(t+\Myfrac{1}{2})}+\zeta^{(t)}\sum\limits_{j\in\mc{N}_i\cup\{i\}}w_{ij}\left(\hat{\mv\theta}_j^{(t+1)}-\hat{\mv\theta}_i^{(t+1)}\right)\).	
		}	
	}
	\Output{\(\mv\theta_i^{(T-1)}\), \(\forall i\in\mathcal{V}\)}
\end{algorithm}

Finally, we make the following assumptions that are widely adopted in the literature on decentralized stochastic optimization \cite{koloskova19decentralized}. 
\begin{assumption}
	Each local empirical risk function \(f_i(\mv\theta)\), \(i\in\mc{V}\), is $L$-smooth and $\mu$-strongly convex, that is, for all \(\mv\theta_1\in\mb{R}^{d\times 1}\) and \(\mv\theta_2\in\mb{R}^{d\times 1}\), it satisfies the inequalities
	\begin{align}
	f_i(\mv\theta_1)\le f_i(\mv\theta_2)+\nabla f_i(\mv\theta_2)^T(\mv\theta_1-\mv\theta_2)+\frac{L}{2}\left\|\mv\theta_1-\mv\theta_2\right\|^2, \label{eq:L-smooth}
	\end{align} and
	\begin{align}
	f_i(\mv\theta_1)\ge f_i(\mv\theta_2)+\nabla f_i(\mv\theta_2)^T(\mv\theta_1-\mv\theta_2)+\frac{\mu}{2}\left\|\mv\theta_1-\mv\theta_2\right\|^2. \label{eq:mu-strongly convex}
	\end{align}\label{assump:mu-L convexity}
\end{assumption}

\begin{assumption}
	The variance of the mini-batch gradient \(\hat{\nabla}f_i(\mv\theta_i)\) is bounded as
	\begin{align}
	\mb{E}_{\mc{D}_i^{(t)}}[\|\hat{\nabla}f_i(\mv\theta_i)-\nabla f_i(\mv\theta_i)\|^2] \le \sigma_i^2, \label{assump:bounded variance}
	\end{align} 
	and its expected Euclidean norm is bounded as
	\begin{align}
	\mb{E}_{\mc{D}_i^{(t)}}[\|\hat{\nabla}f_i(\mv\theta_i)\|^2]\le G^2, \label{assump:bounded norm}
	\end{align}
	where the expectation \(\mb{E}_{\mc{D}_i^{(t)}}[\cdot]\) is taken over the selection of a mini-batch \(\mathcal{D}_i^{(t)}\subseteq\mc{D}_i\). 
\end{assumption}

\subsection{Communication Model}
\begin{figure}[htp]
	\centering
	\includegraphics[width=2.8in]{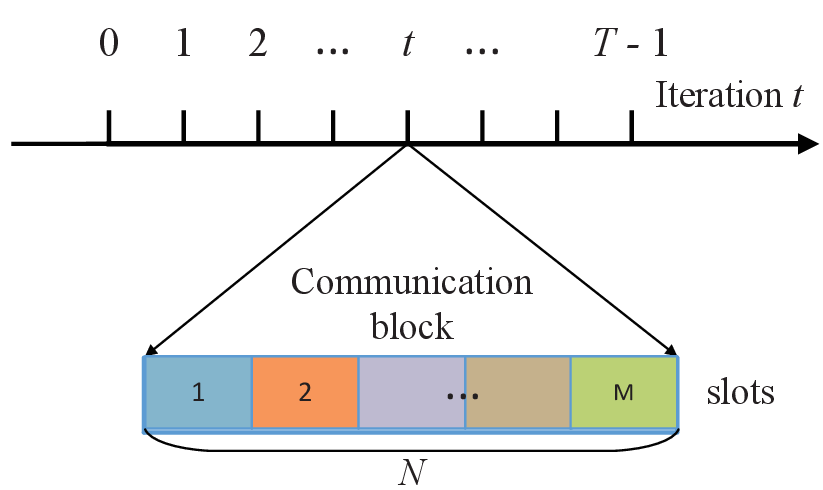}
	\vspace{-.10in}
	\caption{Timeline of training iterations and communication blocks: A communication block of $N$ channel uses, divided into $M$ slots, is employed to exchange the compressed difference between model parameters among neighboring devices.}\label{fig:training timeline}
	\vspace{-.20in}
\end{figure}

As seen in Fig. \ref{fig:training timeline}, at the end of every iteration $t$, communication takes place within one communication block of a total number $N$ of channel uses spanning over $M$ equal-length slots, denoted by \(\mc{S}=\{1,\ldots,M\}\). Slow fading remains constant across all iterations, and is binary, determining whether a link is blocked ar not. A link \((i,j)\in\mc{E}\) is by definition not \emph{blocked}, while all the other links \((i,j)\notin\mc{E}\) are blocked. We assume that the connectivity graph \(\mc{G}(\mc{V},\mc{E})\) with all the unblocked links as edges satisfies the following assumption. 
\begin{assumption}
	Graph \(\mc{G}(\mc{V},\mc{E})\) is a connected graph. \label{assump:connected graph} 
\end{assumption}
For all unblocked links \((i,j)\in\mc{E}\), the channel coefficient between device $i$ and $j$ is modelled as 
\begin{align}
h_{ij}^{\prime(t)}\triangleq \sqrt{A_0}\left(\frac{d_{ij}}{d_0}\right)^{-\frac{\gamma}{2}}h_{ij}^{(t)}, \label{eq:channel coefficient}
\end{align} where the small-scale fading coefficient \(h_{ij}^{(t)}\sim\mathcal{CN}(0,1)\) remains unchanged within one communication block and varies independently across blocks,  and the path loss gain \( \sqrt{A_0}(\Myfrac{d_0}{d_{ij}})^{\Myfrac{\gamma}{2}}\) is constant across all iterations, where \(A_0\) is the average channel power gain at reference distance $d_0$; \(d_{ij}\) is the distance between device $i$ and $j$; and \(\gamma\) is the path loss exponent factor. 

Each device is subject to an energy constraint of $NP^{(t)}$ per communication block. If a device is active for \(M^\prime\le M\) slots, the energy per symbol is hence given by \(\tfrac{NP^{(t)}}{M^\prime\Myfrac{N}{M}}=P^{(t)}\tfrac{M}{M^\prime}\). The mean-square power of the added white Gaussian noise (AWGN) is denoted as \(N_0\).

\subsection{Compression}\label{subsec:compression}
In this subsection, we describe the assumed compression operator \(\mc{C}(\cdot)\) and decompression operator \(\mc{D}(\cdot)\) that are used in the update \eqref{eq:consensus updates}.
We specifically adopt \emph{random linear coding (RLC)} compression \cite{abdi20analog}. Let \(\mv A^{(t)}=\tfrac{1}{\sqrt{m}}\mv H\mv R^{(t)}\) be the linear encoding matrix, where \(\mv H\in\{\pm 1\}^{m\times d}\) with \(m\le d\) is a partial Hadamard matrix with mutually orthogonal rows, i.e., \(\tfrac{1}{d}\mv H\mv H^T=\mv I\); and \(\mv R^{(t)}\in\mb{R}^{d\times d}\) is a diagonal matrix with its diagonal entries \([\mv R^{(t)}]_{ii}\triangleq r^{(t)}_i\) drawn from uniform distributions such that \(\Pr(r^{(t)}_i=1)=\Pr(r^{(t)}_i=-1)=0.5\), for all \(i=1,\ldots,d\). The compression operator is given by the linear projection \(\mc{C}^{(t)}(\mv u)=\mv A^{(t)}\mv u\), while decoding takes place as \(\mc{D}^{(t)}(\mv v)=\tfrac{m}{d}(\mv A^{(t)})^T\mv v\). The complexity of the encoding/decoding operations is linear with order \(\bigO(md)\), and thus easy to implement even on IoT devices. The concatenation of the compression and decompression operators, namely,  \(\mc{D}^{(t)}(\mc{C}^{(t)}(\mv u))=\tfrac{m}{d}(\mv A^{(t)})^T\mv A^{(t)}\mv u\), satisfies the {\color{black}\emph{compression operator}} condition \cite{koloskova19decentralized,abdi20analog} 
\begin{align}
\mb{E}\left\|\mv u-\frac{m}{d}(\mv A^{(t)})^T\mv A^{(t)}\mv u\right\|^2=\left(1-\frac{m}{d}\right)\|\mv u\|^2,\; \mbox{for all}\ \mv u\in\mb{R}^{d\times 1}. \label{eq:standard compression operator}
\end{align}
We note that the random matrices \(\{\mv R^{(t)}\}\) need to be shared among devices such that the same random sequence \(\{\mv A^{(t)}\}\) is agreed upon by all devices. {\color{black} In practice, sharing the random sequence \(\{\mv A^{(t)}\}\) can be done offline by tasking a randomly selected device to flood $d$ seeds through the network during a calibration phase. Since this phase takes place only once, its communication overhead can be neglected. Finally, by \eqref{eq:standard compression operator}, the quality of signal reconstruction in terms of the mean-square error (MSE) of RLC can be easily quantified, making RLC a natural choice to develop all analysis framework for wireless implementations of FL, as elaborated on in Section \ref{sec:Convergence Analysis for Digital Transmission} and Section \ref{sec:Convergence Analysis For Analog Transmission}.

\section{Digital and Analog Transmission Protocols}\label{sec:Digital and Analog Transmisison Protocols}
In this section, we describe digital and analog wireless implementations of the decentralized learning algorithm reviewed in the previous section. The implementations are meant to serve as prototypical templates for the deployment of decentralized learning. In practice, specific scheduling strategies are in need to allocate transmission slots as seen in Fig. \ref{fig:training timeline} to devices in collaboration.

\subsection{Digital Transmission}\label{subsec:Digital Transmission}
In digital transmission protocol, devices represent their model updates as digital messages for transmission. At iteration \(t\), each device \(i\in\mc{V}\) broadcast to all its neighbors using one dedicated time slot of the communication block decided by the scheduling policy in place.  
\subsubsection{Scheduling}
{\color{black}The analysis to be developed in Section \ref{sec:Convergence Analysis for Digital Transmission} applies to any scheduling protocol that satisfies the following conditions: \emph{(i)} no two connected devices transmit in the same slot due to the half-duplex transmission constraints; and \emph{(ii)} no two devices connected to the same device transmit in the same slot, so as not to cause interference at their common neighbor. To design a scheduling scheme meeting these properties, one can construct an auxiliary graph \(\mathcal{G}^d(\mathcal{V},\mathcal{E}^d)\) with degree $\Delta^d<K$ such that the edge set \(\mathcal{E}^d\supseteq\mathcal{E}\) includes not only the original edges in \(\mathcal{E}\), but also one edge for each pair of nodes sharing one or more common neighbors. One can then carry out vertex-coloring on the auxiliary graph \(\mathcal{G}^d(\mathcal{V},\mathcal{E}^d)\) using a classical greedy algorithm, such that any two nodes connected by an edge are assigned distinct colors \cite[Algorithm G]{Husfeldt15graph}. Scheduling proceeds by assigning the nodes with the same color to the same slot (see \cite{ozfatura20decentralized} for an example).} {\color{black}The complexity of this centralized scheduling scheme is of the order \(\bigO(K^2)\) \cite{Husfeldt15graph}. There are also decentralized implementations of vertex-coloring that could be applied to this setting with the complexity of \(\bigO(\Delta^dK)\) \cite{chak20coloring}.}

\subsubsection{Transmission}
The number \(B_i^{(t)}\) of bits that device \(i\in\mc{V}\) can successfully broadcast to its neighbors during a slot allocated to it by the scheduling algorithm is limited by the neighboring device with the worst channel power gain. Accordingly, we have 
\begin{align}
B_i^{(t)} =\frac{N}{M}\log_2\left(1+\frac{P^{(t)}}{N_0}M\min\limits_{j\in\mathcal{N}_i}\vert h_{ij}^{\prime(t)} \vert^2\right). \label{eq:B_i^{(t)}}
\end{align}
{\color{black}We recall that, in \eqref{eq:B_i^{(t)}}, the number $M$ of time slots per iteration is decided by the scheduling scheme.}

To quantize the encoded vector \(\mv A_i^{(t)}(\mv\theta_i^{(t+\Myfrac{1}{2})}-\hat{\mv\theta}_i^{(t)})\in\mb{R}^{m\times 1}\) into \(B_i^{(t)}\) bits, we employ a simple per-element $b$-bit quantizer \(\mc{Q}_b(\cdot)\) with chip-level precision so that $b=64$ or $b=32$ is for double-precision or single-precision floating-point, respectively, according to IEEE $754$ standard. Communication constraints thus impose the inequality \(bm_i^{(t)}\le B_i^{(t)}\), which is satisfied by setting \(m_i^{(t)}=\lfloor\Myfrac{B_i^{(t)}}{b}\rfloor\). Based on the (received) quantized signal, each device $i\in\mc{V}$ updates the estimated model parameters of its own as well as of its neighbors in \(j\in\{i\}\cup\mc{N}_i\) as (cf.~\eqref{eq:local estimates})
\begin{align}
\hat{\mv\theta}_j^{(t+1)}=\hat{\mv\theta}_j^{(t)}+\frac{m_j^{(t)}}{d}(\mv A_j^{(t)})^T\mc{Q}_b\left(\mv A_j^{(t)}(\mv\theta_j^{(t+\Myfrac{1}{2})}-\hat{\mv\theta}_j^{(t)})\right). \label{eq:local estimates for digital transmissions}
\end{align}
{\color{black}In order to implement update \eqref{eq:local estimates for digital transmissions}, each node $j\in\mc{V}$ and its neighbors in set $\mc{N}_j$ can share  \emph{a priori} a common sequence of (pseudo-)random matrix \(\tilde{\mv A}_j^{(t)}\in\mb{R}^{m\times d}\) as described in Section \ref{subsec:compression}. If node $j$ sends its current value \(m_j^{(t)}\) to all neighbors and if \(m_j^{(t)}\le m\), all nodes can thus select the same submatrix \(\mv A_j^{(t)}\) from \(\tilde{\mv A}_j^{(t)}\) to evaluate \eqref{eq:local estimates for digital transmissions}. The described digital implementation is summarized in Appendix \ref{appendix:algorithm for digital implementation}.

\subsection{Analog Transmission}\label{subsec:Analog Transmission}
With analog transmission, devices directly transmit their respective updated parameters by mapping analog signals to channel uses, without the need for digitization. As studied in \cite{zhu20broadband}, in addition to broadcast as in digital transmission, it is also useful to schedule all devices that share a common neighbor for simultaneous transmission in order to enable AirComp. Specifically, time slots required to be scheduled in pairs. In the first slot of each pair, one or more center nodes receive a superposition of the signals simultaneously transmitted by all their respective neighbors. The center nodes are referred to as AirComp receivers. In the second slot, the center nodes serve as broadcast transmitters communicating to all their neighbors. The total number $M$ of time slots is thus given by twice as the number $n$ of pairs of time slots, which is specified by the scheduling policy in use. 

\subsubsection{Scheduling}
{\color{black}The considered analog transmission protocols can accommodate any scheduling policy that satisfies the following two principles: \emph{(i)} no two connected nodes are scheduled as AirComp receivers in the same time slot due to half-duplex transmissions; and \emph{(ii)} no two nodes sharing a same neighboring node are scheduled as AirComp receivers in the same time slot. The second condition implies that, when a common neighboring node transmits to enable AirComp, the node's signal transmitted to one of the neighbors will not cause interference to the other. There are potentially many feasible scheduling schemes satisfying the above constraints, and we provide in Appendix \ref{appendix:a scheduling strategy for the analog implementation} a sequential scheduling policy for the purpose of illustration. The scheduling policy described therein aims at selecting as many non-interfering star-based sub-networks as possible in one pair of time slots.} 

To elaborate, we will use the following notation. For each device \(i\in\mc{V}\), we define a set \(\mc{S}_i^{\rm Tx}\subseteq\mc{S}\) of transmission slots, with \(\mc{S}_i^{\rm Tx}=\mc{S}_i^{\rm BT}\cup\mc{S}_i^{\rm AT}\) partitioned into disjoint subsets \(\mc{S}_i^{\rm BT}\) and \(\mc{S}_i^{\rm AT}\) (\(\mc{S}_i^{\rm BT}\cap\mc{S}_i^{\rm AT}=\emptyset\)). Subset \(\mc{S}_i^{\rm BT}\) denotes the set of transmission slots in which device $i$ broadcasts to its neighbors, and \(\mc{S}_i^{\rm AT}\) denotes the set of slots in which device $i$ transmits to enable AirComp. Similarly, we define the set \(\mc{S}_i^{\rm Rx}\subseteq\mc{S}\) of receiving slots for device $i$ as \(\mc{S}_i^{\rm Rx}=\mc{S}_i^{\rm BR}\cup\mc{S}_i^{\rm AR}\) with \(\mc{S}_i^{\rm BR}\cap\mc{S}_i^{\rm AR}=\emptyset\), where \(\mc{S}_i^{\rm BR}\) and \(\mc{S}_i^{\rm AR}\) denote the sets of receiving slots in which device $i\in\mc{V}$ receives from a transmitter in broadcast and AirComp modes, respectively. A sequential scheduling policy that satisfies the conditions \emph{(i)} and \emph{(ii)} listed in Section \ref{subsec:Analog Transmission}.1) is described as follows. 

{\color{black}First, we carry out greedy coloring on the auxiliary graph associated with the original connectivity graph \(\mathcal{G}^{(1)}=\mathcal{G}(\mathcal{V},\mathcal{E})\) as described in Section \ref{subsec:Digital Transmission}. Next, defining \(d_c^{(1)}\) as the sum of the degrees of all nodes that have been assigned the same color $c$ at the first iteration, we set all nodes assigned color $c^\ast = \arg\max_c\{d_c^{(1)}\}$ in \(\mathcal{G}^{(1)}\) as the center nodes, which compose the set \(\mc{N}_{c^\ast}^{(1)}\). In the first slot, the nodes in \(\mc{N}_{c^\ast}^{(1)}\) receive combined signals transmitted by their neighbors in \(\mathcal{G}^{(1)}\); and in the subsequent second slot, the same set of nodes in \(\mc{N}_{c^\ast}{(1)}\) broadcast their respective signal to their neighbors. As a result, the first slot is concurrently in set \(\mc{S}_i^{\rm AR}\) and sets \(\mc{S}_j^{\rm AT}\), for all nodes \(j\in\mc{N}_i\). Conversely, the second slot is concurrently in set \(\mc{S}_i^{\rm BT}\) and sets \(\mc{S}_j^{\rm BR}\) for all nodes \(j\in\mc{N}_i\). The center nodes in \(\mc{N}_{c^\ast}^{(1)}\) and their connected edges, along with any nodes disconnected from \(\mathcal{G}^{(1)}\), are then removed to produce the residue graph \(\mathcal{G}^{(2)}\). The overall procedure is repeated until the residue graph \(\mathcal{G}^{(n)}\) ($n\ge1$) becomes empty, and is summarized in Algorithm \ref{alg:scheduling for analog transmissions}. {\color{black}Intuitively, the more edges are removed at each iteration, the fewer transmission slots will be needed. Since \(d_c^{(n)}\) corresponds to the number of removable edges at iteration $n$, we schedule all nodes with color $c^\ast = \arg\max_c\{d_c^{(n)}\}$ as the center nodes to reduce the total number of required slots and improve spectral efficiency.} We illustrate the outcome of this scheduling policy on the connectivity graph in Fig. \ref{fig:system model} as an example (see Fig. \ref{fig:an example for analog scheduling} in Appendix \ref{appendix:a scheduling strategy for the analog implementation}).
	
{\color{black}Since the number of nodes in the residue graph reduces by at least one at each iteration and the maximum number of iterations is no more than $\Delta^d$, the overall complexity of both centralized and decentralized implementations of the sequential scheduling policy (c.f.~Algorithm \ref{alg:scheduling for analog transmissions}) proves to be $\bigO(\Delta^dK^2)$ \cite{chak20coloring}.} 

\subsubsection{Transmission}
We now describe the transmitted and the received signals in each pair of slots of the communication protocol.

{\it Odd slots}: All devices \(i\in\mc{V}\) operating in AirComp mode for a center node $j$ in an odd slot \(s\in\mc{S}_i^{\rm AT}\cap\mc{S}_j^{\rm AR}\) concurrently transmit the signals by pre-compensating the channel as
\begin{align}
\mv x_{ij}^{(t,s)}=\frac{\sqrt{\gamma_j^{(t,s)}}}{h_{ij}^{\prime(t)}}w_{ji}\mv A^{(t)}(\mv\theta_i^{(t+\Myfrac{1}{2})}-\hat{\mv\theta}_i^{(t)}), \label{eq:AT's transmitted signal}
\end{align} where \(\gamma_j^{(t,s)}\) is a power scaling factor for channel alignment at device $j$. {\color{black} The channel coefficient \(h_{ij}^{\prime(t)}\) needs to be acquired at each device $i$ operating in AirComp. To this end, at the beginning of each paired slots in the schedule, devices operating in AirComp can estimate channel through a pilot sent by their associated center node assuming channel reciprocity.} The receiving center node, device $j$ obtains
\begin{align}
\mv y_j^{(t,s)}=\sqrt{\gamma_j^{(t,s)}}\sum\limits_{i\in\mc{N}_j^{(s)}}w_{ji}\mv A^{(t)}(\mv\theta_i^{(t+\Myfrac{1}{2})}-\hat{\mv\theta}_i^{(t)})+\mv n_j^{(t,s)},  \label{eq:AR's received signal}
\end{align} where \(\mc{N}_j^{(s)}\) is the neighboring set of device $j$ operating in AirComp at slot $s$, and \(\mv n_j^{(t,s)}\sim\mc{CN}(\mv 0, N_0\mv I)\) is the received AWGN at slot $s$ of iteration $t$. Device $j$ estimates the \emph{combined} model parameters \(\sum\limits_{i\in\mc{N}_j^{(s)}}w_{ji}(\mv\theta_i^{(t+\Myfrac{1}{2})}-\hat{\mv\theta}_i^{(t)})\) via the linear estimator
\begin{align}
\hat{\mv y}_j^{(t,s)}=\frac{m}{d}(\mv A^{(t)})^T\Re\left\{\Myfrac{\mv y_j^{(t,s)}}{\sqrt{\gamma_j^{(t,s)}}}\right\}. \label{eq:AR's estimated model parameter}
\end{align}

{\it Even slots}: Any device \(i\in\mc{V}\) operating in broadcast mode in an even slot \(s\in\mc{S}_i^{\rm BT}\) transmits a signal
\begin{align}
\mv x_i^{(t,s)}=\sqrt{\alpha_i^{(t,s)}}\mv A^{(t)}(\mv\theta_i^{(t+\Myfrac{1}{2})}-\hat{\mv\theta}_i^{(t)}), \label{eq:BT's transmitted signal}
\end{align} where \(\alpha_i^{(t,s)}\) is device $i$'s transmitting power scaling factor in slot $s\in\mc{S}_i^{\rm BT}$ of iteration $t$. Each neighboring device \(j\in\mc{N}_i^{(s)}\), with \(s\in\mc{S}_j^{\rm BR}\), receives from device $i$ the signal
\begin{align}
\mv y_{ij}^{(t,s)}=\sqrt{\alpha_i^{(t,s)}}h_{ij}^{\prime(t)}\mv A^{(t)}(\mv\theta_i^{(t+\Myfrac{1}{2})}-\hat{\mv\theta}_i^{(t)})+\mv n_j^{(t,s)}, \label{eq:BR's received signal} 
\end{align} 
where \(\mv n_j^{(t,s)}\sim\mc{CN}(\mv 0, N_0\mv I)\) is the received AWGN. Device $j$ estimates the signal \(\mv\theta_i^{(t+\Myfrac{1}{2})}-\hat{\mv\theta}_i^{(t)}\) via the linear estimator
\begin{align}
\hat{\mv y}_{ij}^{(t,s)}=w_{ji}\frac{m}{d}(\mv A^{(t)})^T\Re\left\{\frac{\mv y_{ij}^{(t,s)}}{\sqrt{\alpha_{i}^{(t,s)}}h_{ij}^{'(t)}}\right\}, \label{eq:BR's estimated model parameter} 
\end{align} where \(\Re\{\cdot\}\) denotes the real part of its argument.

Next, device $j\in\mc{V}$ updates its estimate of the combined model parameters from all neighboring devices in \(\mc{N}_j\) by aggregating the estimates obtained at all receiving slots in set \(\mc{S}_j^{\rm Rx}=\mc{S}_j^{\rm BR}\cup\mc{S}_j^{\rm AR}\) as
\begin{align}
\hat{\mv y}_j^{(t+1)}=\hat{\mv y}_j^{(t)}+\sum\limits_{s\in\mc{S}_j^{\rm AR}}\hat{\mv y}_j^{(t,s)}+\sum\limits_{s\in\mc{S}_j^{\rm BR}}\hat{\mv y}_{i_sj}^{(t,s)}, \label{eq:estimate of the combined models}
\end{align} where node $i_s\in\mc{N}_j$ is the node that transmits in broadcast mode in slot \(s\in\mc{S}_j^{\rm BR}\).   The initial estimate of the combined model parameters is given by \(\hat{\mv y}_i^{(0)}=\mv 0\), \(\forall i\in\mc{V}\). 

The power scaling parameters \(\gamma_j^{(t,s)}\) (cf.~\eqref{eq:AT's transmitted signal}) and \(\alpha_{i}^{(t,s)}\) (cf.~\eqref{eq:BT's transmitted signal}) for \(s\in\mc{S}_i^{\rm Tx}\) need to be properly chosen such that the power consumed by device $i\in\mc{V}$ per communication block satisfies
\begin{align}
\sum\limits_{s\in\mc{S}_i^{\rm AT}}\mb{E}[\|\mv x_{ij_s}^{(t,s)}\|^2]+\sum\limits_{s\in\mc{S}_i^{\rm BT}}\mb{E}[\|\mv x_{i}^{(t,s)}\|^2]\le NP^{(t)},\; \forall i\in\mc{V}, \label{C:power constraint per commun. block} 
\end{align} where node $j_s$ is the center node connected to node $i$ in slot \(s\in\mc{S}_i^{\rm AT}\). Applying a simple equal power policy across different transmission slots of a device \(i\in\mc{V}\) for all communication blocks, we have (cf.~\eqref{C:power constraint per commun. block})
\begin{align}
\mb{E}[\|\mv x_{ij}^{(t,s)}\|^2] & \le\Myfrac{NP^{(t)}}{|\mc{S}_i^{\rm Tx}|},\; \forall s\in\mc{S}_i^{\rm AT}, \label{C:power constraint per AT's slot}\\
\mb{E}[\|\mv x_{i}^{(t,s)}\|^2] & \le\Myfrac{NP^{(t)}}{|\mc{S}_i^{\rm Tx}|},\; \forall s\in\mc{S}_i^{\rm BT}. \label{C:power constraint per BT's slot} 
\end{align}

{\color{black}By substituting \eqref{eq:AT's transmitted signal} for $\mv x_{ij}^{(t,s)}$ in \eqref{C:power constraint per AT's slot}, if follows that we have the inequality 
\begin{align}
	\gamma_j^{(t,s)}\le NP^{(t)}\frac{\vert h_{ij}^{\prime(t)}\vert^2}{|\mc{S}_i^{\rm Tx}|\|\mv\theta_i^{(t+1/2)}-\hat{\mv\theta}_i^{(t)}\|^2w_{ji}^2}, \label{eq:gamma_j^{(t,s)}}
\end{align} 
which implies that the power scaling factor \(\gamma_j^{(t,s)}\) for channel alignment at device $j$ is chosen as \(\gamma_j^{(t,s)}=\min_{i\in\mc{N}_j^{(s)}}v_{ij}^{(t)}\), where \(v_{ij}^{(t)}=\Myfrac{\vert h_{ij}^{\prime(t)}\vert^2}{(|\mc{S}_i^{\rm Tx}|\|\mv\theta_i^{(t+1/2)}-\hat{\mv\theta}_i^{(t)}\|^2w_{ji}^2)}\). As a result, the power scaling factor \(\gamma_j^{(t,s)}\) can be acquired in the initial channel estimation phase by finding the minimum of \(\mc{N}_j^{(s)}\) of non-negative values at the center node $j$. When \(|\mc{N}_j^{(s)}|\) is sufficiently small, each node $i\in\mc{N}_j^{(s)}$ can calculate $v_{ij}^{(t)}$ based on its estimated channel and send its value as feedback to the center node $j$. When the underlying connectivity is dense, and hence \(|\mc{N}_j^{(s)}|\) is large, this approach may entail an excessive communication overhead. In such cases, an estimate of \(\gamma_j^{(t,s)}\) can be found by AirComp-assisted wireless sensing following \cite[Section 5]{abari16AirComp}. Assuming a target estimation error of $\epsilon$, the required communication overhead amounts to $\lceil\log_2(\Myfrac{v_{\max}}{2\epsilon})\rceil$ rounds of power detection, where $v_{\max}$ denotes the maximum range of \(\gamma_j^{(t,s)}\). Then, the center node $j$ shares \(\gamma_j^{(t,s)}\) with all nodes in \(\mc{N}_j^{(s)}\) by broadcasting.} 

Similarly, by substituting \eqref{eq:BT's transmitted signal} for $\mv x_i^{(t,s)}$ in \eqref{C:power constraint per BT's slot}, the power scaling factor \(\alpha_i^{(t,s)}\) is expressed as 
\begin{align}
	\alpha_i^{(t,s)}=\Myfrac{NP^{(t)}}{\|\mv\theta_i^{(t+1/2)}-\hat{\mv\theta}_i^{(t)}\|^2}, \label{eq:alpha_i_{(t,s)}}
\end{align} which requires negligible communication overhead.

Furthermore, each device $j\in\mc{V}$ needs to update the estimate of its own model parameter \(\hat{\mv\theta}_j^{(t+1)}\) as
\begin{align}
\hat{\mv\theta}_j^{(t+1)}=\hat{\mv\theta}_j^{(t)}+\frac{m}{d}(\mv A^{(t)})^T\mv A^{(t)}(\mv\theta_j^{(t+\Myfrac{1}{2})}-\hat{\mv\theta}_j^{(t)}). \label{eq:estimate of one's own model} 
\end{align}
Finally, device $j\in\mc{V}$ approximates update \eqref{eq:consensus updates} as
\begin{align}
\mv\theta_j^{(t+1)}=\mv\theta_j^{(t+\Myfrac{1}{2})}+\zeta^{(t)}\left(w_{jj}\hat{\mv\theta}_j^{(t+1)}+\hat{\mv y}_j^{(t+1)}-\hat{\mv\theta}_j^{(t+1)}\right). \label{eq:consensus updates for analog scheme}
\end{align}


To sum up, the proposed analog implementation is presented in Appendix \ref{appendix:algorithm for analog implementation}}. 

\section{Convergence Analysis For Digital Transmission}\label{sec:Convergence Analysis for Digital Transmission}
In this section, we derive convergence properties of the general class of digital transmission protocols presented in Section \ref{subsec:Digital Transmission}. The analysis holds for any fixed transmission schedule, which determines the number $M$ of slots. We start by recalling that, at each iteration $t$, update \eqref{eq:local estimates for digital transmissions} is carried out by device \(i\in\mc{V}\) for all nodes \(j\in\{i\}\cup\mc{N}_i\). In \eqref{eq:local estimates for digital transmissions}, the concatenation of compression, quantization, and decompression yields an output vector \(\tfrac{m_j^{(t)}}{d}(\mv A_j^{(t)})^T\) \(\mc{Q}_b(\mv A_j^{(t)}\mv u_j^{(t)})\) for the input vector \(\mv u_j^{(t)}=\mv\theta_j^{(t+\Myfrac{1}{2})}-\hat{\mv\theta}_j^{(t)}\). The number \(m_j^{(t)}=\lfloor\Myfrac{B_j^{(t)}}{b}\rfloor\) of rows of matrix \(\mv A_j^{(t)}\) at iteration $t$ depends on the current rate \eqref{eq:B_i^{(t)}} supported by the fading channels between device $j$ and its neighbors. Taking the randomness of the fading realizations into account, the counterpart of the compression operator \eqref{eq:standard compression operator} under digital transmission is given by the following lemma.
{\color{black}\begin{lemma}
On average over RLC, the MSE for the concatenation of compression, quantization, and decompression under digital transmission satisfies
\begin{align}
\mb{E}\left\|\mv u-\frac{m_i^{(t)}}{d}(\mv A_i^{(t)})^T\mc{Q}_b(\mv A_i^{(t)}\mv u)\right\|^2\le\left(1-\omega^{(t)}\right)\|\mv u\|^2, \label{eq:digital compression operator}
\end{align}
for all \(\mv u\in\mb{R}^{d\times 1}\) and for all \(i\in\mc{V}\), where we have \(\omega^{(t)}=\min_{i\in\mc{V}}\omega_i^{(t)}\) with \(\omega_i^{(t)}=\tfrac{m_i^{(t)}}{d}\) and \(m_i^{(t)}\) denoting the number of rows of \(\mv A_i^{(t)}\).  
\label{lemma:MSE for estimation in digital transmissions}
\end{lemma}
}
\begin{IEEEproof}
	Please refer to Appendix \ref{appendix:proof of MSE for estimation in digital transmissions}.
\end{IEEEproof}

{\color{black}By \eqref{eq:digital compression operator}, the parameter \(\omega=\min_{t\in\{0,\ldots,T-1\}}\{\omega^{(t)}\}\), where \(\omega^{(t)}\in[0,1]\), is a measure of the quality of the reconstruction of the model difference used in update \eqref{eq:local estimates for digital transmissions}.} Supposing static channel conditions in which the transmission rate \eqref{eq:B_i^{(t)}} remains constant over iterations, we have \(m_i^{(t)}=m_i\), \(\forall i\in\mc{V}\). In these conditions,  the right-hand side (RHS) of \eqref{eq:digital compression operator} can reduce to \((1-\tfrac{m_i}{d})\|\mv u\|^2\) (see \cite[Appendix C]{xing21FL}), which is exactly the RHS of \eqref{eq:standard compression operator} given \(m=m_i\) and \(\mv A^{(t)}=\mv A_i^{(t)}\). {\color{black}Assuming fading channel conditions, the following corollary quantifies the quality of model reconstruction on average over Rayleigh-fading channels.}
{\color{black}\begin{corollary}
On average over RLC and the Rayleigh-fading based channel model described in section \ref{sec:System Model}, the MSE for the RLC compression operator (c.f.~\eqref{eq:standard compression operator}) under digital transmission satisfies 
\begin{align}
	\mb{E}\left\|\mv u-\frac{m_i^{(t)}}{d}(\mv A_i^{(t)})^T\mc{Q}_b(\mv A_i^{(t)}\mv u)\right\|^2\le\left(1-\frac{1}{d}\min\limits_{i\in\mc{V}}\sum_{n=1}^dG_i^{(t)}(n)\right)\|\mv u\|^2, \label{eq:digital compression operator averaged over channel fading}
\end{align}
for all \(\mv u\in\mb{R}^{d\times 1}\) and for all \(i\in\mc{V}\), where the function \(G_i^{(t)}(n)\!:\mb{Z}_{+}\mapsto\mb{R}\), \(i\in\mc{V}\), defined as
\begin{align}
G_i^{(t)}(n)=\exp\left(-\frac{N_0}{P^{(t)}}\frac{1}{MA_0}\left(2^{nb\frac{M}{N}}-1\right)\sum\limits_{j\in\mc{N}_i}\left(\frac{d_{ij}}{d_0}\right)^r\right). \label{eq:G_i}
\end{align}\label{corollary:MSE for estimation in Rayleigh fading}
\end{corollary}
}
\begin{IEEEproof}
Please refer to Appendix \ref{appendix:proof of MSE for estimation in Rayleigh fading}. 	
\end{IEEEproof}
Note that the MSE for RLC given by the RHS of \eqref{eq:digital compression operator averaged over channel fading} decreases with \(P^{(t)}\). In particular, when the transmission power \(P^{(t)}\to\infty\), it is seen in \eqref{eq:G_i} that \(G_i(n)\to 1\), thus leading to zero mean-square estimation error.

With Lemma \ref{lemma:MSE for estimation in digital transmissions}, the convergence properties of the digital protocol can be quantified in a manner similar to \cite[Theorem 4]{koloskova19decentralized}. To this end, we define the following topology-related parameters dependent on the mixing matrix \(\mv W\): the spectral gap \(\delta=1-\|\tfrac{\mv1\mv 1^T}{K}-\mv W\|\); the parameter \(\beta=\|\mv I-\mv W\|_2\); and the function \(p(\delta,\omega)=\tfrac{\delta^2\omega}{2(16\delta+\delta^2+4\beta^2+2\delta\beta^2-8\delta\omega)}\) that depends on the spectral gap \(\delta\) and {\color{black} on the model-difference estimation quality $\omega$}. Then, the convergence of the digital implementation is provided by the following theorem.
\begin{theorem}[Optimality Gap for Digital Transmission {\cite[Theorem 19]{koloskova19decentralized}}]
For learning rate \(\eta^{(t)}=\frac{3.25}{\mu}\tfrac{1}{t+a}\) with \(a\ge\max\{\tfrac{5}{p(\delta,\omega)},\tfrac{13L}{\mu}\}\), consensus step size \(\zeta^{(t)}=\tfrac{2p(\delta,\omega)}{\delta}\triangleq\zeta_0(\delta,\omega)\), and  {\color{black}fixed fading realizations \(\{h_{ij}^{\prime(t)}\}\), on average over SGD and RLC}, Algorithm~\ref{alg:digital implementation} yields an optimality gap satisfying
\begin{multline}
\mb E[F(\tilde{\mv\theta}_T)]-F^\ast\le\mb  E\left[\frac{1}{S_T}\sum_{t=0}^{T-1}w^{(t)}F(\bar{\mv\theta}^{(t)})\right]-F^\ast\\ \le\underbrace{\frac{\mu}{3.25}\frac{a^3-3.25a^2}{S_T}v_e^{(0)} + \frac{1.625(2a+T)T}{\mu S_T}\frac{{\bar\sigma}^2}{K}}_{\rm centralized\ error}+\underbrace{\frac{158.45\times 24LT}{\mu^2(p(\delta,\omega))^2S_T}G^2}_{\rm consensus\ error}, \label{eq:convergence rate for digital scheme}
\end{multline}	where \(w^{(t)}=(a+t)^2\); \(S_T=\sum_{t=0}^{T-1}w^{(t)}\); \(\tilde{\mv\theta}_T=\tfrac{1}{S_T}\sum_{t=0}^{T-1}w^{(t)}\bar{\mv\theta}^{(t)}\) is the weighted sum of the average iterate \(\bar{\mv\theta}^{(t)}=\tfrac{1}{K}\sum_{i\in\mathcal{V}}\mv\theta_i^{(t)}\) across all the communication rounds; {\color{black}\(F^\ast\) denotes the optimum objective value for problem (P0); \(\mu\) is the parameter for $\mu$-strongly convex function \(F(\mv\theta)\);} \(v_e^{(0)}=\|\bar{\mv\theta}^{(0)}-\mv\theta^\ast\|^2\) measures the initial distance to the optimal model parameter; and  \(\bar\sigma^2=\tfrac{1}{K}\sum_{i\in\mathcal{V}}\sigma_i^2\) is the average of the mini-batch gradient variance over all devices. \label{theorem:optimality gap for digital scheme}
\end{theorem} 

\begin{remark}
{\color{black}The labels ``centralized error'' and ``consensus error'' in \eqref{eq:convergence rate for digital scheme} refer to decomposition of the upper bound on optimality gap into a term that accounts for the performance of the average model \(\bar{\mv\theta}^{(t)}=\tfrac{1}{K}\sum_{i\in\mc{V}}\mv\theta_i^{(t)}\) and {\color{black}is inevitable even in centralized training due to the use of estimates of the gradients in SGD} --- the ``centralized error'', and a term that measures the disagreement among agents --- the ``consensus error''.} To gain insights into how wireless resources, channel conditions, and topology of the connectivity graph affect the performance of the digital wireless implementation, we can rewrite \eqref{eq:convergence rate for digital scheme} using relevant parameters in ``big O'' notation as \cite[Theorem 4]{koloskova19decentralized}
\begin{align}
\mb E[F(\tilde{\mv\theta}_T)]-F^\ast\le\bigO\left(\frac{\bar\sigma^2}{\mu KT}\right)+\bigO\left(\frac{\bar\sigma^2}{\mu KT^2}+\frac{LG^2}{\mu^2(p(\delta,\omega))^2T^2}\right)+\bigO\left(\frac{\mu}{T^3}\right). \label{eq:convergence rate in bigO for digital scheme}
\end{align} 
This result shows that when the total number of iterations $T$ is sufficiently large, the optimality gap \eqref{eq:convergence rate in bigO for digital scheme} behaves as \(\bigO\left(\tfrac{\bar\sigma^2}{\mu KT}\right)\), which recovers the convergence rate of centralized SGD with ideal communications. However, when the wireless communication resources are limited, and hence $NT\ll\infty$, the second term, scaling as \(\bigO\left(\tfrac{1}{T^2}\right)\), becomes equally important in \eqref{eq:convergence rate in bigO for digital scheme}, demonstrating the impact of the topology via \(\delta\) and \(\beta\), as well as the effect of the quality of digital transmission via \(\omega\). Since the function \(p(\delta,\omega)\) is monotonically increasing with \(\delta\in[0,1]\) and \(\omega\in[0,1]\), the second term in \eqref{eq:convergence rate in bigO for digital scheme} decreases with \(\delta\) and \(\omega\). This implies that convergence is improved for more connected graphs with larger $\delta$ \cite{xiao04fast}, and for smaller estimation errors with larger \(\omega\). \label{remark:optimality gap for digital scheme}
\end{remark}

{\color{black}\subsection{Numerical Illustration}\label{subsec:Numerical Examples for Analysis of Digital Scheme}
In this subsection, we corroborate the analysis by numerically evaluating each constituent term in the upper bounds \eqref{eq:convergence rate for digital scheme} on the optimality gap. We consider a setup consisting of $K=20$ devices that are located at randomly and independently selected distances in the interval \([50, 200]\) m away from a center position, with all angles (in radius) uniformly distributed in the interval \([0, 2\pi]\). The connectivity graph, accounting for the impact of slow fading, is modelled as: {\em (i)} a complete graph; {\em (ii)} a planar $5\times 4$ grid graph; {\em (iii)} a planar $5\times 4$ grid graph with torus wrapping; or {\em (iv)} a star graph as in conventional FL. We set the strong-convexity parameter as \(\mu=0.0002\), and the smoothness factor as \(L=0.16\). {\color{black}We adopt the following standard choice of $\mv W$: \(w_{ij}=\alpha\) for all \(j\in\mc{N}_i\), \(w_{ii}=1-|\mc{N}_i|\alpha\), and \(w_{ij}=0\) otherwise.} We set \(\alpha=\Myfrac{2}{(\lambda_1(\mv L)+\lambda_{K-1}(\mv L))}\), where $\mv L$ is the Laplacian of the connectivity graph. We plot the upper bound at iteration $t$ normalized by the corresponding value at $t=0$, hence evaluating the improvement in the expected optimality gap. {\color{black}The purpose of employing such normalized metrics is to standardize numerical values, and hence easier to compare.} The SNR is defined as the received SNR, averaged over small-scale channel fading, at a distance of $125$ meters (m) away from the deployment center.\footnote{We employ a benchmark distance of $125$ m to make the received SNR fall into range of values of practical interest.} Other parameters are set as \(d_0=1\) m, \(A_0=10^{-3.35}\), \(\gamma=3.76\) and \(N_0=-169\) dBm. 

We apply the greedy vertex-coloring algorithm \cite[Algorithm G]{Husfeldt15graph} in order to determine the number of slots. Fig. \ref{fig:digital optimality gap upper bounds vs SNR} plots separately the centralized and the consensus errors, as well as the overall error in \eqref{eq:convergence rate for digital scheme}, for different number of iterations $t=2000$ and $t=5000$ under a planar topology with torus wrapping. The centralized error does not depend on the SNR, and it decreases at the fastest rate over iterations. The consensus error decreases with the received SNR due to the improvement in the parameter $\omega$ that characterizes the quality of model reconstruction (cf.~\eqref{eq:convergence rate for digital scheme}). As a result, the consensus error dominates the optimality gap until the received SNR level increases to a sufficiently large value dependent on the iteration $t$. Furthermore, the overall optimality gap approaches the centralized error as the SNR increases. 
\begin{figure}[htp]
	\centering
	\subfigure[$t=2000$. \label{subfig:digital optimality gap upper bounds vs SNR at small T}]{\includegraphics[width=4.0in]{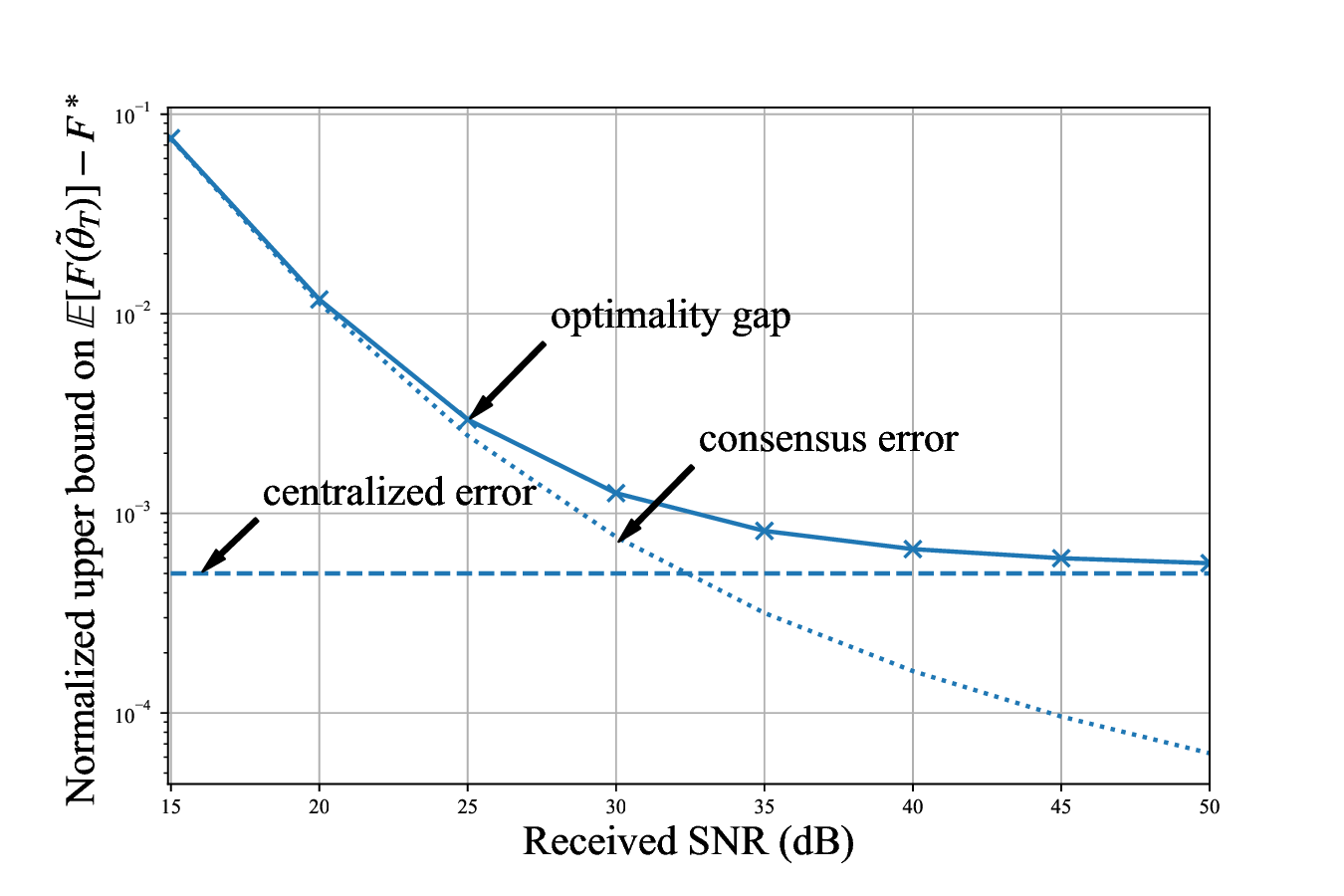}}
	\subfigure[$t=5000$. \label{subfig:digital optimality gap upper bounds vs SNR at large T}]{\includegraphics[width=4.0in]{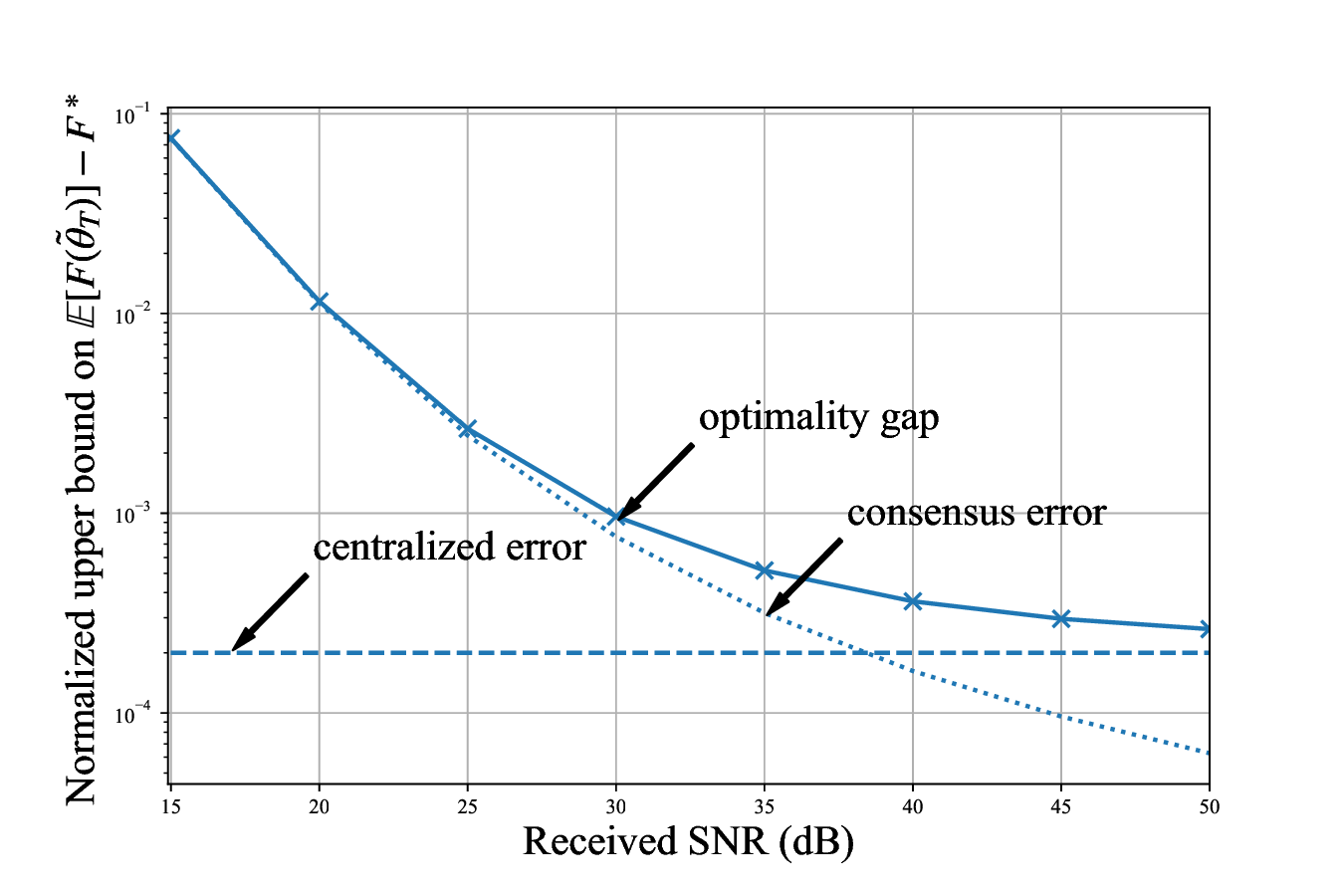}}
	\caption{Normalized upper bounds \eqref{eq:convergence rate for digital scheme} on the optimality gap versus the received SNR levels for a planar grid graph with torus wrapping and $N=10^4$ ($a=4\times10^7$).}\label{fig:digital optimality gap upper bounds vs SNR}
	\vspace{-.20in}
\end{figure}

We now turn to an analysis of the impact of the topology of the connectivity graph on the convergence for digital transmission. To this end, we set the same received SNR for all devices ignoring the impact of path loss in order to isolate the impact of different topologies. {\color{black}We employ TDMA-based scheduling that assigns only one device as the transmitter at one slot such that there are equal number $M=K$ of slots for all topologies.} Fig. \ref{fig:digital optimality gap upper bounds vs topologies} reports the optimality gap, along with the constituent errors in \eqref{eq:convergence rate for digital scheme}. The optimality gap decreases with the spectral gap $\delta$ of the connectivity graph, which equals $\delta=1$, $0.31$, $0.103$, and $0.095$, for complete graph, planar graph with and without torus wrapping, and star topology, respectively. {\color{black}Fig. \ref{fig:digital optimality gap upper bounds vs topologies} suggests that, under the constant-weight mixing strategy, if the underlying connectivity graph is denser, the nodes perform better model aggregation, thus achieving consensus faster (smaller ``consensus error'').} Fig. \ref{fig:digital optimality gap upper bounds vs topologies} shows that the consensus error contributes very little to the overall error for densely connected graphs, such as the complete graph and the planar grid with torus wrapping, while it becomes dominant in less densely connected graphs, such as the planar grid and the star graph.
\begin{figure}[htp]
	\vspace{-.20in}
	\centering
	\includegraphics[width=4.2in]{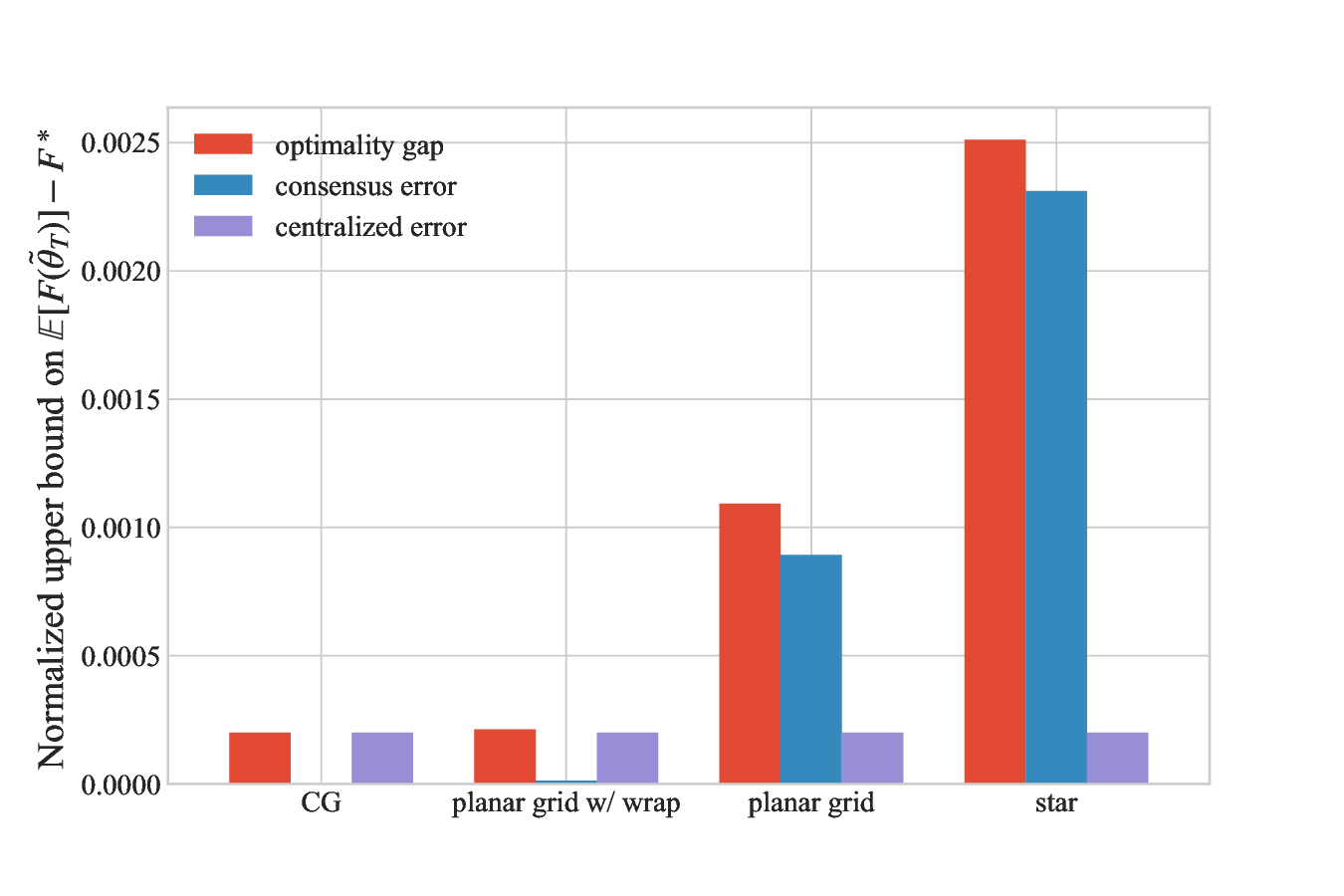}
	\vspace{-.10in}
	\caption{Normalized upper bounds \eqref{eq:convergence rate for digital scheme} on the optimality gap for different topologies of the connectivity graph with $M=K$, SNR$=30$ dB, and $N=10^4$ for $t=5000$ ($a=1.3\times10^8$).}
	\label{fig:digital optimality gap upper bounds vs topologies}
	\vspace{-.20in}
\end{figure} 
}

\section{Convergence Analysis For Analog Transmission}\label{sec:Convergence Analysis For Analog Transmission}
In this section, we derive convergence properties for the general class of analog transmission schemes described in Section \ref{subsec:Analog Transmission}. The analysis holds for any fixed scheduling scheme operating in pairs of slots as described in Section \ref{subsec:Analog Transmission}. We start by noting that, while for the digital implementation, the number $m_j^{(t)}=\lfloor\Myfrac{B_j^{(t)}}{b}\rfloor$ rows of matrix \(\mv A_j^{(t)}\) for device $j\in\mc{V}$ depends on the fading channels between device $j$ and its neighbors, for the analog implementation, the number $m$ of rows of matrix \(\mv A^{(t)}\) is fixed as the number \(m=\Myfrac{N}{M}\) of available channel uses in each slot of the communication block. {\color{black}Therefore, for analog communication, we can directly use \eqref{eq:standard compression operator} to quantify the quality of the estimate of the model difference used in update \eqref{eq:estimate of one's own model}. However, due to the presence of channel noise, update \eqref{eq:estimate of the combined models} for the combined parameters of neighboring devices does not satisfy \(\bar{\mv\theta}^{(t+1)}=\bar{\mv\theta}^{(t+\Myfrac{1}{2})}\).} This calls for a novel derivation of convergence properties that does not follow from \cite[Theorem 19]{koloskova19decentralized}. 

Next, we relate the update in \eqref{eq:consensus updates for analog scheme}, which is subject to Gaussian noise, to the noiseless update in \eqref{eq:consensus updates} in the following lemma. 
\begin{lemma}
The consensus update \eqref{eq:consensus updates for analog scheme} for analog implementation is equivalent to
\begin{align}
\mv\theta_i^{(t+1)}=\mv\theta_i^{(t+\Myfrac{1}{2})}+\zeta^{(t)}\sum\limits_{j\in\mc{N}_i\cup\{i\}}w_{ij}\left(\hat{\mv\theta}_j^{(t+1)}-\hat{\mv\theta}_i^{(t+1)}\right)+\zeta^{(t)}\sum_{\tau=0}^t\frac{m}{d}(\mv A^{(\tau)})^T\tilde{\mv n}_i^{(\tau)}, \label{eq:recast consensus updates for analog scheme}
\end{align}
where \(\tilde{\mv n}_i^{(t)}\sim\mc{N}(0,\tilde N_{0i}^{(t)}\mv I)\) is the effective noise with power
\begin{align}
\tilde N_{0i}^{(t)}=\frac{1}{2}\frac{N_0}{NP^{(t)}}\Bigg(\sum\limits_{s\in\mc{S}_i^{\rm AR}}\max\limits_{j\in\mc{N}_i^{(s)}}\Bigg\{|\mc{S}_j^{\rm Tx}|w_{ij}^2\frac{\|\mv u_j^{(t)}\|^2}{\big|h_{ij}^{\prime(t)}\big|^2}\Bigg\}+\sum\limits_{s\in\mc{S}_i^{\rm BR}}|\mc{S}_{j_s}^{\rm Tx}|w_{ij_s}^2\frac{\|\mv u_{j_s}^{(t)}\|^2}{\big|h_{ij_s}^{\prime(t)}\big|^2}\Bigg), \label{eq:variance of the effective noise}
\end{align} where we denoted \(\mv u_j^{(t)}=\mv\theta_j^{(t+\Myfrac{1}{2})}-\hat{\mv\theta}_j^{(t)}\).
\label{lemma:reformulation of the noisy consensus updates}
\end{lemma}
\begin{IEEEproof}
Please refer to Appendix \ref{appendix:proof of reformulation of the noisy consensus updates}.
\end{IEEEproof}

Comparing \eqref{eq:recast consensus updates for analog scheme} with \eqref{eq:consensus updates} reveals that consensus updates \eqref{eq:consensus updates for analog scheme} for analog implementation are noisy approximation of those for ideal communication used by Algorithm \ref{alg:standard implementation} and digital implementation used by Algorithm \ref{alg:digital implementation}. Note, in particular, that by \eqref{eq:recast consensus updates for analog scheme}, unlike Algorithms \ref{alg:standard implementation}-\ref{alg:digital implementation}, we no longer have the preservation of the average of the model parameters across the network. In fact, the average \(\bar{\mv\theta}^{(t+1)}=\Myfrac{1}{K}\sum_{i\in\mc{V}}\mv\theta_i^{(t+1)}\) of the model parameters, obtained by averaging over \(i\in\mc{V}\) on both sides of \eqref{eq:recast consensus updates for analog scheme}, is corrupted by Gaussian noise as
\begin{align}
\bar{\mv\theta}^{(t+1)}=\bar{\mv\theta}^{(t+\Myfrac{1}{2})}+\zeta^{(t)}\frac{1}{K}\sum_{i\in\mc{V}}\sum_{\tau=0}^t\frac{m}{d}(\mv A^{(\tau)})^T\tilde{\mv n}_i^{(\tau)}. \label{eq:model average for analog scheme}
\end{align} 

Next, we investigate how the noisy consensus updates given by \eqref{eq:recast consensus updates for analog scheme} affect the convergence of Algorithm \ref{alg:analog implementation}. In the proof of Theorem 19 in \cite[Appendix D]{koloskova19decentralized}, the key step leading to the final convergence result is to construct an error sequence defined as
\begin{align}
e^{(t)}=\underbrace{\sum\limits_{i\in\mc{V}}\mb{E}\|\bar{\mv\theta}^{(t)}-\mv\theta_i^{(t)}\|^2}_{\rm consensus\  error}+\underbrace{\sum\limits_{i\in\mc{V}}\mb{E}\|\hat{\mv\theta}_i^{(t+1)}-\mv\theta_i^{(t+\Myfrac{1}{2})}\|^2}_{\rm compression\ error}, \label{eq:def of error sequence}
\end{align} 
where the first term measures the consensus error, while the second term accounts for the impact of compression. It is shown in \cite[Appendix D]{koloskova19decentralized} that the above two sub-terms are coupled with each other via a recursive relation that yields the main result summarized in the previous section. However, as discussed, their approach cannot be directly applied to analog transmission due to the fact that the consensus preserving property no longer holds, i.e., \(\bar{\mv\theta}^{(t+1)}\neq\bar{\mv\theta}^{(t+\Myfrac{1}{2})}\), as seen from \eqref{eq:model average for analog scheme}. To address this challenge, we need a new upper bound on \eqref{eq:def of error sequence}. To this end, we first introduce the function  
\begin{align}
p^{(t)}(\delta,\omega)=\min\{\tilde p^{(t)}(\delta,\omega), p(\delta,\omega)\}, \label{eq:p(t)}
\end{align} where
\begin{align}
\tilde p^{(t)}(\delta,\omega)=\frac{\delta\zeta_0(\delta,\omega)}{\sqrt[\leftroot{-2}\uproot{1}4]{\tilde N_{0,T}}\Myfrac{t}{a^\prime}+1}-\left(\frac{\delta^2}{4}+\frac{2}{\omega}\beta^2\right)\frac{(\zeta_0(\delta,\omega))^2}{\big(\sqrt[\leftroot{-2}\uproot{1}4]{\tilde N_{0,T}}\Myfrac{t}{a^\prime}+1\big)^2}, \label{eq:tilde p(t)}
\end{align}
with \(p(\delta,\omega)\) and \(\zeta_0(\delta,\omega)\) defined in Theorem \ref{theorem:optimality gap for digital scheme}. We also denote 
\begin{align}
\tilde N_{0,T}=\max\limits_{t\in\{0,\ldots,T-1\}}\Big\{\sum_{i\in\mc{V}}\tilde N_{0i}^{(t)}\Big\}, \label{eq:the maximum noise variance}
\end{align}
and \(\omega=\Myfrac{m}{d}\). We will see that the function \(p^{(t)}(\delta,\omega)\) plays an analogous role to $p(\delta,\omega)$ for digital transmission in that it contributes to the decaying rate of the error sequence defined in \eqref{eq:def of error sequence}. Specifically, both \(\tilde p^{(t)}(\delta,\omega)\) and \(p(\delta,\omega)\) are increasing functions of \(\delta\) and \(\omega\).

{\color{black}A key ingredient of the proposed approach is to design a consensus step size that is adaptive to the iteration index $t$. This is because a consensus step size \(\zeta^{(t)}=\zeta_0\), as used in \cite{koloskova19decentralized,koloskova20arbitrary} for ideal communication, causes the accumulated channel noise term in \eqref{eq:model average for analog scheme} to grow with $t$, leading to a possible divergence of the upper bound on \(e^{(t)}\) (see (68) in \cite[Appendix E]{xing21FL} for details). To suppress the error growth induced by channel noise, we propose an adaptive consensus step size \(\zeta^{(t)}=\tfrac{\zeta_0(\delta,\omega)}{\Myfrac{\sqrt[\leftroot{-4}\uproot{1}4]{\tilde N_{0,T}}t}{a^\prime}+1}\), based on which we have the following lemma.}
\begin{lemma}
	For learning rate \(\eta^{(t)}=\tfrac{3.25}{\mu}\tfrac{1}{t+a}\) with \(a\ge\max\{\tfrac{5}{p^{(t)}(\delta,\omega)}, \tfrac{13L}{\mu}\}\), adaptive consensus step size \(\zeta^{(t)}=\tfrac{\zeta_0(\delta,\omega)}{\Myfrac{\sqrt[\leftroot{-4}\uproot{1}4]{\tilde N_{0,T}}t}{a^\prime}+1}\) with \(a^\prime>a\sqrt[\leftroot{-2}\uproot{4}4]{\tilde N_{0,T}}\), and \(T<\infty\), on average over RLC, the error sequence in \eqref{eq:def of error sequence} satisfies
	\begin{align}
	e^{(t+1)}\le \left(1-\frac{p^{(t)}(\delta,\omega)}{2}\right)e^{(t)}+\frac{1}{p^{(t)}(\delta,\omega)}(\eta^{(t)})^2\left(24KG^2+A(\delta,\omega)\sqrt[\leftroot{-2}\uproot{4}4]{\tilde N_0}\right), \label{eq:recursive upper bound for the error sequence}
	\end{align} where \(A(\delta,\omega)=\delta(\zeta_0(\delta,\omega)a^\prime)^3(2-\omega)\omega^2d(\frac{\mu}{3.25})^2\) is a function depending on the spectral gap \(\delta\) of graph \(\mc{G}(\mc{V},\mc{E})\) and the estimation quality \(\omega\). 
	\label{lemma:recursive upper-bound for the error sequence}
\end{lemma}
{\color{black}\begin{IEEEproof}
We provide herein an outline of the proof highlighting the key steps. First, to understand the dynamics of the error sequence $e^{(t)}$, we need to revisit the respective upper bounds on the one-step dynamics of its constituents, i.e., the consensus error \(\sum_{i\in\mc{V}}\mb{E}\|\bar{\mv\theta}^{(t)}-\mv\theta_i^{(t)}\|^2\) (c.f.~\cite[Lemma 17]{koloskova19decentralized}) and the compression error \(\sum_{i\in\mc{V}}\mb{E}\|\hat{\mv\theta}_i^{(t+1)}-\mv\theta_i^{(t+\Myfrac{1}{2})}\|^2\) (c.f.~\cite[Lemma 18]{koloskova19decentralized}). Unlike the setting studied in \cite{koloskova19decentralized}, both terms are affected by the effective noise power \(\sum_{\tau=0}^t\tilde{N}_0^{(\tau)}\) accumulated up to iteration $t$. Specifically, this noise term appears in the bounds multiplied by \((\zeta^{(t)})^2\). Therefore, the consensus rate \(\zeta^{(t)}\) should be properly designed so as to balance the conflicting requirements of reducing the impact of the accumulated channel noise and accelerating the consensus updates. Combining the bounds on consensus and compression errors produces an upper bound on the one-step dynamics of the overall error sequence $e^{(t)}$, which encompasses the first term in the RHS of \eqref{eq:recursive upper bound for the error sequence} and an additional term that is the sum of a function of \(\eta^{(t)}\) and a function of \(\zeta^{(t)}\). Finally, by judiciously designing the sequence \(p^{(t)}(\delta,\omega)\) to transform the function of \(\zeta^{(t)}\) into an upper bound on the function of \(\eta^{(t)}\), we recover \eqref{eq:recursive upper bound for the error sequence}. For the full version of this proof, please refer to Appendix \ref{appendix:proof of recursive upper-bound for the error sequence}.	
\end{IEEEproof}}
\begin{remark}
When communication is ideal, i.e., \(\tilde N_{0,T}\to 0\), the function \(p^{(t)}(\delta,\omega)\) reduces to \(p(\delta,\omega)\), and the upper bound \eqref{eq:recursive upper bound for the error sequence} reduces to the result derived in \cite[Lemma 21]{koloskova19decentralized}.
\end{remark}

By leveraging Lemma \ref{lemma:recursive upper-bound for the error sequence}, the convergence of the analog implementation scheme is revealed as follows.

\begin{theorem}[Optimality Gap for Analog Transmission] For a given total number $T$ of iterations, learning rate \(\eta^{(t)}=\tfrac{3.25}{\mu}\tfrac{1}{t+a}\) with \(a\ge\max\{\tfrac{5}{p^{(t)}(\delta,\omega)}, \tfrac{13L}{\mu}\}\), adaptive consensus step size \(\zeta^{(t)}=\tfrac{\zeta_0(\delta,\omega)}{\Myfrac{\sqrt[\leftroot{-4}\uproot{1}4]{\tilde N_{0,T}}t}{a^\prime}+1}\) with \(a^\prime>a\sqrt[\leftroot{-2}\uproot{4}4]{\tilde N_{0,T}}\), and {\color{black}fixed fading realizations \(\{h_{ij}^{\prime(t)}\}\), on average over SGD, RLC and channel noise}, Algorithm \ref{alg:analog implementation} yields an optimality gap satisfying
\begin{multline}
\mb E[F(\tilde{\mv\theta}_T)]-F^\ast\le\mb  E\left[\frac{1}{S_T}\sum_{t=0}^{T-1}w_tF(\bar{\mv\theta}^{(t)})\right]-F^\ast
\le \underbrace{\frac{\mu}{3.25}\frac{a^3-3.25a^2}{S_T}v_e^{(0)}+\frac{1.625(2a+T)T}{\mu S_T}\frac{\bar\sigma^2}{K}}_{\rm centralized\ error}+\\
\underbrace{\frac{158.45\times 24G^2LT}{\mu^2(p^{(T)}(\delta,\omega))^2 S_T}}_{\rm noiseless\ consensus\  error}+\underbrace{\frac{158.45\frac{A(\delta,\omega)}{K}\sqrt[\leftroot{-2}\uproot{4}4]{\tilde N_{0,T}}LT}{\mu^2(p^{(T)}(\delta,\omega))^2 S_T}+\frac{1}{K^2}D(\delta,\omega)\sqrt{\tilde N_{0,T}}}_{\rm AWGN\ error}, \label{eq:convergence rate for analog scheme}
\end{multline} where \(D(\delta,\omega)=\omega^2d\tfrac{\mu}{3.25}(\zeta_0(\delta,\omega)a^\prime)^2\). \label{theorem:optimality gap for analog scheme}
\end{theorem}
{\color{black}\begin{IEEEproof}
	We provide herein an outline of the proof highlighting the key steps. First, we characterize the exact upper bound on the error sequence $e^{(t)}$ based on Lemma \ref{lemma:recursive upper-bound for the error sequence} by leveraging \cite[Lemma 22]{koloskova19decentralized}. Next, we provide an upper bound on the dynamics of the distance to the optimal solution \(\mb{E}\|\bar{\mv\theta}^{(t+1)}-\mv\theta^\ast\|^2\) in terms of the optimality gap \(\mb{E}[F(\bar{\mv\theta}^{(t)})]-F^\ast\) and of polynomials of the learning rate \(\eta^{(t)}\). Unlike \cite{koloskova19decentralized}, this bound is also corrupted by the product of \((\zeta^{(t)})^2\) and the effective noise power \(\sum_{\tau=0}^t\tilde{N}_0^{(\tau)}\) accumulated up to iteration $t$. Finally, by properly designing the consensus rate \(\zeta^{(t)}\), we arrive at an upper bound on the modified time-average of the optimality gap \(\mb{E}[F(\bar{\mv\theta}^{(t)})]-F^\ast\) by using a variant of \cite[Lemma 3.3]{stich18sparsified}. For the full version of this proof, please refer to Appendix \ref{appendix:proof of optimality gap for analog scheme}.	
\end{IEEEproof}}
\begin{remark}
{\color{black}Similar to analysis given in Section \ref{sec:Convergence Analysis for Digital Transmission}, we decompose the upper bound on the optimality gap \eqref{eq:convergence rate for analog scheme} into different terms. The ``centralized error'' carries the same meaning as that in \eqref{eq:convergence rate for digital scheme}; the ``noiseless consensus error'' quantifies the disagreement among agents in the absence of communication noise, i.e., \(\tilde N_{0i}^{(t)}=0\) (cf.~\eqref{eq:variance of the effective noise}); and the ``AWGN error'' accounts for the impact of Gaussian noise on the consensus updates in \eqref{eq:consensus updates for analog scheme}.} 
To gain insights into how wireless resources, channel conditions, and topology of the connectivity graph affect the performance of the analog wireless implementation, we can write \eqref{eq:convergence rate for analog scheme} in the following form
\begin{multline}
\mb E[F(\tilde{\mv\theta}_T)]-F^\ast\le\bigO\Bigg(\frac{\sqrt{\tilde N_{0,T}}}{K^2}\Bigg)+\bigO\Bigg(\frac{\bar\sigma^2}{\mu KT^2}+\frac{\left(24G^2+\frac{A(\delta,\omega)}{K}\sqrt[\leftroot{-2}\uproot{4}4]{\tilde N_{0,T}}\right)L}{\mu^2(p^{(T)}(\delta,\omega))^2T^2}\Bigg)+\\
\bigO\left(\frac{\bar\sigma^2}{\mu KT}\right)+\bigO\left(\frac{\mu}{T^3}\right). \label{eq:convergence rate in bigO for analog scheme}
\end{multline}
{\color{black}Compared with \eqref{eq:convergence rate in bigO for digital scheme}, the upper bound \eqref{eq:convergence rate in bigO for analog scheme} reveals that, even if \(T\to\infty\), there is a non-vanishing term \(\bigO\left(\tfrac{\sqrt{\tilde N_{0,T}}}{K^2}\right)\), as well as a term scaling as \(\bigO\left(\tfrac{\sqrt[\leftroot{-2}\uproot{4}4]{\tilde N_{0,T}}}{(p^{(T)}(\delta,\omega))^2T^2}\right)\) that may not vanish either. We also note that \(\tilde N_{0,T}\) is non-decreasing with $T$ by its definition \eqref{eq:the maximum noise variance}. This highlights the significant impact of the topology parameters $\delta$ and $\beta$, as well as of the quality of the reconstruction $\omega=\Myfrac{m}{d}$ via function \(p^{(T)}(\delta,\omega)\).} Furthermore, as the effective noise power \(\tilde N_{0i}^{(t)}\) in \eqref{eq:variance of the effective noise} for \(i\in\mc{V}\) decreases with the transmitting power \(P^{(t)}\), the convergence rate improves with \(P^{(t)}\). In particular, when \(P^{(t)}\to\infty\), we have \(\tilde N_{0,T} = 0\), and therefore \eqref{eq:convergence rate in bigO for analog scheme} reduces exactly to the corresponding expression for the noiseless case in \eqref{eq:convergence rate in bigO for digital scheme}. \label{remark:optimality gap for analog scheme}
\end{remark}

{\color{black}\begin{remark}
Theorem \ref{theorem:optimality gap for analog scheme} assumes smooth and strongly convex loss functions. Existing results such as \cite[Theorem 4.1]{koloskova20arbitrary} on smooth and non-convex loss functions are based on constant learning step size, and they cannot be applied to our setting. This is because constant learning step sizes yield increasing upper bounds on the error sequence \(e^{(t)}\) (c.f.~\eqref{eq:exact upper bound for the total error}), thus leading to possible divergence of the time average of the expected gradient norms \(\tfrac{1}{T}\sum_{t=0}^{T-1}\mb{E}\|\nabla F(\bar{\mv\theta}^{(t)})\|^2\), which is often adopted to analyze the convergence to stationary point for non-convex objectives. Accordingly, one of the key challenges in analyzing convergence bounds for analog wireless implementations in non-convex settings is to jointly design the diminishing learning step size \(\eta^{(t)}\) and the decreasing consensus rate \(\zeta^{(t)}\). This is left for future work. 
\end{remark}}

{\color{black}\subsection{Numerical Illustration}\label{subsec:Numerical Examples for Analysis of Analog Scheme}
In this subsection, we elaborate on the results obtained from the analysis above by following the approach in Section \ref{subsec:Numerical Examples for Analysis of Digital Scheme}. For the scheduling scheme, we apply the sequential scheduling policy proposed in Algorithm \ref{alg:scheduling for analog transmissions} (see Appendix \ref{appendix:a scheduling strategy for the analog implementation}). Fig. \ref{fig:analog optimality gap upper bounds vs SNR} plots the upper bounds and the individual terms in \eqref{eq:convergence rate for analog scheme} as in Fig. \ref{fig:digital optimality gap upper bounds vs SNR} for digital communication under the planar grid topology with torus wrapping. The centralized error and the noiseless consensus error are independent of SNR. Note that the latter depends on the parameter \(\omega=\tfrac{\Myfrac{N}{M}}{d}\) that quantifies the quality of the estimate in \eqref{eq:standard compression operator}, which is independent of the SNR. As a result, the impact of SNR on the optimality gap is only through the AWGN error, which dominates the other terms when the SNR level is sufficiently small, here around $25$ dB, while it  becomes negligible when the SNR is large enough, here lager than $40$ dB.
\begin{figure}[htp]
	\centering
	\subfigure[$t=2000$. \label{subfig:analog  optimality gap upper bounds vs SNR at small T}]{\includegraphics[width=4.0in]{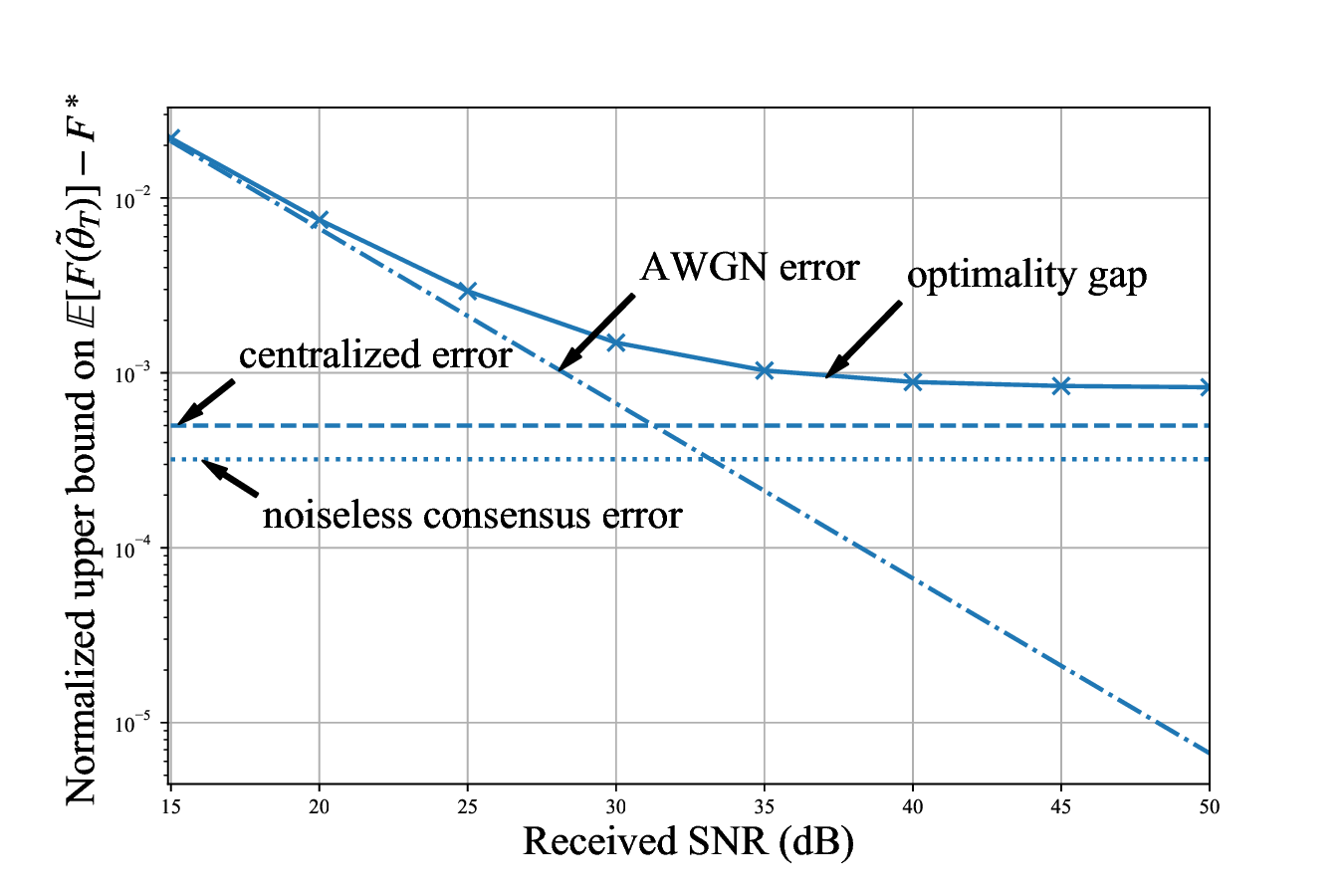}}
	\subfigure[$t=5000$. \label{subfig:analog optimality gap upper bounds vs SNR at large T}]{\includegraphics[width=4.0in]{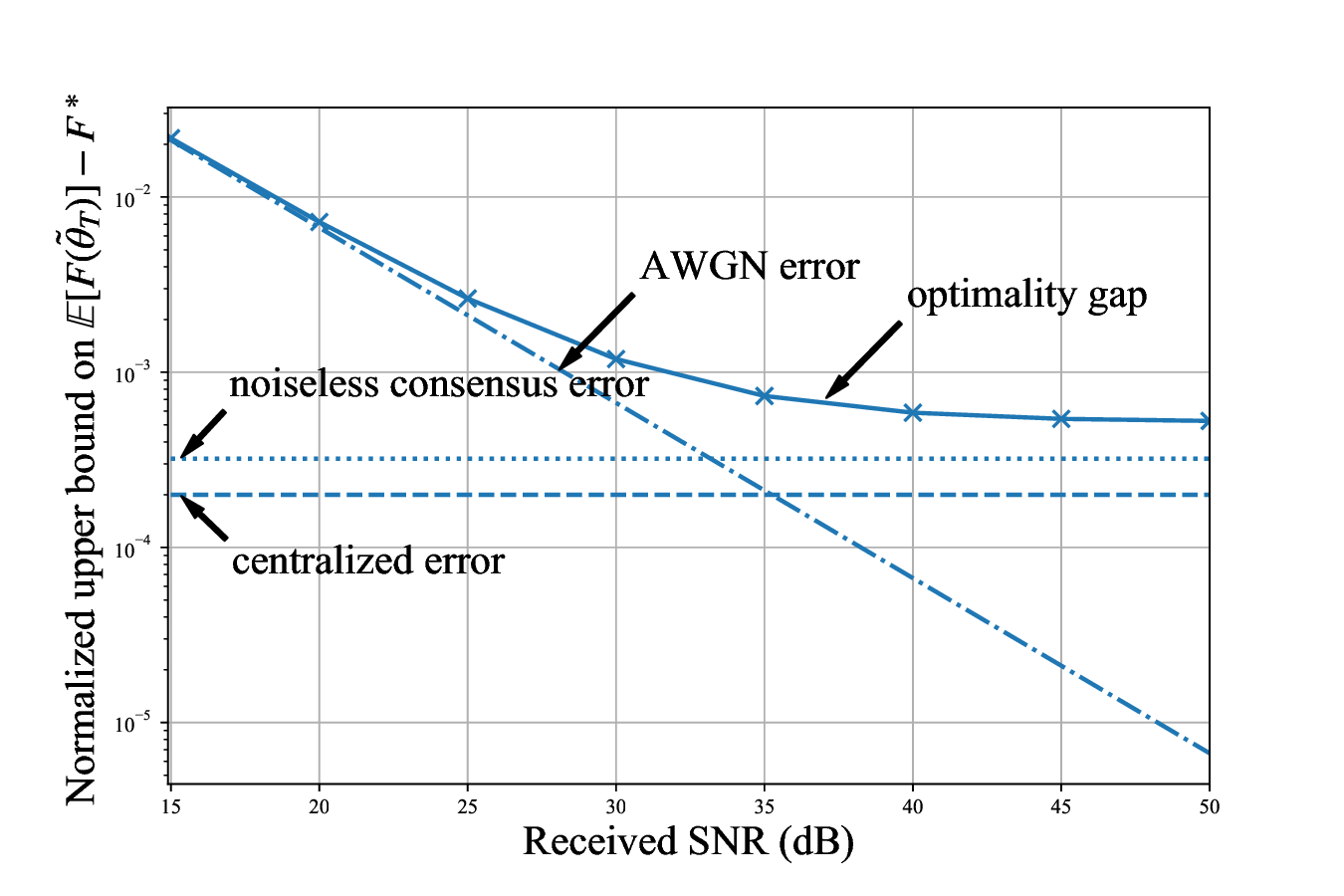}}
	\caption{Normalized upper bounds \eqref{eq:convergence rate for analog scheme} on the optimality gap versus the received SNR levels for a planar grid graph with torus wrapping and $N=10^4$ ($a=7\times10^6$, $a^\prime=a\sqrt[\leftroot{-2}\uproot{4}4]{\tilde N_{0,T}}$, and $\tilde N_{0,T}=10^{-5}$).}\label{fig:analog optimality gap upper bounds vs SNR}
	\vspace{-.20in}
\end{figure}

Next, we study the impact of the topology of the connectivity graph on the convergence for analog transmission in Fig. \ref{fig:analog optimality gap upper bounds vs topologies}. The general conclusions are analogous to Fig. \ref{fig:digital optimality gap upper bounds vs topologies}, which illustrates the corresponding results for digital communication. In particular, the noiseless consensus error is shown to increase when the connectivity graph is less densely connected, i.e., with smaller $\delta$. In contrast,  the AWGN error is less sensitive to a change in connectivity, and it becomes dominant for sufficiently large number of iterations such as $t=5000$.
\begin{figure}[htp]
	\vspace{-.20in}
	\centering
	\includegraphics[width=4.2in]{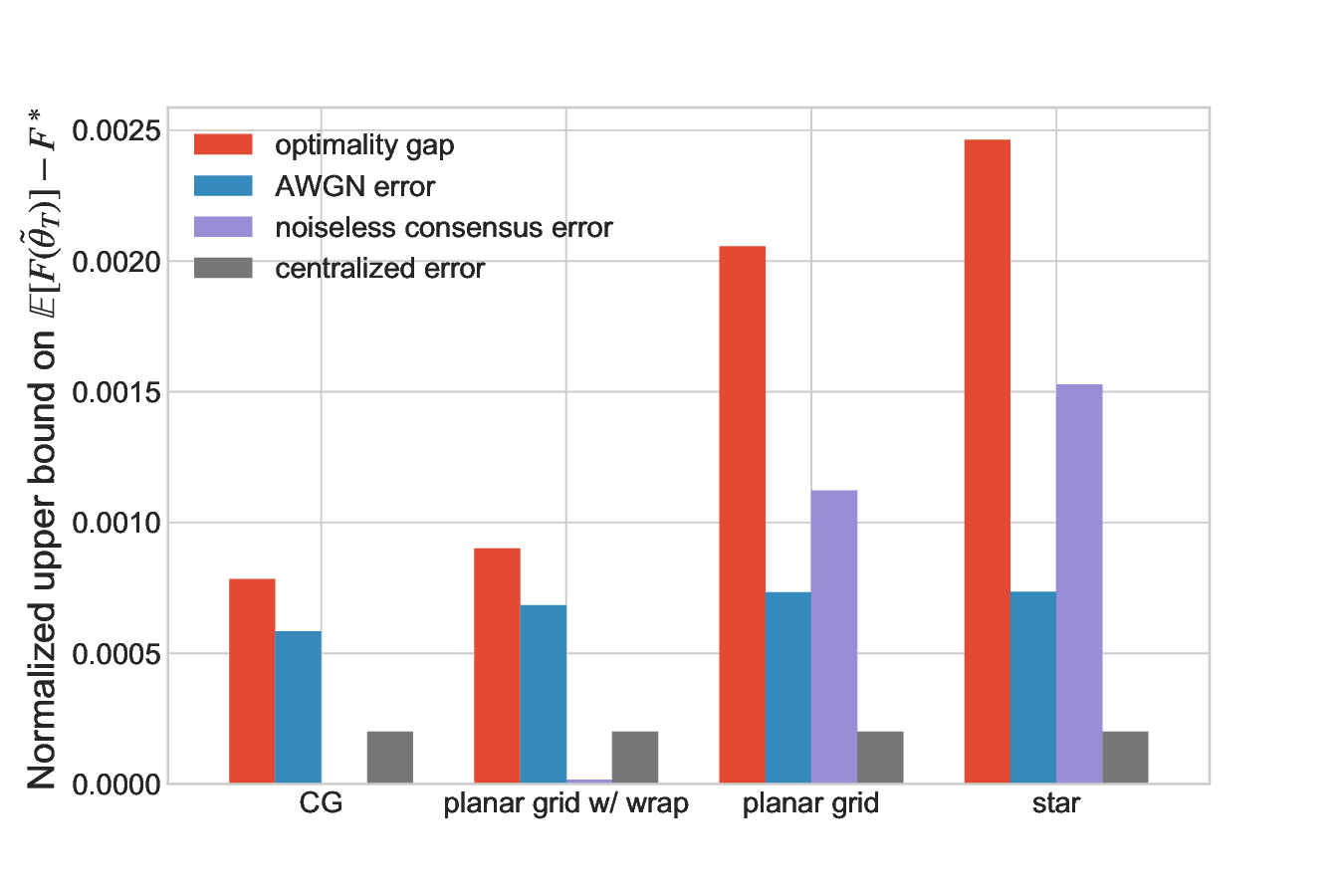}
	\vspace{-.10in}
	\caption{Normalized upper bounds \eqref{eq:convergence rate for analog scheme} on the optimality gap for different topologies of the connectivity graph with $M=K$, SNR$=30$ dB, and $N=10^4$ for $t=5000$ ($a=3\times10^7$, $a^\prime=2a\sqrt[\leftroot{-2}\uproot{4}4]{\tilde N_{0,T}}$, and $\tilde N_{0,T}=10^{-5}$ corresponding to $30$ dB received SNR).}
	\label{fig:analog optimality gap upper bounds vs topologies}
	\vspace{-.20in}
\end{figure}     
}

\section{Numerical Experiments}\label{sec:Numerical Experiments}
In this section, we corroborate the analysis developed in Sections \ref{sec:Convergence Analysis for Digital Transmission} and \ref{sec:Convergence Analysis For Analog Transmission} by evaluating the empirical performance of the digital and analog wireless implementations over a wireless D2D network. We consider the learning task of image classification over the Fashion-MNIST dataset \cite{xiao17fashion} that consists of $28\times 28$ images divided into $C=10$ classes. There are $60,000$ training data samples and $10,000$ test data samples, which are equally divided among classes. Each device \(i\in\mc{V}\) has data samples from at least six different classes, with the number of missing classes being uniformly selected in the set \(\{0,1,2,3,4\}\). An equal number \(x_i\) of data samples are then selected across the available classes at device $i$, such that the total number \(\sum_{i\in\mc{V}}\sum_{n=1}^{10}\mathbbm{1}_{n,i}x_i\) of training samples in use are maximized, where \(\mathbbm{1}_{n,i}\) is an indicator function denoting whether class $n\in\{1,\ldots,10\}$ is available at device $i$ ($\mathbbm{1}_{n,i}=1$) or not ($\mathbbm{1}_{n,i}=0$). All devices share a softmax regression model. We adopt the standard cross-entropy loss with $\ell_2$ regularization \(f_i(\mv\theta)=-\tfrac{1}{|\mathcal{D}_i|}\sum_{\mv\xi\in\mc{D}_i}\sum_{n\in\Omega_i}y_{i,n}\log(\mathrm{softmax}_n(\mv a_{i}(\mv\theta,\mv\xi)))+\tfrac{\mu}{2}\|\mv\theta\|^2\), where \(\Omega_i\) denotes the set of available classes on device \(i\in\mc{V}\); \(y_{i,n}\in\{0,1\}\) is the one-hot encoded label corresponding to the $n$th class for data sample \(\mv\xi\in\mc{D}_i\); the $n$the entry of  \(\mathrm{softmax}\) is defined as  \(\mathrm{softmax}_n(\mv z)=\tfrac{e^{z_n}}{\sum_{n=1}^{C}e^{z_n}}\), where \(\mv z=(z_1, \ldots, z_C)^T\); and \(\mv a_{i}(\mv\theta,\mv\xi)\) is the vector consisting of the logits for device $i$ with its $n$th entry corresponding to class $n$. The SGD is executed with mini-batch size of \(|\mathcal{D}_i^{(t)}|=64\), and we add momentum to all updates with a factor of $0.9$. In line with the local empirical risk function defined above, the strong-convexity parameter is set as $\mu=0.002$. The smoothness factor $L$ is numerically computed as the largest eigenvalue of the data Gramian matrix. 

{\color{black}The optimal value \(F^\ast\) used for quantifying the empirical optimality gap is numerically obtained by the standard decentralized SGD (applying Algorithm \ref{alg:standard implementation} with \(\mc{C}=\mc{D}=\mv I\) for sufficiently large $T$), with the hyper parameters \(a>0\) for the learning rate \(\eta^{(t)}=\tfrac{3.25}{\mu}\tfrac{1}{t+a}\) and the (constant) consensus step size \(\zeta_0\in(0,1]\) optimized via grid search. {\color{black}We consider $K=20$ devices with the connectivity graph modelled as a planar $5\times 4$ grid graph with torus wrapping or a chain graph.} All the other parameters for simulations are set as in Section \ref{subsec:Numerical Examples for Analysis of Digital Scheme} unless specified otherwise. As benchmarks, we consider decentralized learning with ideal communications, i.e., applying Algorithm \ref{alg:standard implementation} with \(\mc{C}^{(t)}=\mc{D}^{(t)}=\mv I\), as well as independent learning that carries out training based solely on local data with no communications among devices.\footnote{The source code implementing numerical experiments is available at \url{https://github.com/Fuzzy-Face/JSAC_FL}.}  

We start by studying the impact of the proposed adaptive consensus rate on the convergence for analog transmission. We recall that the analysis in Section \ref{sec:Convergence Analysis For Analog Transmission} has revealed that a consensus rate \(\zeta^{(t)}=\tfrac{\zeta_0}{\Myfrac{t}{1000}+1}\) is necessary in order to ensure convergence. In line with the analysis, Fig. \ref{fig:analog optimality gap for constant vs adaptive consensus rate} shows that the optimality gap with fixed consensus rates diverges as a function of the number of iteration $t$, while it converges with the proposed adaptive consensus rate. The figure also  demonstrates the advantages brought by communications w.r.t local training with no communications. 
\begin{figure}[ht]
	\centering
	\includegraphics[width=4.2in]{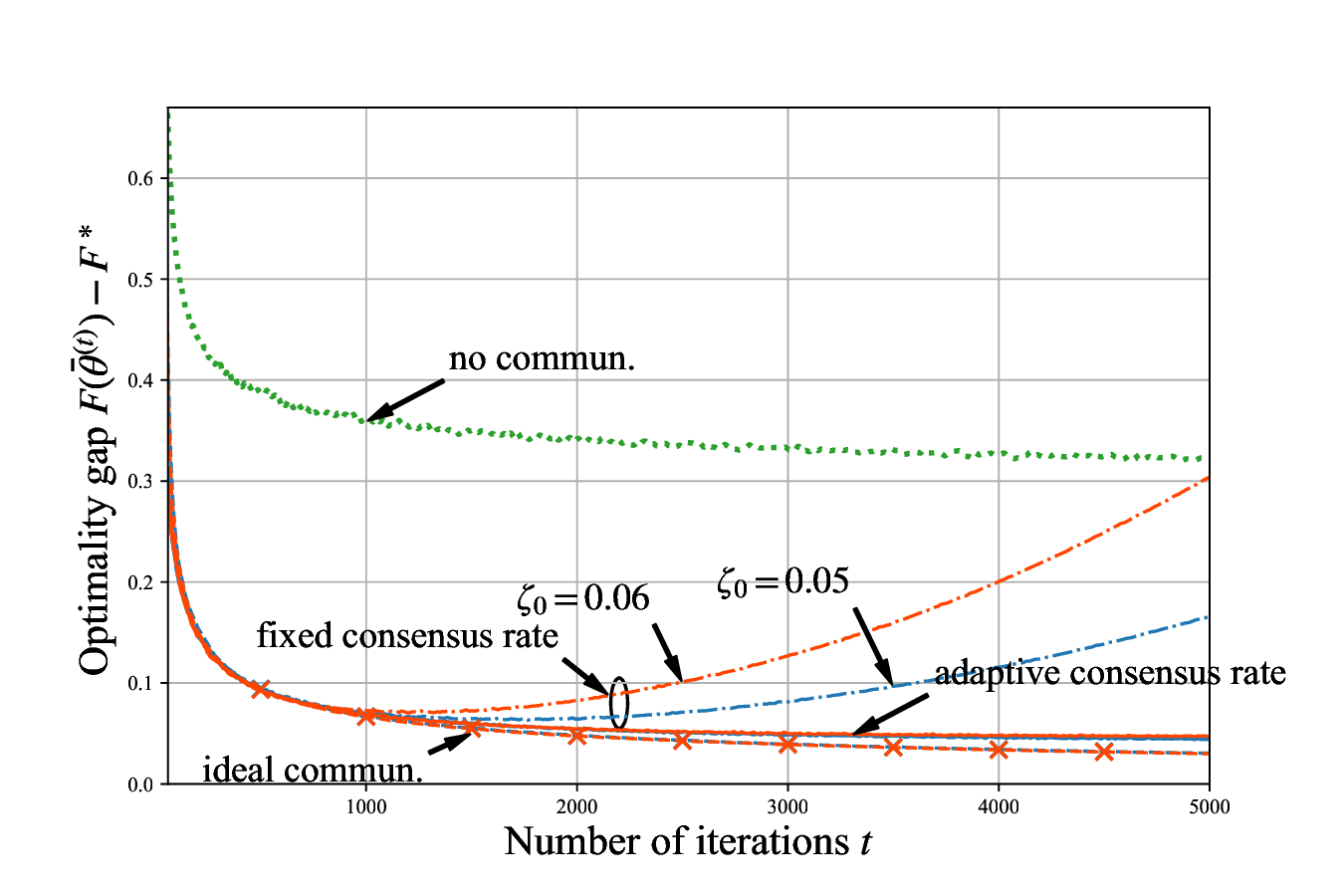}
	\caption{Empirical optimality gap versus the number of iterations $t$ for analog transmission over a planar grid graph with torus wrapping, with SNR$=30$ dB and $N=8000$ ($a=200$).}
	\label{fig:analog optimality gap for constant vs adaptive consensus rate}
	\vspace{-.20in}
\end{figure}

Next, we provide a performance comparison between digital and analog wireless implementations in terms of the analytical and the empirical upper bounds on the optimality gap. We plot the optimality gap normalized by its value evaluated at iteration $t=200$ and $t=0$ for the analytical and the empirical results, respectively. {\color{black}For the empirical results, we set $\eta^{(t)}=\tfrac{3.25}{\mu}\tfrac{1}{t+200}$ for all schemes; $\zeta^{(t)}=0.001$ for the schemes of ideal, digital and no communication; and $\zeta^{(t)}=\tfrac{0.001}{\Myfrac{t}{d}+1}$ with $d=35355,\ 42045,\ 50000,\ 62872,\ \mbox{and}\ 74767$ for $N=500,\ 1000,\ 2000,\ 5000,\ \mbox{and}\ 10000$, respectively, for the analog implementation.} 
Fig. \ref{fig:digital vs analog} compares digital and analog implementations, benchmarked by ideal communication and no communication. The analytical bounds shown in Fig. \ref{subfig:comparison between upper bounds of digital and analog schemes} are in general agreement with the empirical results shown in Fig. \ref{subfig:comparison between empirical optimality gaps of digital and analog schemes}, demonstrating the practical relevance of the theory developed in this paper. In particular, the analysis correctly predicts the advantages of the analog implementation for sufficiently small number of channel uses $N$, and the marginal benefits of the digital implementation in the complementary regime of a large number of channel uses, e.g., when we have $N\ge 5000$. 
\begin{figure}[!ht]
	\centering
	\subfigure[Normalized upper bounds on the optimality gap for different schemes. \label{subfig:comparison between upper bounds of digital and analog schemes}]{\includegraphics[width=4.2in]{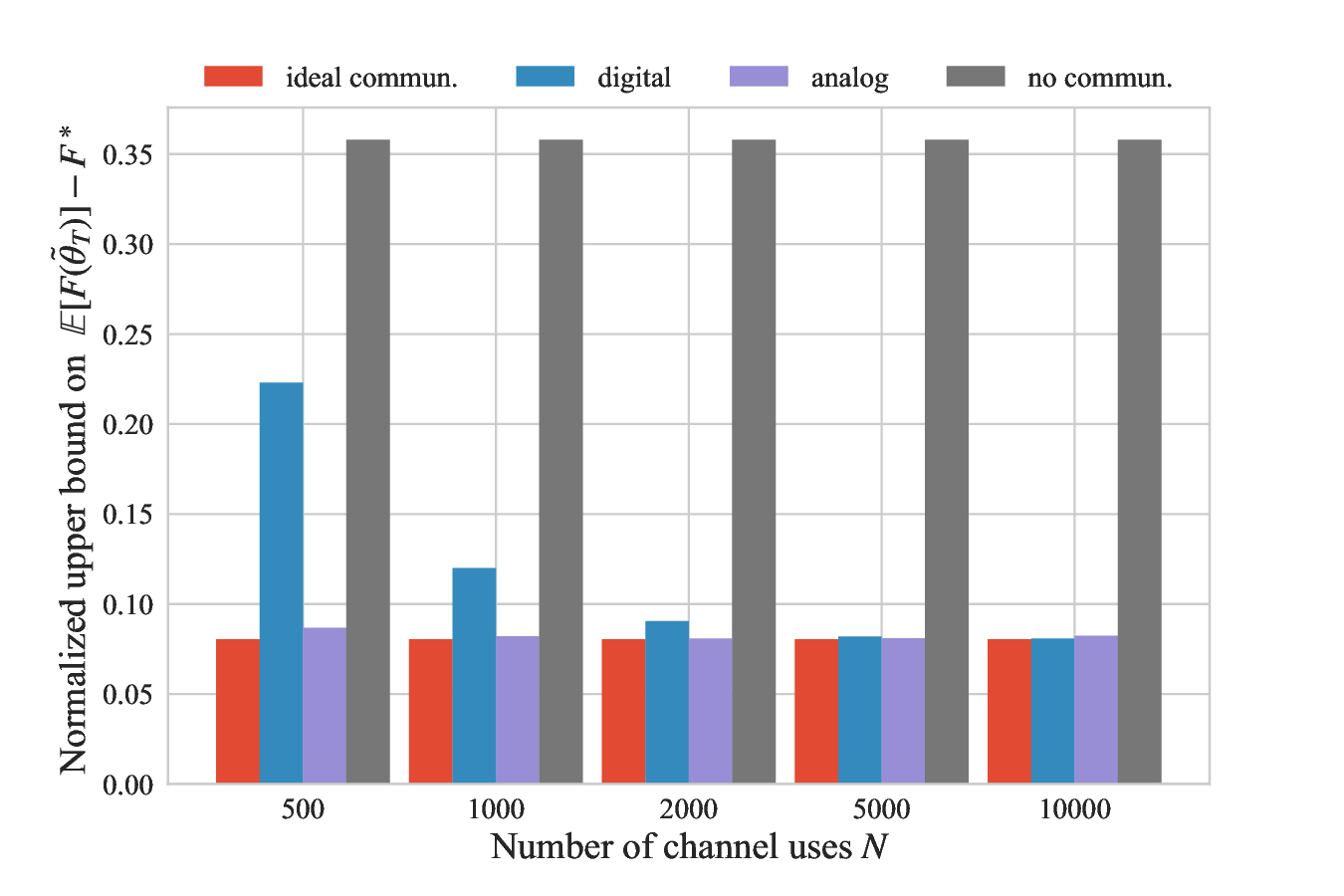}}
	\subfigure[Normalized empirical optimality gap for different schemes. \label{subfig:comparison between empirical optimality gaps of digital and analog schemes}]{\includegraphics[width=4.2in]{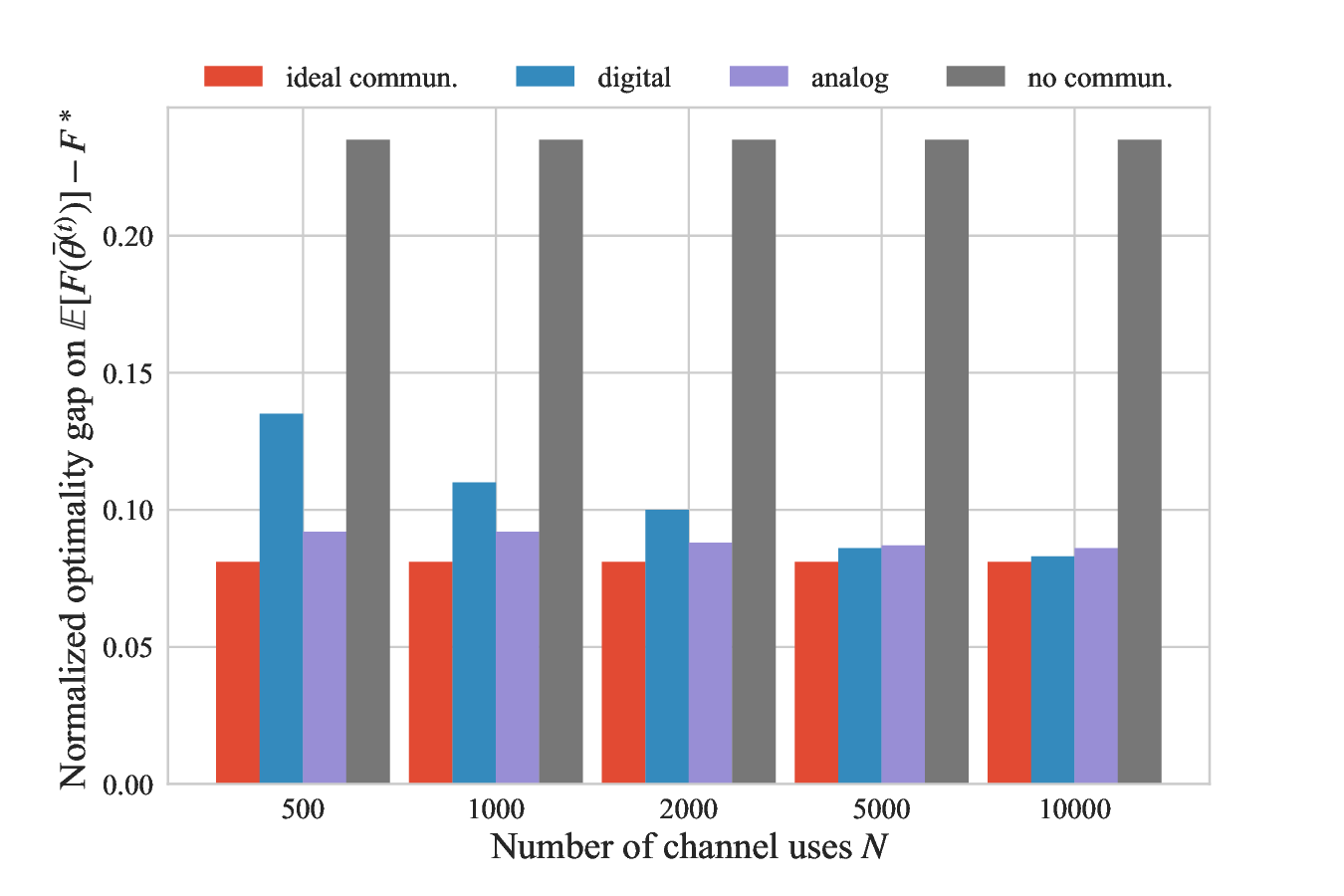}}
	\caption{{\color{black}Performance comparison of analog and digital implementations in terms of normalized (upper bounds on) optimality gap for iteration $t=2500$ for different number $N$ of channel uses with SNR set to $20$ dB over a chain graph.}}\label{fig:digital vs analog}
	\vspace{-.20in}
\end{figure}
   
\section{Conclusions}\label{sec:Conclusions}
This paper has initiated studies on the communication-efficient implementations of DSGD algorithms for wireless FL in fully decentralized architectures. Specifically, we have proposed generic digital and analog transmission protocols tailored to FL over wireless D2D networks by enabling broadcasting for digital transmission, and both broadcasting and AirComp for analog transmission. We also adopted a practically favourable linear compression scheme, RLC, to reduce communication burden, and implemented a consensus step size adaptive to the training iteration. For both implementations, we developed rigorous analysis framework in term of their convergence properties, characterizing the impact of the connectivity topology, the quality of transmission, and/or the channel noise on the optimality gap. Empirical experiments on an image-classification task verified the analytical results as well as the importance of an adaptive consensus step size.

{\color{black}There are several important research directions beyond the scope of this paper that may warrant more research.} {\color{black}First, the channel-inversion based analog AirComp transmission can cause channel noise enhancement, which could be mitigated by designing optimal power control policies beyond channel inversion \cite{cao21feel}. Secondly, it is also important to extend the results in this work to more general learning settings lifting the assumption of $\mu$-$L$ convexity. Moreover, it would be interesting to apply more advanced (compression)-based gossip algorithms \cite{sun21DeFedAvg, vogels20low-rank} for decentralized wireless FL. It is worth emphasizing that the adaptations of these schemes to wireless FL in a D2D architecture would require novel design to cope with the presence of channel impairments. Last but not the least, to address the ``stragglers'' issues in decentralized wireless FL, effective device scheduling, possibly combined with asynchronous training, would be worth investigating along with the corresponding convergence analysis leveraging random graph theory \cite{wang19MATCHA}.}     

\appendices
\section{Algorithm for Digital Implementation}\label{appendix:algorithm for digital implementation}
\begin{algorithm}[htp]
	\vspace{1pt}\caption{{Digital Wireless Implementation} \vspace{1pt}}\label{alg:digital implementation}
	\SetKwInOut{Input}{Input}
	\SetKwInOut{Output}{Output}
	\Input{Consensus step size \(\zeta^{(t)}\), SGD learning step size \(\eta^{(t)}\), connectivity graph \(\mc{G}(\mc{V},\mc{E})\) and mixing matrix \(\mv W\)}
	Initialize at each node \(i\in\mathcal{V}\): \(\mv\theta_i^{(0)}\), \(\hat{\mv\theta}_j^{(0)}=\mv 0\), \(\forall j\in\mathcal{N}_i\bigcup\{i\}\)\;
	\For{\(t=0, 1, \ldots,T-1\) }{
		\For(in parallel){\emph{each device \(i\in\mc{V}\)}}{
			update \(\mv\theta_i^{(t+\Myfrac{1}{2})}=\mv\theta_i^{(t)}-\eta^{(t)}\hat\nabla f_i(\mv\theta_i^{(t)})\)\;
		}
		\For{\emph{slot \(s=1,\ldots,M\)}}{
			\For(in parallel){\emph{each scheduled device $i$ at slot $s$}}{
				broadcast \(\mc{Q}_b\left(\mv A_i^{(t)}(\mv\theta_i^{(t+\Myfrac{1}{2})}-\hat{\mv\theta}_i^{(t)})\right)\)\;
			}
			\For(in parallel){\emph{each neighboring  device \(j\in\mc{N}_i\)}}{
				receive \(\mc{Q}_b\left(\mv A_j^{(t)}(\mv\theta_j^{(t+\Myfrac{1}{2})}-\hat{\mv\theta}_j^{(t)})\right)\)\;
			}
		}
		\For(in parallel){\emph{each device \(i\in\mc{V}\)}}{
			update \eqref{eq:local estimates for digital transmissions} and \eqref{eq:consensus updates}.
		}	
	}
	\Output{\(\mv\theta_i^{(T-1)}\), \(\forall i\in\mathcal{V}\)}
\end{algorithm}

\section{A Scheduling Strategy for the Analog Implementation}\label{appendix:a scheduling strategy for the analog implementation}
\begin{algorithm}[htp]
	\vspace{1pt}{\color{black}\caption{Scheduling Policy for Analog Transmission\vspace{1pt}}}\label{alg:scheduling for analog transmissions}
	\SetKwInOut{Input}{Input}
	\SetKwInOut{Output}{Output}
	\Input{$n=0$, graph \(\mc{G}^{(1)}=\mc{G}(\mc{V},\mc{E})\)}
	Initialize for each node \(i\in\mc{N}_i\): \(\mc{S}_i^{\rm AT}=\mc{S}_i^{\rm AR}=\mc{S}_i^{\rm BT}=\mc{S}_i^{\rm BR}=\emptyset\)\;
	\Repeat{\(\mc{G}^{(n)}=\emptyset\)}{
		update $n=n+1$\;
		color the auxilary graph associated with graph \(\mc{G}^{(n)}\) as described in Section \ref{subsec:Digital Transmission}\;
		choose $c^\ast = \arg\max_c\{d_c^{(n)}\}$\;
		Construct a set \(\mc{N}_{c^\ast}^{(n)}\) of center nodes composed of all nodes in color $c^\ast$\;
		\ForEach{\emph{center node $i\in\mc{N}_{c^\ast}^{(n)}$}}{
			update \(\mc{S}_i^{\rm AR}\leftarrow\mc{S}_i^{\rm AR}\bigcup\{2n-1\}\)\;
			update \(\mc{S}_i^{\rm BT}\leftarrow\mc{S}_i^{\rm BT}\bigcup\{2n\}\)\;
		}
		\ForEach{\emph{center node $i$'s neighbors \(j\in\mc{N}_i\) in graph \(\mc{G}^{(n)}\)}}{
			update \(\mc{S}_j^{\rm AT}\leftarrow\mc{S}_j^{\rm AT}\bigcup\{2n-1\}\)\;
			update \(\mc{S}_j^{\rm BR}\leftarrow\mc{S}_j^{\rm BR}\bigcup\{2n\}\)\;
		}
		update \(\mc{V}=\mc{V}\setminus\big(\mc{N}_{c^\ast}^{(n)}\bigcup\{\mbox{ disconnected node from}\, \mc{G}^{(n)}\}\big)\)\;  update \(\mc{E}=\mc{E}\setminus\big(\bigcup_{i\in\mc{N}_{c^\ast}^{(n)}}\big(\bigcup_{j\in\mc{N}_i}\{(i,j)\}\big)\big)\)\;
		update graph \(\mc{G}^{(n+1)}=\mc{G}(\mc{V},\mc{E})\).
	}
	\Output{\(M=2n\), \(\{\mc{S}_i^{\rm AT}, \mc{S}_i^{\rm AR}, \mc{S}_i^{\rm BT}, \mc{S}_i^{\rm BR}\}\), \(\forall i\in\mc{V}\)}
\end{algorithm}

\begin{figure}[htp]
	\centering
	\subfigure{\includegraphics[width=5.0in]{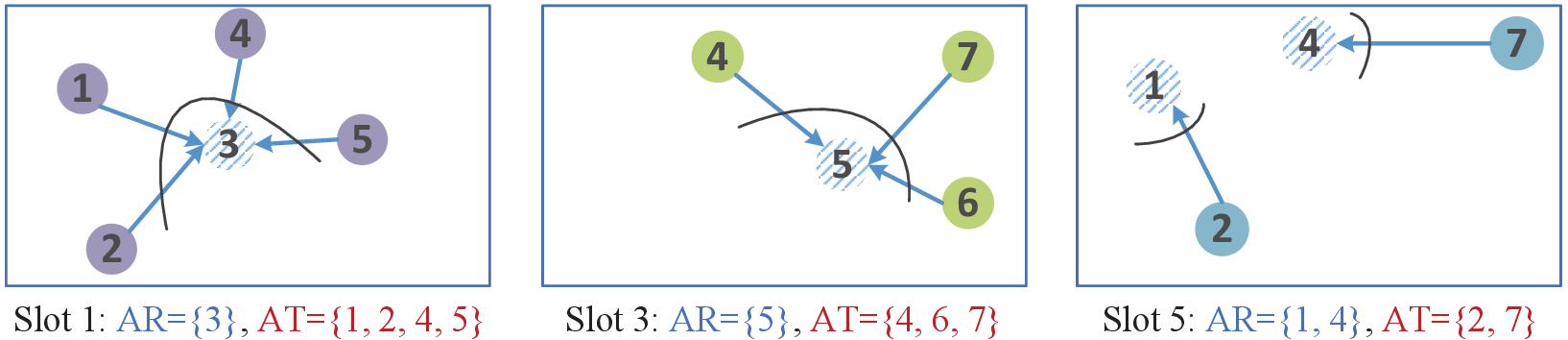}}
	\vfill
	\subfigure{\includegraphics[width=5.0in]{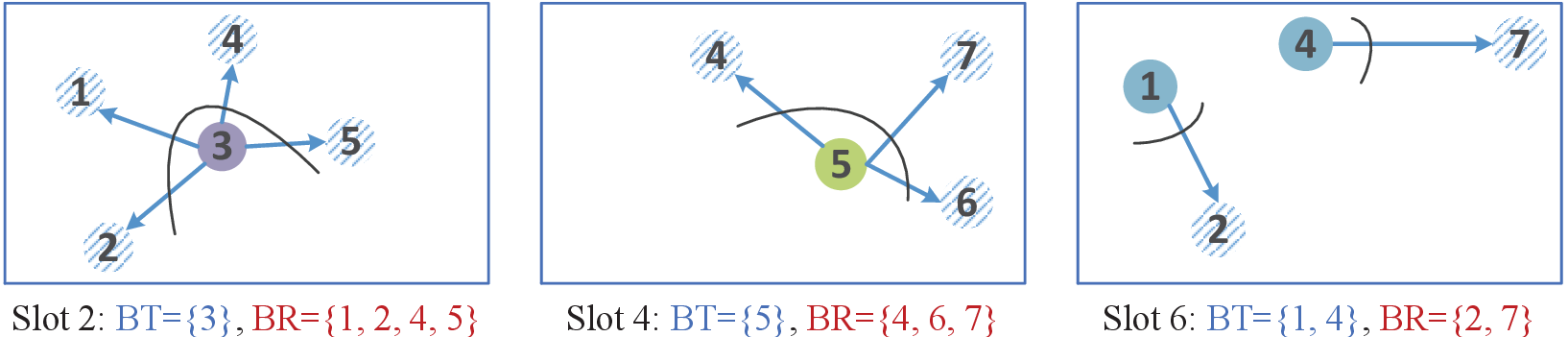}}
	\vspace{-0.1in}
	\caption{\color{black}The outcome of the scheduling policy in Algorithm \ref{alg:scheduling for analog transmissions}, yields $M=6$ transmission slots and the following sets: \(\mc{S}_1^{\rm AT}=\{1\}\), \(\mc{S}_1^{\rm BT}=\{6\}\), \(\mc{S}_1^{\rm AR}=\{5\}\), \(\mc{S}_1^{\rm BR}=\{2\}\); \(\mc{S}_2^{\rm AT}=\{1,5\}\), \(\mc{S}_2^{\rm BR}=\{2,6\}\); \(\mc{S}_3^{\rm BT}=\{2\}\), \(\mc{S}_3^{\rm AR}=\{1\}\); \(\mc{S}_4^{\rm AT}=\{1,3\}\), \(\mc{S}_4^{\rm BT}=\{6\}\), \(\mc{S}_4^{\rm AR}=\{5\}\), \(\mc{S}_4^{\rm BR}=\{2,4\}\); \(\mc{S}_5^{\rm AT}=\{1\}\), \(\mc{S}_5^{\rm BT}=\{4\}\), \(\mc{S}_5^{\rm AR}=\{3\}\), \(\mc{S}_5^{\rm BR}=\{2\}\); \(\mc{S}_6^{\rm AT}=\{3\}\), \(\mc{S}_6^{\rm BR}=\{4\}\); and \(\mc{S}_7^{\rm AT}=\{3,5\}\), \(\mc{S}_7^{\rm BR}=\{4,6\}\).}\label{fig:an example for analog scheduling}
	\vspace{-.2in}
\end{figure}

\section{Algorithm for Analog Implementation}\label{appendix:algorithm for analog implementation}
\begin{algorithm}[htp]
	\caption{{Analog Wireless Implementation}\vspace{1pt}}\label{alg:analog implementation}
	\SetKwInOut{Input}{Input}
	\SetKwInOut{Output}{Output}
	\Input{Consensus step size \(\zeta^{(t)}\), SGD learning step size \(\eta^{(t)}\), connectivity graph \(\mc{G}(\mc{V},\mc{E})\), mixing matrix \(\mv W\), transmission slots \(M=2n\), and subsets \(\{\mc{S}_i^{\rm BT}, \mc{S}_i^{\rm BR}, \mc{S}_i^{\rm AT}, \mc{S}_i^{\rm AR}\}\), \(i\in\mc{V}\)}
	Initialize at each node \(i\in\mathcal{V}\):  \(\mv\theta_i^{(0)}\), \(\hat{\mv\theta}_j^{(0)}=\mv 0\), \(\forall j\in\mc{N}_i\cup\{i\}\),  and \(\hat{\mv{y}}_i^{(0)}=\mv 0\)\;
	\For{\(t=0,1,\ldots,T-1\)}{
		perform Line $3$ - $5$ in Algorithm \ref{alg:digital implementation}\;
		\For({{\it AirComp transmission}}){\emph{slot \(s=1,3,\ldots,2n-1\)}}{
			\For(in parallel){\emph{each node $i$ in \(\{i\in\mc{V}\mid s\in\mc{S}_i^{\rm AT}\}\)}}{
				transmit to a center node $j$ \eqref{eq:AT's transmitted signal}\;
			}
			\For(in parallel){\emph{each center node $j$ in \(\{j\in\mc{V}\mid s\in\mc{S}_j^{\rm AR}\}\)}}{
				receive \eqref{eq:AR's received signal} and estimate \eqref{eq:AR's estimated model parameter}\;
			}
		}
		\For({{\it BC transmission}}){\emph{slot \(s=2,4,\ldots,2n\)}}{
			\For(in parallel){\emph{each node $i$ in \(\{i\in\mc{V}\mid s\in\mc{S}_i^{\rm BT}\}\)}}{
				broadcast \eqref{eq:BT's transmitted signal}\;
			}
			\For(in parallel){\emph{each node $j$ in \(\{j\in\mc{V}\mid s\in\mc{S}_j^{\rm BR}\}\)}}{
				receive \eqref{eq:BR's received signal} and 
				estimate \eqref{eq:BR's estimated model parameter} \; 	
			}
		}
		\For(in parallel){\emph{each device \(j\in\mc{V}\)}}{
			update \eqref{eq:estimate of the combined models}, \eqref{eq:estimate of one's own model} and \eqref{eq:consensus updates for analog scheme}.
		}	
	}
	\Output{\(\mv\theta_i^{(T-1)}\), \(\forall i\in\mathcal{V}\)}
\end{algorithm}

\section{Proof of Lemma~\ref{lemma:MSE for estimation in digital transmissions}}\label{appendix:proof of MSE for estimation in digital transmissions}
First, as per the definition of the linear encoding matrix \(\mv A=\tfrac{1}{\sqrt{m}}\mv H\mv R\), it follows that \(\mv A\mv A^T=\tfrac{1}{m}\mv H\mv R\mv R\mv H^T=\tfrac{1}{m}\mv H\mv H^T=\tfrac{d}{m}\mv I\).\footnote{We will omit the superscript \((\cdot)^{(t)}\) whenever it does not cause ambiguity throughout the appendices.} In addition, by denoting the $l$th row of the partial Hadamard matrix \(\mv H\) by \(\mv h_l^T\in\mb{R}^{1\times d}\), \(l=1,\ldots,m\), \(\mb E[\mv A^T\mv A]\) is given by
\begin{align}
	\mb E[\mv A^T\mv A]
	& = \frac{1}{m}\mb E[\mv R[\mv h_1, \ldots, \mv h_m][\mv h_1, \ldots, \mv h_m]^T\mv R]\notag\\
	& = \frac{1}{m}\mb E[[\mv R\mv h_1, \ldots, \mv R\mv h_m][\mv R\mv h_1, \ldots, \mv R\mv h_m]^T]\notag\\
	& = \frac{1}{m}\sum\limits_{l=1}^m\mb E[(\mv R\mv h_l)(\mv R\mv h_l)^T]\notag\\
	& = \frac{1}{m}\sum\limits_{l=1}^m\mb E\begin{bmatrix}
		r_1^2h_{l,1}^2 & \ldots & r_1r_dh_{l,1}h_{l,d}\\
		\vdots & \ddots & \vdots\\
		r_1r_dh_{l,1}h_{l,d} & \ldots & r_d^2h_{l,d}^2
	\end{bmatrix}\stackrel{(a)}{=}\mv I,\label{eq:expectation of A^TA}
\end{align} where \(h_{l,n}\) denotes the $n$th entry of \(\mv h_l\), \(n=1,\ldots, d\); and \((a)\) is due to the fact that \(r_n^2h_{l,n}^2=1\) for all \(l=1,\ldots, m\), as well as \(\mb E[r_{n_1}r_{n_2}]=0\) for all \(n_1\neq n_2\). 
Then given the number of rows of \(\mv A\), i.e., $m$, fixed, for all \(\mv u\in\mb{R}^{d\times 1}\), on average over RLC, it follows that
\begin{align}
	\mb{E}\left\|\mv u-\frac{m}{d}\mv A^T\mv A\mv u\right\|^2 
	& = \mb E\left[\mv u^T\mv u-\frac{2m}{d}\mv u^T\mv A^T\mv A\mv u+\left(\frac{m}{d}\right)^2\mv u^T\mv A^T\mv A\mv A^T\mv A\mv u\right]\notag\\
	& \stackrel{(a)}{=}\|\mv u\|^2-\frac{m}{d}\mv u^T\mb E[\mv A^T\mv A]\mv u\notag\\
	& \stackrel{(b)}{=}\left(1-\frac{m}{d}\right)\|\mv u\|^2,\label{eq:contraction equality for compression}
\end{align} where \((a)\) comes from \(\mv A\mv A^T=\tfrac{d}{m}\mv I\) and \((b)\) is obtained by plugging \eqref{eq:expectation of A^TA} into \((a)\). As a result, denoting $\tfrac{m_i}{d}$ by $\omega_i$, \(i\in\mc{V}\), we have
\begin{align}
	& \mb{E}\left\|\mv u-\frac{m_i}{d}\mv A_i^T\mc{Q}_b(\mv A_i\mv u)\right\|^2=
	\mb E\left\|\left(\mv u-\frac{m_i}{d}\mv A_i^T\mv A_i\mv u\right)+\left(\frac{m_i}{d}\mv  A_i^T\mv A_i\mv u-\frac{m_i}{d}\mv A_i^T\mc{Q}_b(\mv A_i\mv u)\right)\right\|^2\notag\\
	& \stackrel{(a)}{=} \mb E\left[\left\|\mv u-\frac{m_i}{d}\mv A_i^T\mv A_i\mv u\right\|^2+2\frac{m_i}{d}\left(\mv u-\frac{m_i}{d}\mv A_i^T\mv A_i\mv u\right)^T\mv A_i^T\mv Q_b\mv A_i\mv u+\frac{m_i^2}{d^2}\left\|\mv A_i^T\mv Q_b\mv A_i\mv u\right\|^2\right]\notag\\
	& \stackrel{(b)}{=} \mb E\left\|\mv u-\frac{m_i}{d}\mv A_i^T\mv A_i\mv u\right\|^2+\frac{m_i^2}{d^2}\mb E\left\|\mv A_i^T\mv Q_b\mv A_i\mv u\right\|^2\notag\\
	& \stackrel{(c)}{=} \left(1-\frac{m_i}{d}+\frac{m_i}{d}\epsilon_b^2\right)\|\mv u\|^2\notag\\
	& \stackrel{(d)}{\approx}\left(1-\frac{m_i}{d}\right)\|\mv u\|^2 \label{eq:contraction approx. for compression}\\
	& \le(1-\min\limits_{i\in\mc{V}}\omega_i)\|\mv u\|^2\notag 
\end{align}
where \((a)\) is because of \(\mv A_i\mv u-\mc{Q}_b(\mv A_i\mv u)=\mv Q_b\mv A\mv u\) in accordance with IEEE 754 standard, in which \(\mv Q_b\in\mb{R}^{m_i\times m_i}\) is a diagonal matrix with its $i$th diagonal entry satisfying \(|[\mv Q_b]_{kk}|<\epsilon_b\), \(k=1,\ldots,m_i\); \((b)\) is due to the fact \(\mv A_i\mv A_i^T=\tfrac{d}{m_i}\mv I\); and \((c)\) follows from \eqref{eq:contraction equality for compression} and \(\mb E[\mv A_i^T(\mv Q_b)^2\mv A_i]=\tfrac{1}{m_i}\sum_{k=1}^{m_i}|[\mv Q_b]_{kk}|^2\mv I\le\epsilon_b^2\mv I\). In addition, considering very little value of \(\epsilon_b\) in common double-precision floating number systems  (\(\epsilon_b=2^{-24}\) for $b=32$ and \(\epsilon_b=2^{-53}\) for $b=52$) \cite[ch. 2]{higham02accuracy}, we safely approximate \((c)\) by \((d)\). Lemma~\ref{lemma:MSE for estimation in digital transmissions} is thus proved.

\section{Proof of Corollary~\ref{corollary:MSE for estimation in Rayleigh fading}}\label{appendix:proof of MSE for estimation in Rayleigh fading}
Note that \eqref{eq:contraction approx. for compression} implies \(m_i\le d\), for all \(i\in\mc{V}\). We thus assume that when the channel condition can support more than $d$ rows for the linear encoding matrix \(\mv A_i\), \(\mv A_i\in\mb{R}^{d\times d}\) is kept. That said, the MSE corresponding to cases when \(m_i\ge d\) are all equal to zero. As a result, when \(m_i^{(t)}\) varies with channel state at the $t$th iteration, on average over RLC and the Rayleigh-fading channels, according to the law of iterated expectation,  we have 
\begin{align}
	\mb{E}\left\|\mv u-\frac{m_i}{d}\mv A_i^T\mc{Q}_b(\mv A_i\mv u)\right\|^2
	& = \mb E_{m_i}\left[\mb E{}_{\rm RLC}\left[\left.\kern-\nulldelimiterspace\left\|\mv u-\frac{m_i}{d}\mv A_i^T\mc{Q}_b(\mv A_i\mv u)\right\|^2 \right|m_i\right]\right]\notag\\
	& \stackrel{\eqref{eq:contraction approx. for compression}}{\approx} \mb E_{m_i< d}\left[\left(1-\frac{m_i}{d}\right)\right]\|\mv u\|^2\notag\\
	& = \sum_{n=0}^d\Pr(m_i=n)\left(1-\frac{m_i}{d}\right). \label{eq:contraction property averaged over fading}
\end{align}

Next, it remains to calculate \(\Pr(m_i=n)\) for \(n=0,\ldots,d-1\), \(i\in\mc{V}\). For $n=0$, it follows that
\begin{align}
	\Pr(m_i=0) & =\Pr(\Myfrac{B_i}{b}<1)\notag\\
	& =\Pr\left(\min\limits_{j\in\mc{N}_i}|h_{ij}^\prime|^2<\frac{N_0}{PM}\left(2^{b\frac{M}{N}}-1\right)\right)\notag\\
	&\stackrel{(a)}{=}1-\exp\left(-\frac{N_0}{P}\frac{1}{MA_0}\left(2^{b\frac{M}{N}}-1\right)\sum\limits_{j\in\mc{N}_i}\left(\frac{d_{ij}}{d_0}\right)^r\right)\notag\\
	& \stackrel{\eqref{eq:G_i}}{=} 1-G_i(1),
\end{align}
where \((a)\) follows from the fact that \(|h_{ij}^\prime|^2\), \(j\in\mc{N}_i\), are exponentially and independently distributed with mean value \(A_0(\tfrac{d_0}{d_{ij}})^r\) such that \(\Pr(\min_{j\in\mc{N}_i}|h_{ij}^\prime|^2<x)=1-\exp(-\tfrac{x}{A_0}\sum_{j\in\mc{N}_i}(\tfrac{d_{ij}}{d_0})^r)\). Moreover, for \(n=1,\ldots,d-1\), it is easy to show that \(\Pr(m_i=n)=\Pr(n\le\Myfrac{B_i}{b}<n+1)=G_i(n)-G_i(n+1)\).
To sum up, the probability mass function (PMF) of $m_i$, \(i\in\mc{V}\), is given by 
\begin{align}
	\Pr(m_i=n)=\left\{\begin{array}{ll} 1-G_i(1),& n=0, \\
		G_i(n)-G_i(n+1), & n=1,\ldots,d-1, \\
		G_i(d), & n\ge d.
	\end{array}\right. \label{eq:PMF for m_i}
\end{align}

Finally, Plugging \eqref{eq:PMF for m_i} into \eqref{eq:contraction property averaged over fading}, we arrive at 
\begin{align}
	\mb{E}\left\|\mv u-\frac{m_i}{d}\mv A_i^T\mc{Q}_b(\mv A_i\mv u)\right\|^2 & \approx\sum_{n=0}^d\Pr(m_i=n)\left(1-\frac{m_i}{d}\right)\notag\\
	& = 1-G_i(1)+\sum_{n=1}^{d-1}(G_i(n)-G_i(n+1))\left(1-\frac{n}{d}\right), \label{eq:temp}
\end{align} which, after some manipulation employing telescoping sums, implies \eqref{eq:digital compression operator averaged over channel fading} in Corollary~\ref{corollary:MSE for estimation in Rayleigh fading}.

\section{Proof of Lemma~\ref{lemma:reformulation of the noisy consensus updates}}\label{appendix:proof of reformulation of the noisy consensus updates}
We substitute \eqref{eq:AR's received signal} and \eqref{eq:BR's received signal} into \eqref{eq:AR's estimated model parameter} and \eqref{eq:BR's estimated model parameter}, respectively, and rewrite \eqref{eq:estimate of the combined models} as 
\begin{multline}
	\hat{\mv y}_j^{(t+1)} = \hat{\mv y}_j^{(t)} + \frac{m}{d}(\mv A^{(t)})^T\mv A^{(t)}\Big(\sum\limits_{s\in\mc{S}_j^{\rm AR}}\sum\limits_{i\in\mc{N}_j^{(s)}}w_{ji}(\mv\theta_i^{(t+\Myfrac{1}{2})}-\hat{\mv\theta}_i^{(t)})+\sum\limits_{s\in\mc{S}_j^{\rm BR}}w_{ji_s}(\mv\theta_{i_s}^{(t+\Myfrac{1}{2})}-\hat{\mv\theta}_{i_s}^{(t)})\Big)\\
	+\frac{m}{d}(\mv A^{(t)})^T\tilde{\mv n}_j^{(t)}, \label{eq:rewritten estimate of the combined models}
\end{multline} 
in which \(\tilde{\mv n}_j^{(t)}\) is the effective noise defined as
\begin{align}
	\tilde{\mv n}_j^{(t)}=\sum\limits_{s\in\mc{S}_j^{\rm AR}}\frac{\Re\{\mv n_j^{(t,s)}\}}{\sqrt{\gamma_j^{(t,s)}}}+\sum\limits_{s\in\mc{S}_j^{\rm BR}}\frac{w_{ji_s}}{\sqrt{\alpha_{i_s}^{(t,s)}}}\Re\left\{\frac{\mv n_j^{(t,s)}}{h_{i_sj}^{\prime(t)}}\right\},
\end{align}
which is also Gaussian with zero mean and the covariance matrix calculated as 
\begin{align}
	\mb{E}[\tilde{\mv n}_j^{(t)}(\tilde{\mv n}_j^{(t)})^T] = &  \kern-2pt\sum\limits_{s\in\mc{S}_j^{\rm AR}}\kern-2pt\frac{1}{\gamma_j^{(t,s)}}\mb{E}[\Re\{\mv n_j^{(t,s)}\}(\Re\{\mv n_j^{(t,s)}\})^T]+\kern-2pt\sum\limits_{s\in\mc{S}_j^{\rm BR}}\kern-2pt\frac{w_{ji_s}^2}{\alpha_{i_s}^{(t,s)}}\mb{E}\bigg[\Re\Big\{\frac{\mv n_j^{(t,s)}}{h_{i_sj}^{\prime(t)}}\Big\}\Big(\Re\Big\{\frac{\mv n_j^{(t,s)}}{h_{i_sj}^{\prime(t)}}\Big\}\Big)^T\bigg]\notag\\
	\stackrel{(a)}{=} &  \frac{N_0}{2}\bigg(\sum\limits_{s\in\mc{S}_j^{\rm AR}}\frac{1}{\gamma_j^{(t,s)}}+\sum\limits_{s\in\mc{S}_j^{\rm BR}}\frac{w_{ji_s}^2}{\alpha_{i_s}^{(t,s)}\big|h_{i_sj}^{\prime(t)}\big|^2}\bigg)\mv I, \label{eq:covariance matrix of the effective noise}
\end{align} where $(a)$ is due to the facts that the entries \(\mv n_j^{(t,s)}\) are $i.i.d.$ with each denoted by \(n_{ji}^{(t,s)}\sim\mc{CN}(0,N_0)\), and the real and the image part of \(n_{ji}^{(t,s)}\) are also independent Gaussian random variables with zero mean and variance \(\tfrac{N_0}{2}\), thus leading to \(\sum\limits_{s\in\mc{S}_j^{\rm BR}}\tfrac{w_{ji_s}^2}{\alpha_{i_s}^{(t,s)}}\mb{E}[\Re\{\tfrac{\mv n_j^{(t,s)}}{h_{i_sj}^{\prime(t)}}\}(\Re\{\frac{\mv n_j^{(t,s)}}{h_{i_sj}^{\prime(t)}}\})^T]=\tfrac{N_0}{2|h_{i_sj}^{\prime(t)}|^2}\mv I\). Then specifying \(\alpha_{i_s}^{(t,s)}\) and \(\gamma_j^{(t,s)}\) by \eqref{C:power constraint per BT's slot} and \eqref{C:power constraint per AT's slot}, respectively, we obtain \eqref{eq:variance of the effective noise}.

Furthermore, the scheduling scheme in \ref{alg:scheduling for analog transmissions} suggests that each node is scheduled to receive as a center node for \emph{at most one} slot (cf.~Line 13 in Algorithm \ref{alg:scheduling for analog transmissions}), i.e., \(|\mc{S}_j^{\rm AR}|\le 1\), while it may be scheduled to receive from a broadcast Tx for \emph{multiple} slots. In all cases, the estimate update \eqref{eq:rewritten estimate of the combined models} always aggregates model parameters from all of node $j$'s neighbors. Hence, \eqref{eq:rewritten estimate of the combined models} can be simplified as 
\begin{align}
	\hat{\mv y}_j^{(t+1)} = \hat{\mv y}_j^{(t)} + \frac{m}{d}(\mv A^{(t)})^T\mv A^{(t)}\sum\limits_{i\in\mc{N}_j}w_{ji}\mv u_i^{(t)}+\frac{m}{d}(\mv A^{(t)})^T\tilde{\mv n}_j^{(t)}, \label{eq:recursive estimate of the combined models}
\end{align} 
where \(\mv\theta_i^{(t+\Myfrac{1}{2})}-\hat{\mv\theta}_i^{(t)}\triangleq\mv u_i^{(t)}\). By recursively applying \eqref{eq:recursive estimate of the combined models}, it follows, for any device \(i\in\mc{V}\), that  
\begin{align}
	\hat{\mv y}_i^{(t+1)}=\hat{\mv y}_i^{(0)}+\sum_{\tau=0}^t\left(\frac{m}{d}(\mv A^{(\tau)})^T\mv A^{(\tau)}\sum_{j\in\mc{N}_i}w_{ij}\mv u_j^{(\tau)}\right)+\sum_{\tau=0}^t\frac{m}{d}(\mv A^{(\tau)})^T\tilde{\mv n}_i^{(\tau)}. \label{eq:exact estiamte of the combined models}
\end{align}

On the other hand, applying weighted sum over \(j\in\mc{N}_i\) on both sides of \eqref{eq:estimate of one's own model}, it follows that
\begin{align}
	\sum_{j\in\mc{N}_i}w_{ij}\hat{\mv\theta}_j^{(t+1)}=\sum_{j\in\mc{N}_i}w_{ij}\hat{\mv\theta}_j^{(t)}+\frac{m}{d}(\mv A^{(t)})^T\mv A^{(t)}\sum_{j\in\mc{N}_i}w_{ij}\mv u_j^{(t)}. \label{eq:recursive estimate of the combined hat theta}
\end{align}
Recursively applying \eqref{eq:recursive estimate of the combined hat theta} leads to
\begin{align}
	\sum_{j\in\mc{N}_i}w_{ij}\hat{\mv\theta}_j^{(t+1)}=\sum_{j\in\mc{N}_i}w_{ij}\hat{\mv\theta}_i^{(0)}+\sum_{\tau=0}^t\left(\frac{m}{d}(\mv A^{(\tau)})^T\mv A^{(\tau)}\sum_{j\in\mc{N}_i}w_{ij}\mv u_j^{(\tau)}\right). \label{eq:exact estimate of the combined hat theta}
\end{align}
Then, combining \eqref{eq:exact estiamte of the combined models} and \eqref{eq:exact estimate of the combined hat theta} with the fact \(\hat{\mv y}_i^{(0)}=\sum_{j\in\mc{N}_i}w_{ij}\hat{\mv\theta}_i^{(0)}=\mv 0\), we have 
\begin{align}
	\hat{\mv y}_i^{(t+1)}=\sum_{j\in\mc{N}_i}w_{ij}\hat{\mv\theta}_j^{(t+1)}+\sum_{\tau=0}^t\frac{m}{d}(\mv A^{(\tau)})^T\tilde{\mv n}_i^{(\tau)}. \label{eq:relation between hat y and hat theta}
\end{align}
By substituting \eqref{eq:relation between hat y and hat theta} for \(\hat{\mv y}_i^{(t+1)}\) in device $i$'s consensus update for analog implementation, \eqref{eq:consensus updates for analog scheme} can be recast as \eqref{eq:recast consensus updates for analog scheme}. 

\section{Proof of Lemma~\ref{lemma:recursive upper-bound for the error sequence}}\label{appendix:proof of recursive upper-bound for the error sequence}
To prove Lemma \ref{lemma:recursive upper-bound for the error sequence}, we first provide variants of \cite[Lemma 17 and Lemma 18]{koloskova19decentralized}, where matrix notations are used for the simplicity of notation.
\begin{lemma}[Variant of {\cite[Lemma 17]{koloskova19decentralized}}]
	Denoting \([\mv\theta_1^{(t)},\ldots,\mv\theta_K^{(t)}]\in\mb{R}^{d\times K}\), \([\hat{\mv\theta}_1^{(t)},\ldots,\hat{\mv\theta}_K^{(t)}]\in\mb{R}^{d\times K}\), \([\bar{\mv\theta}^{(t)},\ldots,\bar{\mv\theta}^{(t)}]\in\mb{R}^{d\times K}\) and \([\tilde{\mv n}_1^{(t)},\ldots,\tilde{\mv n}_K^{(t)}]\) \(\in\mb{R}^{d\times K}\), by \(\mv\Theta^{(t)}\), \(\hat{\mv\Theta}^{(t)}\), \(\bar{\mv\Theta}^{(t)}\) and \(\tilde{\mv N}^{(t)}\), respectively, then for consensus step size \(\zeta^{(t)}\ge 0\), mixing matrix \(\mv W\) and any parameter \(\alpha_1>0\), on average over RLC and AWGN, we have
	\begin{multline}
		\mb{E}\|\mv\Theta^{(t+1)}-\bar{\mv\Theta}^{(t+1)}\|_F^2\le(1+\alpha_1)\left(1-\delta\zeta^{(t)}\right)^2\mb {E}\|\mv\Theta^{(t+\Myfrac{1}{2})}-\bar{\mv\Theta}^{(t+\Myfrac{1}{2})}\|_F^2+\\ 
		(1+\alpha_1^{-1})\beta^2(\zeta^{(t)})^2\mb{E}\|\mv\Theta^{(t+\Myfrac{1}{2})}-\hat{\mv\Theta}^{(t+1)}\|_F^2+ (\zeta^{(t)})^2\omega^2d\sum\limits_{\tau=0}^t\tilde N_0^{(\tau)}, \label{eq:variant of Lemma 17}
	\end{multline} where \(\tilde N_0^{(\tau)}=\sum_{i\in\mc{V}}\tilde N_{0i}^{(\tau)}\) is the sum-variance of effective noise over all devices.
	\label{lemma:variant of Lemma 17}
\end{lemma}
\begin{IEEEproof}
	With the fact \((\mv W-\mv I)\Myfrac{\mv 1\mv 1^T}{K}=\mv 0\), rewrite \eqref{eq:recast consensus updates for analog scheme} and \eqref{eq:model average for analog scheme} in matrix forms as \(
	\mv\Theta^{(t+1)}=\mv\Theta^{(t+\Myfrac{1}{2})}+\zeta^{(t)}\hat{\mv\Theta}^{(t+1)}(\mv W-\mv I)+\zeta^{(t)}\omega\sum_{\tau=0}^t(A^{(\tau)})^T\tilde{\mv N}^{(\tau)} \) and \(\bar{\mv\Theta}^{(t+1)}=\bar{\mv\Theta}^{(t+\Myfrac{1}{2})}+\zeta^{(t)}\omega\sum_{\tau=0}^t(A^{(\tau)})^T\tilde{\mv N}^{(\tau)}\Myfrac{\mv 1\mv 1^T}{K}\), respectively. Then, we have
	\begin{align*}
		& \mb{E}\|\mv\Theta^{(t+1)}-\bar{\mv\Theta}^{(t+1)}\|_F^2  \\
		=& \mb{E}\Big\|\mv\Theta^{(t+\Myfrac{1}{2})}-\bar{\mv\Theta}^{(t+\Myfrac{1}{2})}+\zeta^{(t)}\hat{\mv\Theta}^{(t+1)}(\mv W-\mv I)
		+\zeta^{(t)}\omega\sum_{\tau=0}^t(\mv A^{(\tau)})^T\tilde{\mv N}^{(\tau)}\Big(\mv I-\frac{\mv 1\mv 1^T}{K}\Big)\Big\|_F^2\\
		\stackrel{(a)}{=}&  \mb{E}\Big\|\mv\Theta^{(t+\Myfrac{1}{2})}-\bar{\mv\Theta}^{(t+\Myfrac{1}{2})}+\zeta^{(t)}\hat{\mv\Theta}^{(t+1)}(\mv W-\mv I)\Big\|_F^2+(\zeta^{(t)})^2\omega^2\mb{E}\Big\|\sum_{\tau=0}^t(\mv A^{(\tau)})^T\tilde{\mv N}^{(\tau)}\Big(\mv I-\frac{\mv 1\mv 1^T}{K}\Big)\Big\|_F^2\\
		\stackrel{(b)}{\le} &\  \mb{E}\Big\|(\mv\Theta^{(t+\Myfrac{1}{2})}-\bar{\mv\Theta}^{(t+\Myfrac{1}{2})})(\mv I+\zeta^{(t)}(\mv W-\mv I))+\zeta^{(t)}(\hat{\mv\Theta}^{(t+1)}-\mv\Theta^{(t+\Myfrac{1}{2})})(\mv W-\mv I)\Big\|_F^2\\
		& +(\zeta^{(t)})^2\omega^2d\left(1-\frac{1}{K}\right)\sum_{\tau=0}^t\tilde N_0^{(\tau)} 
		\stackrel{(c)}{\le}\ \mbox{RHS of \eqref{eq:variant of Lemma 17}},
	\end{align*}
	where $(a)$ is due to the expectation taken over Gaussian noise and RLC, i.e., \(\mb{E}[(\mv A^{(\tau)})^T\tilde{\mv N}^{(\tau)}]\), is zero;  the first term in $(b)$ is as a result of \(\bar{\mv X}(\mv W-\mv  I)=\mv 0\) for any \(\mv X\in\mb{R}^{d\times K}\); the second term in $(b)$ is based on \(\mv A^{(\tau)}(\mv A^{(\tau)})^T=\tfrac{d}{m}\mv I\) and $i.i.d.$ entries of \(\tilde{\mv n}_i^{(\tau)}\) such that \(\mb{E}\|\tilde{\mv n}_i^{(\tau)}\|^2=m\tilde N_{0i}^{(\tau)}\) with \(\sum_{i\in\mc{V}}\tilde N_{0i}^{(\tau)}\triangleq\tilde N_0^{(\tau)}\); and $(c)$ follows \cite[Lemma 17]{koloskova19decentralized}.
\end{IEEEproof}

\begin{lemma}[Variant of {\cite[Lemma 18]{koloskova19decentralized}}]
	In addition to \(\mv\Theta^{(t)}\), \(\hat{\mv\Theta}^{(t)}\), \(\bar{\mv\Theta}^{(t)}\) and \(\tilde{\mv N}^{(t)}\) denoted as in Lemma \ref{lemma:variant of Lemma 17}, denoting \([\hat{\nabla}f_1(\mv\theta_1^{(t)}),\ldots, \hat{\nabla}f_K(\mv\theta_K^{(t)})]\in\mb{R}^{d\times K}\) by \(\hat{\nabla}F^{(t)}\), then for consensus step size \(\zeta^{(t)}\ge 0\), mixing matrix \(\mv W\) and any parameter \(\alpha_2>0\), on average over RLC and AWGN, we have
	\begin{multline}
		\mb {E}\|\mv\Theta^{(t+\Myfrac{3}{2})}-\hat{\mv\Theta}^{(t+2)}\|_F^2\le(1+\alpha_2^{-1})(1-\omega)\beta^2(\zeta^{(t)})^2
		\mb {E}\|\mv\Theta^{(t+\Myfrac{1}{2})}-\eta^{(t+1)}\hat{\nabla}F^{(t+1)}-\bar{\mv\Theta}^{(t+\Myfrac{1}{2})}\|_F^2+\\
		(1+\alpha_2)(1-\omega)\left(1+\beta\zeta^{(t)}\right)^2\mb {E}\|\mv\Theta^{(t+\Myfrac{1}{2})}-\eta^{(t+1)}\hat{\nabla}F^{(t+1)}-\hat{\mv\Theta}^{(t+1)}\|_F^2+\\
		(\zeta^{(t)})^2(1-\omega)\omega^2d\sum_{\tau=0}^t\tilde N_0^{(\tau)}. \label{eq:variant of Lemma 18}                                                                                                                                               
	\end{multline}
	\label{lemma:variant of Lemma 18}
\end{lemma}
\begin{IEEEproof}
	Given the constant \(\omega=\Myfrac{m}{d}\) in analog implementation, \eqref{eq:estimate of one's own model} implies that, for \(i\in\mc{V}\),
	\begin{align}
		\hat{\mv\Theta}_i^{(t+1)}=\hat{\mv\Theta}_i^{(t)}+\omega(\mv A^{(t)})^T\mv A^{(t)}(\mv\Theta_i^{(t+\Myfrac{1}{2})}-\hat{\mv\Theta}_i^{(t)}). \label{eq:matrix-form estimate of the model}
	\end{align} 
	Therefore, substituting \eqref{eq:matrix-form estimate of the model} for \(\hat{\mv\Theta}^{(t+2)}\), if follows that
	\begin{align*}
		& \mb {E}\|\mv\Theta^{(t+\Myfrac{3}{2})}-\hat{\mv\Theta}^{(t+2)}\|_F^2  
		= \mb {E}\Big\|(\mv\Theta^{(t+\Myfrac{3}{2})}-\hat{\mv\Theta}_i^{(t+1)})-\omega(\mv A^{(t+1)})^T\mv A^{(t+1)}(\mv\Theta_i^{(t+\Myfrac{3}{2})}-\hat{\mv\Theta}_i^{(t+1)})\Big\|_F^2\\ 
		\stackrel{(a)}{=} & (1-\omega)\mb{E}\|\mv\Theta^{(t+\Myfrac{3}{2})}-\hat{\mv\Theta}^{(t+1)}\|_F^2\stackrel{(b)}{=}(1-\omega)\mb{E}\|\mv\Theta^{(t+1)}-\eta^{(t+1)}\hat{\nabla}F^{(t+1)}-\hat{\mv\Theta}^{(t+1)}\|_F^2\\
		\stackrel{{(c)}}{=}&(1-\omega) \mb{E}\Big\|\mv\Theta^{(t+\Myfrac{1}{2})}-\eta^{(t+1)}\hat{\nabla}F^{(t+1)}+\zeta^{(t)}\hat{\mv\Theta}^{(t+1)}(\mv W-\mv I)
		-\hat{\mv\Theta}^{(t+1)}+\zeta^{(t)}\omega\sum_{\tau=0}^t(\mv A^{(\tau)})^T\tilde{\mv N}^{(\tau)}\Big\|_F^2\\
		\stackrel{{(d)}}{\le} & (1-\omega)\Big(\mb{E}\big\|\mv\Theta^{(t+\Myfrac{1}{2})}-\eta^{(t+1)}\hat{\nabla}F^{(t+1)}+\zeta^{(t)}\hat{\mv\Theta}^{(t+1)}(\mv W-\mv I)
		-\hat{\mv\Theta}^{(t+1)}\big\|_F^2+\\
		&\mb{E}\big\|\zeta^{(t)}\omega\sum_{\tau=0}^t(\mv A^{(\tau)})^T\tilde{\mv N}^{(\tau)}\big\|_F^2\Big)\\
		\stackrel{{(e)}}{\le} & (1-\omega)\Big((1+\alpha_2^{-1})\beta^2(\zeta^{(t)})^2
		\mb {E}\|\mv\Theta^{(t+\Myfrac{1}{2})}-\eta^{(t+1)}\hat{\nabla}F^{(t+1)}-\bar{\mv\Theta}^{(t+\Myfrac{1}{2})}\|_F^2+\\
		& (1+\alpha_2)\left(1+\beta\zeta^{(t)}\right)^2\mb {E}\|\mv\Theta^{(t+\Myfrac{1}{2})}-\eta^{(t+1)}\hat{\nabla}F^{(t+1)}-\hat{\mv\Theta}^{(t+1)}\|_F^2+(\zeta^{(t)})^2\omega^2d\sum_{\tau=0}^t\tilde N_0^{(\tau)}\Big),
	\end{align*}
	where \((a)\) is based on \eqref{eq:standard compression operator}; \((b)\) executes one SGD step (cf.~\eqref{eq:local updates}); \((c)\) is due to the analog consensus update (cf.~\eqref{eq:recast consensus updates for analog scheme}); \((d)\) is as a results of zero-mean AWGN; and \((e)\) follows \cite[Lemma 18]{koloskova19decentralized} combining with the noise variance as derived similarly in the proof for Lemma \ref{lemma:variant of Lemma 18}.
\end{IEEEproof}

We elaborate on non-trivial modifications made based on \cite[Appendix C]{koloskova19decentralized} in the sequel.\footnote{We will omit the dependence of functions on $\delta$ and $\omega$ but the iteration index $t$ throughout the appendices, as long as it does not cause any ambiguity in the context.} First, we need to modify two auxiliary functions to be adaptive as below:
\begin{align}
	\eta_1(\zeta^{(t)})= & (1+\alpha_1)(1-\delta\zeta^{(t)})^2+(1+\alpha_2^{-1})(1-\omega)\beta^2(\zeta^{(t)})^2, \label{eq:eta_1(t)}\\
	\xi_1(\zeta^{(t)}) = & (1+\alpha_1^{-1})\beta^2(\zeta^{(t)})^2+(1+\alpha_2)(1-\omega)\left(1+\beta\zeta^{(t)}\right)^2, \label{eq:xi_1(t)}
\end{align} and introduce another auxiliary function defined as
\begin{align}
	\eta_2(x)=\left(\frac{\delta^2}{4}+\frac{2}{\omega}\beta^2\right)x^2-\delta x+1=1-\tilde p(x), \label{eq:eta_2(x)}
\end{align} where 
\begin{align}
	\tilde p(x)=\delta x-\left(\frac{\delta^2}{4}+\frac{2}{\omega}\beta^2\right)x^2. \label{eq:tilde p(x)}
\end{align}
According to the definition of function \(\eta_2(x)\), which is a quadratic convex function decreasing over \([0,x^\ast]\) with \(x^\ast=\min_x\eta_2(x)\), it follows that \(\tilde p(x)\) increases over \([0,x^\ast]\).

Next, we substitute \(\zeta^{(t)}=\tfrac{\zeta_0}{\sqrt[\leftroot{-4}\uproot{1}4]{\tilde N_0}\Myfrac{t}{a^\prime}+1}\) for \(x\) in \(\tilde p(x)\) to obtain function \(\tilde p^{(t)}:\mb{R}_{+}\mapsto\mb{R}\) in \eqref{eq:tilde p(t)}. Note that since $\zeta_0$ given in Theorem \ref{theorem:optimality gap for digital scheme} is given by \(\zeta_0=\lambda^\prime x^\ast\) for a specific choice \(\lambda^\prime=\tfrac{8\beta^2+\delta^2\omega}{2(16\delta+\delta^2+4\beta^2+2\delta\beta^2-8\delta\omega)}\in(0,1]\) (see \cite[(20), (24)]{koloskova19decentralized} for detail), it follows that \(\zeta^{(t)}\le\zeta^{(0)}=\zeta_0\le x^\ast\). As we establish the facts that \(\zeta^{(t)}\in[0,x^\ast]\) decreases over \(t\ge 0\) and \(\tilde p(x)\) increases over \([0,x^\ast]\), \(\tilde p^{(t)}\) turns out to be decreasing over \(t\ge 0\) according to the composition of monotonic functions. Furthermore, based on the relation of \(\eta_1(x)\le\eta_2(x)\) derived in \cite[Apppendix C]{koloskova19decentralized}, which holds for any \(x\) and \(\omega>0\), since 
\begin{align}
	\eta_1(\zeta_0)\le \eta_2(\zeta_0)=1 - \tilde p(\zeta_0)\stackrel{\eqref{eq:tilde p(t)}}{=}1-\tilde p^{(0)}\kern-4pt\stackrel{\mbox{\cite[(24)]{koloskova19decentralized}}}{\le}\kern-4pt1-p(\delta,\omega), 
\end{align} it implies that \(p(\delta,\omega)\le\tilde p^{(0)}\). 

In addition, we have
\begin{align}
	\eta_1(\zeta^{(t)})\le \eta_2(\zeta^{(t)})=1 - \tilde p(\zeta^{(t)})\stackrel{\eqref{eq:tilde p(t)}}{=}1-\tilde p^{(t)}, \label{eq:upper-bound for eta_1(t)}
\end{align} and 
\begin{align}
	\xi_1(\zeta^{(t)})\stackrel{(a)}{\le}\xi_1(\zeta_0)\kern-4pt\stackrel{\mbox{\cite[(26)]{koloskova19decentralized}}}{\le}\kern-4pt1-p(\delta,\omega), \label{eq:upper-bound for xi_1(t)}
\end{align} where \((a)\) is due to the fact that function \(\xi_1(\zeta^{(t)})\) increases over \(\zeta^{(t)}>0\) and \(\zeta^{(t)}\le\zeta_0\). Finally, combining \eqref{eq:upper-bound for eta_1(t)} and \eqref{eq:upper-bound for xi_1(t)}, we arrive at the following variant of \cite[(21)]{koloskova19decentralized}
\begin{align}
	\max\left\{\eta_1(\zeta^{(t)}), \xi_1(\zeta^{(t)})\right\}\le
	 1-\min\left\{\tilde p^{(t)}, p(\delta,\omega)\right\}\stackrel{\eqref{eq:p(t)}}{=}1-p^{(t)}. \label{eq:maximum of eta_1(t) and xi_1(t)}
\end{align}

Now, we are ready to derive the recursive upper-bound given by \eqref{eq:recursive upper bound for the error sequence}. The following inequalities are 
frequently recalled in the sequel. For given matrices \(\mv X\) and \(\mv Y\) of the same size, and a square matrix \(\mv Z\), 
\begin{align}
	\|\mv X+\mv Y\|_F^2\le(1+\alpha)\|\mv X\|_F^2+(1+\alpha^{-1})\|\mv Y\|_F^2,\; \forall\alpha>0, \label{eq:basic inequality for add}
\end{align} and 
\begin{align}
	\|\mv X\mv Z\|_F\le\|\mv X\|_F\|\mv Z\|_2. \label{eq:basic inequality for multi}
\end{align}

By applying one SGD step (cf.~\eqref{eq:local updates}), it follows that, for any given \(\alpha_3>0\), 
\begin{align}
	\mb {E}\|\mv\Theta^{(t+\Myfrac{1}{2})}-\bar{\mv\Theta}^{(t+\Myfrac{1}{2})}\|_F^2 & = \mb {E}\Big\|(\mv\Theta^{(t)}-\bar{\mv\Theta}^{(t)})-\eta^{(t)}\hat{\nabla}F^{(t)}\Big(\mv I-\frac{\mv 1\mv 1^T}{K}\Big)\Big\|_F^2\notag\\
	&\kern-10pt\stackrel{\eqref{eq:basic inequality for add},\eqref{eq:basic inequality for multi}}{\le}\kern-10pt  (1+\alpha_3^{-1})\mb{E}\|\mv\Theta^{(t)}-\bar{\mv\Theta}^{(t)}\|_F^2+(1+\alpha_3)(\eta^{(t)})^2\mb{E}\|\hat{\nabla}F^{(t)}\|_F^2\Big\|\mv I-\frac{\mv 1\mv 1^T}{K}\Big\|\notag\\
	&\kern-4pt\stackrel{\eqref{assump:bounded norm}}{\le}\kern-4pt  (1+\alpha_3^{-1})\mb{E}\|\mv\Theta^{(t)}-\bar{\mv\Theta}^{(t)}\|_F^2+(1+\alpha_3)K(\eta^{(t)})^2G^2. \label{eq:SGD for upper-bound 1-1} 
\end{align} 
Similarly, for the same choice of \(\alpha_3\), it is also true that
\begin{align}
	& \mb{E}\|\mv\Theta^{(t+\Myfrac{1}{2})}-\eta^{(t+1)}\hat{\nabla}F^{(t+1)}-\bar{\mv\Theta}^{(t+\Myfrac{1}{2})}\|_F^2\notag \\
	\stackrel{\eqref{eq:basic inequality for add}}{\le}&  (1+\alpha_3^{-1})\mb{E}\|\mv\Theta^{(t)}-\bar{\mv\Theta}^{(t)}\|_F^2+(1+\alpha_3)\mb{E}\Big\|\eta^{(t)}\hat{\nabla}F^{(t)}\Big(\mv I-\frac{\mv 1\mv 1^T}{K}\Big)+\eta^{(t+1)}\hat{\nabla}F^{(t+1)}\Big\|_F^2\notag\\
	\stackrel{\eqref{assump:bounded norm}}{\le}&   (1+\alpha_3^{-1})\mb{E}\|\mv\Theta^{(t)}-\bar{\mv\Theta}^{(t)}\|_F^2+6(1+\alpha_3)K(\eta^{(t)})^2G^2, \label{eq:SGD for upper-bound 2-1} 
\end{align} and
\begin{multline}
	\mb {E}\|\mv\Theta^{(t+\Myfrac{1}{2})}-\eta^{(t+1)}\hat{\nabla}F^{(t+1)}-\hat{\mv\Theta}^{(t+1)}\|_F^2\stackrel{\eqref{eq:basic inequality for add},\eqref{assump:bounded norm}}{\le} (1+\alpha_3^{-1})\|\mv\Theta^{(t+\Myfrac{1}{2})}-\hat{\mv\Theta}^{(t+1)}\|_F^2+\\
	(1+\alpha_3)K(\eta^{(t)})^2G^2. \label{eq:upper-bound for SGD 2-2}
\end{multline}

Then, combining \eqref{eq:variant of Lemma 17} and \eqref{eq:variant of Lemma 18}, where relevant terms are substituted by \eqref{eq:SGD for upper-bound 1-1}-\eqref{eq:upper-bound for SGD 2-2}, after some manipulations, we have
\begin{multline}
	e^{(t+1)}\stackrel{\eqref{eq:def of error sequence}}{\le} \eta_1(\zeta^{(t)})(1+\alpha_3^{-1})\mb{E}\|\mv\Theta^{(t)}-\bar{\mv\Theta}^{(t)}\|_F^2+\xi_1(\zeta^{(t)})(1+\alpha_3^{-1})\|\mv\Theta^{(t+\Myfrac{1}{2})}-\hat{\mv\Theta}^{(t+1)}\|_F^2 +\\
	(\eta_1(\zeta^{(t)})+\xi_1(\zeta^{(t)}))(1+\alpha_3)(\eta^{(t)})^26KG^2+
	(\zeta^{(t)})^2(2-\omega)\omega^2d\sum_{\tau=0}^t\tilde N_0^{(\tau)}\\
	\le\underbrace{\max\left\{\eta_1(\zeta^{(t)}), \xi_1(\zeta^{(t)})\right\}(1+\alpha_3^{-1})e^{(t)}+12\max\left\{\eta_1(\zeta^{(t)}), \xi_1(\zeta^{(t)})\right\}(1+\alpha_3)(\eta^{(t)})^2KG^2}_{\rm part\ I}+\\
	\underbrace{(\zeta^{(t)})^2(2-\omega)\omega^2d\sum_{\tau=0}^t\tilde N_0^{(\tau)}}_{\rm part\ II}.\label{eq:immediate recursive upper bound for the error sequence}
\end{multline}
Note that by using \eqref{eq:maximum of eta_1(t) and xi_1(t)} and choosing \(\alpha_1=\tfrac{\delta\zeta^{(t)}}{2}\), \(\alpha_2=\tfrac{\omega}{2}\) and \(\alpha_3=\tfrac{p^{(t)}}{2}\) \cite[(20), Lemma 21]{koloskova19decentralized}, part I in \eqref{eq:immediate recursive upper bound for the error sequence} proves to be \((1-\tfrac{p^{(t)}}{2})e^{(t)}+\tfrac{2}{p^{(t)}}(\eta^{(t)})^212KG^2\). Furthermore, looking into part II, we have
\begin{align}
	(\zeta^{(t)})^2\sum_{\tau=0}^t\tilde N_0^{(\tau)}\stackrel{(a)}{\le} & \frac{(\zeta^{(t)})^2}{p^{(t)}}p^{(t)}t\tilde N_0
	\, \stackrel{(b)}{\le} \frac{(\zeta^{(t)})^2}{p^{(t)}}\tilde p^{(t)}\Big(\sqrt[\leftroot{-2}\uproot{4}4]{\tilde N_0}\Myfrac{t}{a^\prime}\Big)a^\prime\tilde N_0^{\frac{3}{4}}\notag\\
	\stackrel{(c)}{\le} & \frac{\delta\zeta_0}{p^{(t)}}(\zeta^{(t)})^2\frac{1}{\sqrt[\leftroot{-2}\uproot{1}4]{\tilde N_0}\Myfrac{t}{a^\prime}+1}\Big(\sqrt[\leftroot{-2}\uproot{4}4]{\tilde N_0}\Myfrac{t}{a^\prime}+1\Big)a^\prime\tilde N_0^{\frac{3}{4}}
	\stackrel{(d)}{=}\,  \frac{\delta\zeta_0}{p^{(t)}}\left(\frac{\zeta_0\Myfrac{a^\prime}{\sqrt[\leftroot{-2}\uproot{1}4]{\tilde N_0}}}{t+\Myfrac{a^\prime}{\sqrt[\leftroot{-2}\uproot{1}4]{\tilde N_0}}}\right)^2a^\prime\tilde N_0^{\frac{3}{4}}\notag\\
	= &  \frac{1}{p^{(t)}}\left(\frac{\mu}{3.25}\right)^2\left(\frac{\Myfrac{3.25}{\mu}}{t+\Myfrac{a^\prime}{\sqrt[\leftroot{-2}\uproot{1}4]{\tilde N_0}}}\right)^2\delta(\zeta_0a^{\prime})^3\sqrt[\leftroot{-2}\uproot{4}4]{\tilde N_0}\notag\\ \stackrel{(e)}{\le} & \frac{1}{p^{(t)}}\left(\frac{\mu}{3.25}\right)^2(\eta^{(t)})^2\delta(\zeta_0a^{\prime})^3\sqrt[\leftroot{-2}\uproot{4}4]{\tilde N_0}, \label{eq:upper-bound for recast part II}
\end{align} 
where $(a)$ is due to \(\tilde N_{0,T}=\max_{t\in\{0,\ldots,T-1\}}\{\tilde N_0^{(t)}\}\); $(b)$ is as a result of \(p^{(t)}\le\tilde p^{(t)}\) (cf.~\eqref{eq:p(t)}); $(c)$ is because of \(\tilde p^{(t)}\le\tfrac{\delta\zeta_0}{\sqrt[\leftroot{-4}\uproot{1}4]{\tilde N_0}\Myfrac{t}{a^\prime}+1}\) (cf.~\eqref{eq:tilde p(t)}); $(d)$ is by definition of   \(\zeta^{(t)}=\tfrac{\zeta_0}{\sqrt[\leftroot{-4}\uproot{1}4]{\tilde N_0}\Myfrac{t}{a^\prime}+1}\); and $(e)$ results from \(\eta^{(t)}=\tfrac{3.25}{\mu}\tfrac{1}{t+a}\) with \(a<\Myfrac{a^\prime}{\sqrt[\leftroot{-4}\uproot{4}4]{\tilde N_0}}\).

Finally, plugging the above results for part I and part II into \eqref{eq:immediate recursive upper bound for the error sequence}, we complete the proof for Lemma \ref{lemma:recursive upper-bound for the error sequence}.

\section{Proof of Theorem~\ref{theorem:optimality gap for analog scheme}}\label{appendix:proof of optimality gap for analog scheme}
Since \(\tilde p^{(t)}\) is decreasing over \(t\ge0\), by definition of \(p^{(t)}(\delta,\omega)\) in \eqref{eq:p(t)}, \(p^{(t)}(\delta,\omega)\) proves to be non-increasing over \(t\ge0\). Hence, \(p^{(t)}(\delta,\omega)\ge p^{(T)}(\delta,\omega)\), and we can further upper bound the RHS of \eqref{eq:recursive upper bound for the error sequence} by replacing \(p^{(t)}(\delta,\omega)\) with \(p^{(T)}(\delta,\omega)\). Based on Lemma \ref{lemma:recursive upper-bound for the error sequence}, we have the exact upper bound for \(e^{(t)}\) given by \cite[Lemma 22]{koloskova19decentralized}
\begin{align}
e^{(t)}\le\frac{10}{(p^{(T)}(\delta,\omega))^2}(\eta^{(t)})^2\left(24KG^2+A(\delta,\omega)\sqrt[\leftroot{-2}\uproot{4}4]{\tilde N_0}\right). \label{eq:exact upper bound for the total error}
\end{align}
Then, by definition of $e^{(t)}$ (cf.~\eqref{eq:def of error sequence}), we have 
\begin{align}
\sum\limits_{i\in\mc{V}}\mb{E}\|\bar{\mv\theta}^{(t)}-\mv\theta_i^{(t)}\|^2\le \frac{10}{(p^{(T)}(\delta,\omega))^2}(\eta^{(t)})^2\left(24KG^2+A(\delta,\omega)\sqrt[\leftroot{-2}\uproot{4}4]{\tilde N_0}\right). \label{eq:exact upper bound for the consensus error}
\end{align}

Next, to prove Theorem \ref{theorem:optimality gap for analog scheme}, we need to revisit \cite[Lemma 20]{koloskova19decentralized} as follows. 
\begin{align*} 
\mb{E}\|\bar{\mv\theta}^{(t+1)}-\mv\theta^\ast\|^2  & \stackrel{\eqref{eq:local updates},\ \eqref{eq:model average for analog scheme}}{=} \mb{E}\Big\|\bar{\mv\theta}^{(t)}-\eta^{(t)}\frac{1}{K}\sum_{i\in\mc{V}}\hat\nabla f_i(\mv\theta_i^{(t)})+\zeta^{(t)}\frac{1}{K}\sum_{i\in\mc{V}}\sum_{\tau=0}^t\frac{m}{d}(\mv A^{(\tau)})^T\tilde{\mv n}_i^{(\tau)}-\mv\theta^\ast\Big\|^2\notag\\
 & = \underbrace{\mb{E}\Big\|\bar{\mv\theta}^{(t)}-\eta^{(t)}\frac{1}{K}\sum_{i\in\mc{V}}\hat\nabla f_i(\mv\theta_i^{(t)})-\mv\theta^\ast\Big\|^2}_{\rm part\ I}+\underbrace{\mb{E}\Big\|\zeta^{(t)}\frac{1}{K}\sum_{i\in\mc{V}}\sum_{\tau=0}^t\frac{m}{d}(\mv A^{(\tau)})^T\tilde{\mv n}_i^{(\tau)}\Big\|^2}_{\rm part\ II}.
\end{align*}
As for Part I, \cite[Lemma 20]{koloskova19decentralized} can be directly applied, and for part II, we need to relate it to learning rate \(\eta^{(t)}\) as
\begin{align}
& \mb{E}\Big\|\zeta^{(t)}\frac{1}{K}\sum_{i\in\mc{V}}\sum_{\tau=0}^t\frac{m}{d}(\mv A^{(\tau)})^T\tilde{\mv n}_i^{(\tau)}\Big\|^2 = \frac{1}{K^2}\frac{m^2}{d^2}(\zeta^{(t)})^2\sum_{\tau=0}^t\sum_{i\in\mc{V}}\mb{E}\|(\mv A^{(\tau)})^T\tilde{\mv n}_i^{(\tau)}\|^2\notag\\
\stackrel{(a)}{=} &\frac{1}{K^2}\frac{m^2}{d}(\zeta^{(t)})^2\sum_{\tau=0}^t\tilde N_0^{(\tau)}
\stackrel{\eqref{eq:the maximum noise variance}}{\le} \frac{1}{K^2}\frac{m^2}{d}(\zeta^{(t)})^2t\tilde N_{0,T}\le\notag\\
& {\kern-14pt} \frac{1}{K^2}\frac{m^2}{d}\zeta^{(t)}\frac{\zeta_0(\delta,\omega)}{\sqrt[\leftroot{-2}\uproot{1}4]{\tilde N_0}\Myfrac{t}{a^\prime}+1}\Big(\sqrt[\leftroot{-2}\uproot{4}4]{\tilde N_{0,T}}\Myfrac{t}{a^\prime}+1\Big)a^\prime\tilde N_{0,T}^{\frac{3}{4}}
=  \frac{1}{K^2}\frac{m^2}{d}\frac{\mu}{3.25}\frac{\Myfrac{3.25}{\mu}}{t+\Myfrac{a^\prime}{\sqrt[\leftroot{-2}\uproot{1}4]{\tilde N_0}}}(\zeta_0(\delta,\omega)a^\prime)^2\sqrt{\tilde N_{0,T}}\notag\\
\stackrel{(b)}{\le} &  \frac{1}{K^2}\omega^2d\frac{\mu}{3.25}\eta^{(t)}(\zeta_0(\delta,\omega)a^\prime)^2\sqrt{\tilde N_{0,T}}, \label{eq:upper bound for biased noise term}
\end{align} where $(a)$ follows from \(\mv A^{(\tau)}(\mv A^{(\tau)})^T=\tfrac{d}{m}\mv I\) and  \(\mb{E}\|\tilde{\mv n}_i^{(\tau)}\|^2=m\tilde N_{0i}^{(\tau)}\); and $(b)$ is due to \(a<\Myfrac{a^\prime}{\sqrt[\leftroot{-4}\uproot{4}4]{\tilde N_0}}\) and \(\omega=\Myfrac{m}{d}\), as well as definition of \(\eta^{(t)}=\tfrac{3.25}{\mu}\tfrac{1}{t+a}\). With part II replaced by the RHS of \eqref{eq:upper bound for biased noise term}, we obtain the following lemma.
\begin{lemma}[Variant of {\cite[Lemma 20]{koloskova19decentralized}}] Denoting the optimal solution to \(\mathrm{(P0)}\) by \(\mv\theta^\ast\) and the corresponding objective value \(F(\mv\theta^\ast)\) by \(F^\ast\), the average of iterates \(\bar{\mv\theta}^{(t+1)}\) satisfies
	\begin{multline}
	\mb{E}\|\bar{\mv\theta}^{(t+1)}-\mv\theta^\ast\|^2\le(1-\eta^{(t)}\mu)\mb{E}\|\bar{\mv\theta}^{(t)}-\mv\theta^\ast\|^2+\eta^{(t)}\frac{1}{K^2}\omega^2d\frac{\mu}{3.25}(\zeta_0(\delta,\omega)a^\prime)^2\sqrt{\tilde N_{0,T}}+(\eta^{(t)})^2\frac{\bar\sigma^2}{K}\\
	+2\eta^{(t)}(2\eta^{(t)}L-1)(\mb{E}[F(\bar{\mv\theta}^{(t)})]-F^\ast)+
	\eta^{(t)}\frac{2\eta^{(t)}L^2+L}{K}\sum\limits_{i\in\mc{V}}\mb{E}\|\bar{\mv\theta}^{(t)}-\mv\theta_i^{(t)}\|^2. \label{eq:recursive upper bound for the distance to the optimal model parameter}
	\end{multline}\label{lemma:recursive upper bound for the distance to the optimal model parameter}	
\end{lemma} 

Since \(\eta^{(t)}\le\eta^{(0)}=\tfrac{3.25}{\mu a}\), and \(a\ge\tfrac{13L}{\mu}\), it follows that \(\eta^{(t)}\le\tfrac{1}{4L}\) thus leading to \(2L\eta^{(t)}-1\le-0.5\) and \(2\eta^{(t)}L^2+L\le 1.5L\). With this fact and the upper bound on the consensus error given by \eqref{eq:exact upper bound for the consensus error}, \eqref{eq:recursive upper bound for the distance to the optimal model parameter} implies that
\begin{multline}
	v_e^{(t+1)}\le(1-\eta^{(t)}\mu)v_e^{(t)}-\eta^{(t)}f_e^{(t)}+\eta^{(t)}\frac{1}{K^2}\omega^2d\frac{\mu}{3.25}(\zeta_0(\delta,\omega)a^\prime)^2\sqrt{\tilde N_{0,T}}+\\
	(\eta^{(t)})^2\frac{\bar\sigma^2}{K}+(\eta^{(t)})^3\frac{15L}{(p^{(T)}(\delta,\omega))^2}\left(24G^2+\frac{A(\delta,\omega)}{K}\sqrt[\leftroot{-2}\uproot{4}4]{\tilde N_{0,T}}\right), \label{eq:recursive upper bound for the distance to the optimal model parameter in polynomials of learning rate}
\end{multline} where \(v_e^{(t)}=\mb{E}\|\bar{\mv\theta}^{(t)}-\mv\theta^\ast\|^2\) and \(f_e^{(t)}=\mb{E}[F(\bar{\mv\theta}^{(t)})]-F^\ast\) measure, on average over RLC, the distance to the optimal solution and the optimality gap to to the objective value for problem (P0), respectively. Note that the standard result in \cite[Lemma 3.3]{stich18sparsified} is not applicable to \eqref{eq:recursive upper bound for the distance to the optimal model parameter in polynomials of learning rate} to characterize the optimality gap due to the absence of linear terms w.r.t \(\eta^{(t)}\). To capture the performance of the optimality-gap sequence \(\{f_e^{(t)}\}\), we need the following lemma.
\begin{lemma}[Variant of {\cite[Lemma 3.3]{stich18sparsified}}]
	For non-negative sequences \(\{v_e^{(t)}\}\) and \(\{f_e^{(t)}\}\), \(\eta^{(t)}=\tfrac{3.25}{\mu}\tfrac{1}{t+a}\) with \(\mu>0\) and \(a>1\), and constants \(A,\ B,\ C\ge 0\), where \(t=0,\ldots,T-1\), such that
	\begin{align}
	v_e^{(t+1)}\le (1-\eta^{(t)}\mu)v_e^{(t)}+\eta^{(t)}A+(\eta^{(t)})^2B+(\eta^{(t)})^3C-\eta^{(t)}f_e^{(t)}, \label{eq:sufficient condition for Stich's lemma}
	\end{align} we have
	\begin{align}
	\frac{1}{S_T}\sum_{t=0}^{T-1}w^{(t)}f_e^{(t)}\le\frac{\mu}{3.25}\frac{a^3-3.25a^2}{S_T}v_e^{(0)}+A+\frac{1.625(2a+T)T}{\mu S_T}B+\frac{3.25^2T}{\mu^2S_T}C, \label{eq:Stich's lemma} 
	\end{align} where \(w^{(t)}=(a+t)^2\) and \(S_T=\sum_{t=0}^{T-1}w^{(t)}\). \label{lemma:Stich's lemma}
\end{lemma}
\begin{IEEEproof}
	Following similar procedures as that for proving \cite[Lemma 3.3]{stich18sparsified}, first, multiplying \(\tfrac{w^{(t)}}{\eta^{(t)}}\) with both sides of \eqref{eq:sufficient condition for Stich's lemma}, we have 
	\begin{align}
		v_e^{(t+1)}\frac{w^{(t)}}{\eta^{(t)}}\le (1-\eta^{(t)}\mu)\frac{w^{(t)}}{\eta^{(t)}}v_e^{(t)}+w^{(t)}A+w^{(t)}\eta^{(t)}B+w^{(t)}(\eta^{(t)})^2C-w^{(t)}f_e^{(t)}. \label{eq:equivalently sufficient condition for Stich's lemma}
	\end{align}
	To obtain the same relation \((1-\eta^{(t)}\mu)\tfrac{w^{(t)}}{\eta^{(t)}}\le\tfrac{w^{(t-1)}}{\eta^{(t-1)}}\) as in the original proof, it is equivalent to have \((c-3)(a+t)^2+3(a+t)-1\ge 0\) for all $t$. It is thus sufficient to have \({(c-3)(a+t)^2+3(a+t)-1}|{}_{t=0}=(c-3)a^2+3a-1\ge 0\) for a choice of \(c>3\), which is satisfied by, e.g., \(c=3.25\) and any parameter \(a\ge \tfrac{13L}{\mu}\). ( By definition, \(L\ge\mu\) and thus \(a\ge 13\).) Next, by letting \(t=T-1\) and recursively applying \eqref{eq:equivalently sufficient condition for Stich's lemma}, it follows that
	\begin{align}
		\sum_{t=0}^{T-1}w^{(t)}f_e^{(t)}\le(1-\eta^{(0)}\mu) \frac{w^{(0)}}{\eta^{(0)}}v_e^{(0)}+\sum_{t=0}^{T-1}w^{(t)}A+\sum_{t=0}^{T-1}w^{(t)}\eta^{(t)}B+\sum_{t=0}^{T-1}w^{(t)}(\eta^{(t)})^2C. \label{eq:exact expression for the sufficient condition of Stich's lemma}
	\end{align}
	As a result, plugging \(\eta^{(0)}=\tfrac{3.25}{\mu a}\) and dividing both sides of \eqref{eq:exact expression for the sufficient condition of Stich's lemma} by \(S_T\), we obtain \eqref{eq:Stich's lemma}. 
\end{IEEEproof}

According to Lemma \ref{lemma:Stich's lemma}, for constants  \(A=\tfrac{1}{K^2}\omega^2d\frac{\mu}{3.25}(\zeta_0(\delta,\omega)a^\prime)^2\sqrt{\tilde N_{0,T}}\), \(B=\tfrac{\bar\sigma^2}{K}\) and \(C=\frac{15L}{(p^{(T)}(\delta,\omega))^2}(24G^2+\sqrt[\leftroot{-2}\uproot{4}4]{\tilde N_{0,T}}\Myfrac{A(\delta,\omega)}{K})\), as well as \(f_e^{(t)}=\mb{E}[F(\bar{\mv\theta}^{(t)})]-F^\ast\),  \eqref{eq:recursive upper bound for the distance to the optimal model parameter in polynomials of learning rate} implies that
\begin{multline}
\frac{1}{S_T}\sum_{t=0}^{T-1}w^{(t)}\mb{E}[F(\bar{\mv\theta}^{(t)})]-F^\ast
\le\frac{\mu}{3.25}\frac{a^3-3.25a^2}{S_T}v_e^{(0)}+
\frac{158.45\left(24G^2+\frac{A(\delta,\omega)}{K}\sqrt[\leftroot{-2}\uproot{4}4]{\tilde N_{0,T}}\right)LT}{\mu^2(p^{(T)}(\delta,\omega))^2 S_T}+\\
\frac{1.625T(T+2a)}{\mu S_T}\frac{\bar\sigma^2}{K}+\frac{1}{K^2}C(\delta,\omega)\sqrt{\tilde N_{0,T}}. \label{eq:optimality gap forf analog scheme}
\end{multline} 
Combined \eqref{eq:optimality gap forf analog scheme} with \(F(\tilde{\mv\theta}_T)=F(\tfrac{1}{S_T}\sum_{t=0}^{T-1}w^{(t)}\bar{\mv\theta}^{(t)})\le\tfrac{1}{S_T}\sum_{t=0}^{T-1}w^{(t)}F(\bar{\mv\theta}^{(t)})\), which is due to Jensen's inequality, Theorem \ref{theorem:optimality gap for analog scheme} is thus proved.

\bibliographystyle{IEEEtran}
\bibliography{DL_ref}
\end{document}